# Next Generation Radiogenomics Sequencing for Prediction of EGFR and KRAS Mutation Status in NSCLC Patients Using Multimodal Imaging and Machine Learning Approaches

Isaac Shiri[1,2], Hassan Maleki[1,3], Ghasem Hajianfar[1,2], Hamid Abdollahi[1,4], Saeed Ashrafinia[5,6], Mathieu Hatt[7], Mehrdad Oveisi[1,8], Arman Rahmim[5,9,10]

1. Department of Biomedical and Health Informatics, Rajaie Cardiovascular Medical and Research Center, Iran University of Medical Science, Tehran, Iran
2. Research Center for Molecular and Cellular Imaging, Tehran University of Medical Sciences, Tehran, Iran
3. Department of Computer Science and Engineering, Shahid Beheshti University, Tehran, Iran
4. Department of Medical Physics, School of Medicine, Iran University of Medical Science, Tehran, Iran
5. Department of Radiology and Radiological Science, Johns Hopkins University, Baltimore MD, USA
6. Department of Electrical and Computer Engineering, Johns Hopkins University, Baltimore MD, USA
7. INSERM, UMR 1101, LaTIM, Univ Brest, F-29238, Brest, France
8. Department of Computer Science, University of British Columbia, Vancouver BC, Canada
9. Departments of Radiology and Physics & Astronomy, University of British Columbia, Vancouver BC, Canada
10. Department of Integrative Oncology, BC Cancer Research Centre, Vancouver BC, Canada

**First Author:** Isaac Shiri, MSc

Department of Biomedical and Health Informatics, Rajaie Cardiovascular Medical and Research Center, Iran University of Medical Science, Tehran, Iran
**Phone:** +989190245882
**Email:** Isaac.sh92@gmail.com

**Corresponding Author:** Arman Rahmim, PhD

Associate Professor of Radiology and Physics, University of British Columbia
Senior Scientist & Provincial Medical Imaging Physicist, BC Cancer Agency
BC Cancer Research Center
675 West 10th Ave
Office 5-114
Vancouver, BC, V5Z 1L3
http://rahmimlab.com
**Phone:** 604-675-8262
**Email:** arman.rahmim@ubc.ca

**Conflicts of Interest:** The authors declare no potential conflicts of interest.

# Abstract

## Purpose

Considerable progress has been made in assessment and management of NSCLC patients based on mutation status in the EGFR and KRAS. At the same time, NSCLC management through KRAS and EGFR mutation profiling faces some challenges. In the present work, we aimed to evaluate a comprehensive radiomics framework that enabled prediction of EGFR and KRAS mutation status in NSCLC cancer patients based on low dose CT, diagnostic CT, PET modalities radiomic features and machine learning (ML) algorithms.

## Methods

Our study involved NSCLC cancer patient with 186 PET, and 175 low dose CT and CTD images. More than twenty thousand radiomic features from different image-feature sets were extracted. Conventional clinically used standard uptake value (SUV) parameters were also obtained for PET images. Feature value was normalized to obtain Z-scores, followed by student t-test students for comparison, high correlated features were eliminated and the False discovery rate (FDR) correction were performed and q-value were reported for univariate analysis. Six feature selection methods and twelve classifiers were used to predict gene status in patient. We performed 10-fold cross-validation for model tuning to improve robustness and all model evaluation was reported on independent validation sets (68 patients). The mean area under the receiver operator characteristic (ROC) curve (AUC) was obtained for performance evaluation.

## Results

The best predictive power of conventional PET parameters was achieved by SUVpeak (AUC: 0.69, P-value = 0.0002) and MTV (AUC: 0.55, P-value = 0.0011) for EGFR and KRAS, respectively. Univariate analysis of extracted radiomics features improved prediction power up to AUC: 75 (q-value: 0.003, Short Run Emphasis feature of GLRLM from LOG preprocessed image of PET with sigma value 1.5) and AUC: 0.71 (q-value 0.00005, The Large Dependence Low Gray Level Emphasis from GLDM in LOG preprocessed image of CTD sigma value 5) for EGFR and KRAS, respectively. Furthermore, the machine learning algorithm improved the perdition power up to AUC: 0.82 for EGFR (LOG preprocessed of PET image set with sigma 3 with VT feature selector and SGD classifier) and AUC: 0.83 for KRAS (CT image set with sigma 3.5 with SM feature selector and SGD classifier).

## Conclusion

Our findings demonstrated that non-invasive and reliable radiomics analysis can be successfully used to predict EGFR and KRAS mutation status in NSCLC patients. We demonstrated that radiomic features extracted from different image-feature sets could be used for EGFR and KRAS mutation status prediction in NSCLC patients, and showed that they have more predictive power than conventional imaging parameters.

**Key Words:** Radiogenomics, PET/CT, Machine Learning, NSCLC, KRAS, EGFR

# Introduction

Considerable progress has been made recently in the assessment and management of non-small cell lung cancer (NSCLC) based on mutation status in the epidermal growth factor receptor (EGFR) and Kirsten rat sarcoma viral oncogene (KRAS) genes (1). Ongoing studies on molecular cancer profiling have revealed that EGFR and KRAS are involved in the occurrence, development, invasion, and metastasis of NSCLC (2). Moreover, studies have indicated that mutations in KRAS and EGFR are considered as first lines for clinical decision making in NSCLC treatment and outcome improvement (3). Furthermore, recent studies have identified that NSCLC patients with mutant KRAS tumors fail to benefit from adjuvant chemotherapy, and their disease does not respond to EGFR inhibitors (4). In addition, it was found that outcomes and patterns of failure in genotypic subgroups of NSCLC patients, based on mutations in EGFR or KRAS, can inform the design of future trials integrating targeted therapies (5).

Although KRAS and EGFR mutation profiling is critical in NSCLC management, some studies have raised issues with this approach (6, 7). First, such mutation status captures only a small degree of tumor heterogeneity and does not provide a complete picture for the assessment of tumor characteristics. Secondly, this method depicts low repeatability and is not feasible for all cases (8). Furthermore, the method suffers from invasiveness and patients discomfort (9).

As a recently developed paradigm of advanced medical image quantification, radiomics has garnered significant interests given its cost-effectiveness and reliability to characterize tumor heterogeneity, and has enabled improved assessment of therapy response and prediction of molecular pathways (10-14). Accumulating evidence has identified several radiomic features extracted from computed tomography (CT), magnetic resonance (MR) or positron emission tomography (PET) images as highly correlated with genomic parameters in several cancers (8, 15-18). For NSCLC patients, radiomics studies have shown several CT image-features can predict mutation status in EGFR and KRAS (8). For example, features such as size, edge, lucency and homogeneity extracted from CT images could identify EGFR mutation status (19). Also, some diffusion-weighted (DW) MR image-intensity histogram features, including mean, skewness, and $10^{th}$ and $90^{th}$ percentiles, have been shown to predict EGFR mutation in lung adenocarcinoma (20). On the other hand, Velazquez *et al.* developed radiomics models based on CT image-features and clinical parameters to distinguish between $EGFR^-$ and $EGFR^+$, and $KRAS^+$ and $KRAS^-$ (21). Liu *et al.* also evaluated the ability of CT image-features to predict EGFR mutation status in 298 surgically-resected peripheral lung adenocarcinomas in an Asian cohort of patients, and built a high performance predictive model by using multiple logistic regression algorithms (19). Zhang *et al.* also developed a radiogenomic model based on CT image-features to predict EGFR and KRAS mutations in lung adenocarcinoma patients (22).

Incorporating radiomics for a prediction study requires a multi-step process that involves reliable statistical analyses, such as feature selection and classification, to reduce over-fitting and to build robust predictive or prognostic models (23). A number of machine learning (ML) algorithms can provide robust means to identify a subset of features to combine into a multi-parametric model (24). Although several ML algorithms, alone or in combination, have been used in radiomics analysis for feature selection and classification, there is no "one fits all" approach as performance of various ML workflows have been shown to depend on application and/or type of data (25-27) Previous studies have tested cross-combination of different ML approaches, and have suggested distinct ML algorithms that depict high performance for feature selection and classification (24, 25, 28).

In the present study, we aimed to evaluate a comprehensive multimodal (Diagnostic CT, low dose CT and PET modality) radiomics, univariate analysis and machine learning framework to predict EGFR and KRAS mutation status in NSCLC cancer patients.

# Material and methods

## Radiomics analysis

Our radiomics analysis included the following seven steps (Fig. 1).

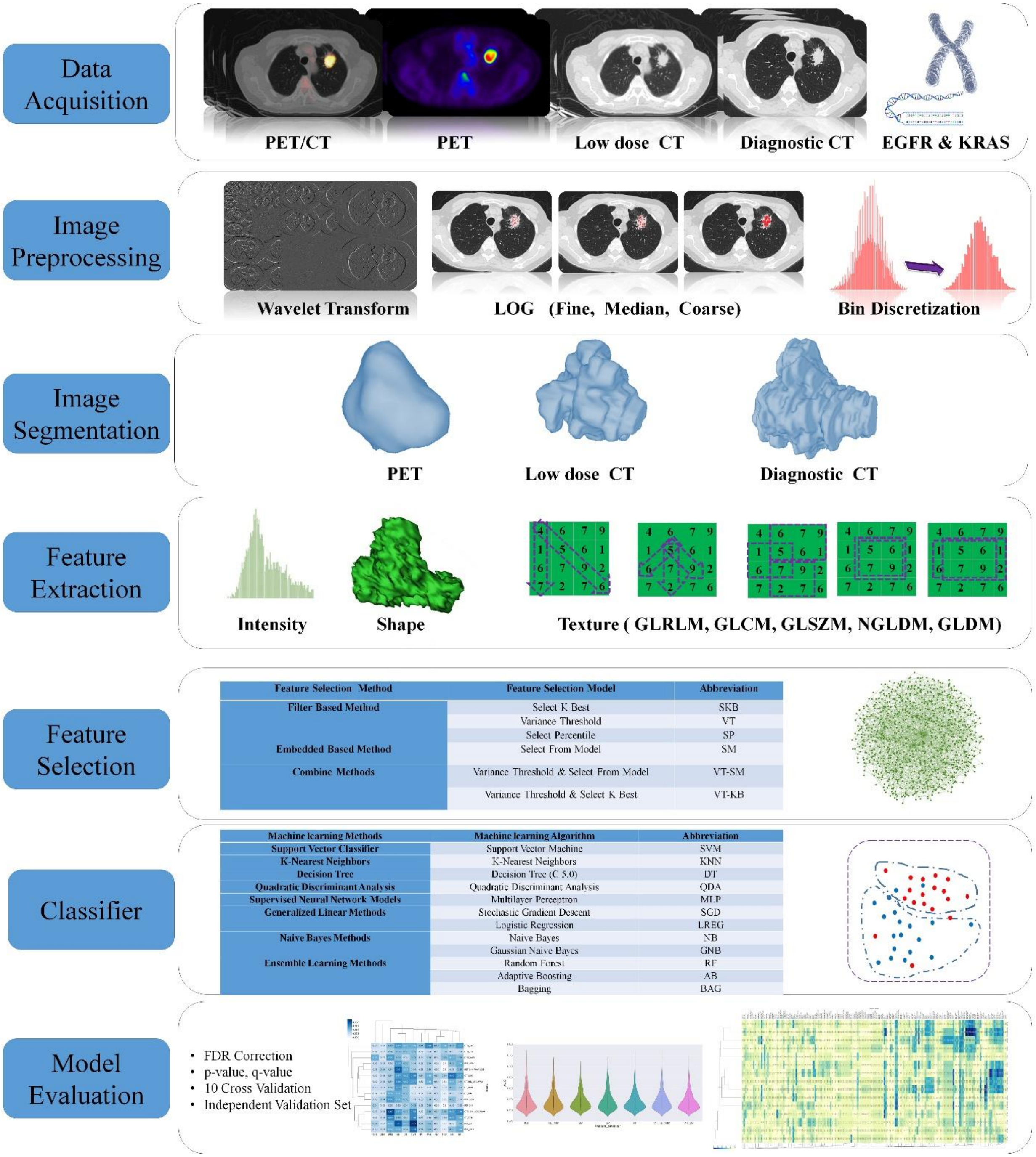

**Figure 1.** The radiogenomics framework employed in the present study

***1- Data collection***

Clinical characteristics of all patients were presented in Table 1.

**Table 1.** Clinical characterization of patients

| Characteristic | | PET | CT | CTD | Validation PET, CT, CTD |
|---|---|---|---|---|---|
| **Patient NO.** | | 186 | 175 | 175 | 68 |
| **Age, mean(SD)** | | 68.6 (9.11) | 69.2 (8.85) | 69.2 (8.85) | 69.1 (8.8) |
| **Weight (Kg), mean(SD)** | | 77.0 (18.5) | 77.8 (18.6) | 77.8 (18.6) | 77.6 (18.6) |
| **Sex (%)** | **Male** | 66.7 | 66.3 | 66.3 | 67.1 |
| | **Female** | 33.3 | 33.7 | 33.7 | 32.9 |
| **Smoking status (%)** | **Nonsmoker** | 21 | 21.7 | 21.7 | 21.5 |
| | **Former** | 61.3 | 61.7 | 61.7 | 61.0 |
| | **Current** | 17.7 | 16.5 | 16.5 | 17.3 |
| **Histology (%)** | **Adenocarcinoma** | 79.6 | 82.3 | 82.3 | 81.4 |
| | **Squamous cell carcinoma** | 18.3 | 16 | 16 | 16.8 |
| | **NOS (not otherwise specified)** | 2.2 | 1.7 | 1.7 | 1.8 |
| **T stage (%)** | **T1** | 36.1 | 38.2 | 38.2 | 37.1 |
| | **T2** | 30.6 | 28 | 28 | 29.3 |
| | **T3** | 11.3 | 9.1 | 9.1 | 9.6 |
| | **T4** | 3.8 | 2.9 | 2.9 | 3 |
| **N stage (%)** | **N0** | 66.7 | 63.4 | 63.4 | 65.3 |
| | **N1** | 8.1 | 6.9 | 6.9 | 7.2 |
| | **N2** | 9.7 | 9.7 | 9.7 | 10.2 |
| **M stage (%)** | **M0** | 82.3 | 78.3 | 78.3 | 80.8 |
| | **M1** | 2.2 | 1.7 | 1.7 | 1.8 |
| **AJCC Staging (%)** | **I** | 50 | 50.3 | 50.3 | 51.5 |
| | **II** | 19.4 | 16 | 16 | 16.8 |
| | **III** | 12.9 | 12 | 12 | 12.6 |
| | **IIII** | 2.2 | 1.7 | 1.7 | 1.8 |
| **Histo-pathological Grade (%)** | **G1** | 15.6 | 14.9 | 14.9 | 15 |
| | **G2** | 39.8 | 38.9 | 38.9 | 40.1 |
| | **G3** | 17.7 | 15.4 | 15.4 | 16.2 |
| | **other** | 11.3 | 10.8 | 10.8 | 11.4 |
| **Lymph vascular invasion (%)** | **Absent** | 68.3 | 67.4 | 67.4 | 69.5 |
| | **Present** | 11.3 | 10.3 | 10.3 | 10.8 |
| **EGFR mutation status (%)** | **Mutant** | 18.8 | 21.1 | 21.1 | 19.8 |
| | **Wild-type** | 61.8 | 62.9 | 62.9 | 63.5 |
| | **Not Available** | 18.8 | 16 | 16 | 16.8 |
| **KRAS mutation status (%)** | **Mutant** | 18.3 | 18.3 | 18.3 | 18.6 |
| | **Wild-type** | 61.3 | 65.1 | 65.1 | 64.1 |
| | **Not Available** | 20.4 | 16.6 | 16.6 | 17.4 |
| **PET SUV** | **SUVmax** | 7.5 (6.7) | | | 7.1 (6.3) |
| | **SULmax** | 6 .1(5.2) | | | 5.6 (4.9) |
| | **SUVpeak** | 9.6 (6.5) | | | 8.9 (6.3) |
| | **SULpeak** | 7.4 (4.9) | | | 6.8 (4.8) |
| **Volume mean(SD)** | | 29915 (48293) | | | 19225 (40185) |
| **Recurrence (%)** | | 17.7 | 17.7 | 17.7 | 17.4 |
| **Time to Last Follow-up, mean(SD)** | | 26.9 (21.7) | 27.0 (22.1) | 27.0 (22.1) | 26.8 (22.2) |
| **Death (%)** | | 21 | 20 | 20 | 20.4 |

### *1-1- Imaging*

This study was conducted on 211 NSCLC cancer patients with available imaging and genomic data. Imaging modalities, including diagnostic CT (CTD) and PET/CT (i.e., low-dose CT (CT) used for PET attenuation correction and the PET image) for all patients were obtained (29-32) . All images were acquired with two vendor of imaging system including; GE and SIEMENS. For CTD, 73% of data were acquired with GE Discovery CT750HD and 27% with SIEMENS. Slice Thickness for diagnostics were 0.625–3 mm (median: 1.5 mm) and an X-ray tube current of 124–699 mA (mean 220 mA) at 80–140 kVp (mean 120 kVp), and Spiral Pitch Factor 0.9-1. A GE Discovery D690 PET/CT and GE Discovery PET/CT scanner were used for PET/CT scanning. 18F-FDG Dose and uptake time were 138.90–572.25 MBq (mean: 309.26 MBq) and 23.08–128.90 min (mean: 66.58 minutes), respectively. Each bed position was 1–5-minute acquisition, and iterative Ordered Subset Expectation Maximization (OSEM) reconstruction was used for PET image reconstruction. Slice Thickness for low dose CT were set as 3-5mm, mA Range 50-400, KVP 120-140 and Spiral Pitch Factor 0.9-1. Due to certain problems including image noise, artifacts, absence of images for some sequences and image mis-segmentation, some patients were excluded, and finally 186 PET, as well as 175 low dose CT and CTD images were segmented.

### *1-2- Genomics*

Tumor samples were excised and cutted by surgeon with slice thickness of 3-5 mm and were frozen for 30 minutes. For single nucleotide mutation detection, SNaPshot technology based on dideoxy single-base extension of oligonucleotide primers was performed after multiplex polymerase chain reaction (PCR). To detect EGFR mutations, Exons including 18, 19, 20 and 21 were assessed. In addition, Exon 2 Positions 12 and 13 were examined with amino acid substitution for missense KRAS mutations.

## *2- Image pre-processing*

Prior to feature extraction, all images were pre-processed with interpolation to isotropic voxel spacing to be rotationally invariant, for feature extraction to allow comparison between image data from different samples and scanner and center. Re-sampling to 1×1×1, 1×1×3 and 3×3×3 were performed for CTD, CT and PET images respectively. Laplacian of Gaussian (LOG), wavelet decomposition (WAV) and discretized into 64 bins (BIN64) preprocessing were performed to generate different set of features. For LOG filter, different sigma values were used to extract fine, medium and coarse features; specifically, they ranged from 0.5 to 5 with 0.5 steps. Wavelet filtering, yields 8 decompositions per level (all possible combinations of applying either a High or a Low pass filter in each of the three dimensions including HHH, HHL, HLH, HLL, LHH, LHL, LLH and LLL). Preprocessing steps (including discretization, LOG and wavelet) were performed on all intensity, histogram and textural features.

## *3- Image segmentation*

All segmentations in PET images were performed using OSIRIX® (33). Lesions were delineated manually on PET imaged. CT images were segmented via automatic region growing using 3D-Slicer (34), and edited and verified by an experienced radiologist. In total we provide the 186 PET, and 175 low dose CT and CTD segmentation region.

## *4- Feature extraction*

In the next step, several features from different feature-classes were extracted. The classes include first order statistics (19 FOS features), shape-based (16 Shape features), gray level co-occurrence matrix (GLCM 23 features), gray level run length matrix (16 GLRLM features), gray level size zone matrix (16 GLSZM features), neighboring gray tone difference matrix (5 NGTDM features), and gray level dependence matrix (14 GLDM features). Image feature extraction was performed by the open-source python library PyRadiomics (35). Radiomics features calculated by this package are in compliance with feature definitions as described by the image biomarker standardization initiative (IBSI) that ensures the harmonization and reproducibility of calculated radiomic features, and thus facilitates the reproducibility of this study (35, 36). More detail on the feature classes is provided in Table 2. In addition to the radiomic features, we also extracted conventional clinical PET biomarkers including metabolic tumor volume (MTV), standard uptake values ($SUV_{max}$, $SUV_{peak}$), including normalization to lean body mass ($SUL_{max}$, $SUL_{peak}$).

**Table 2.** Radiomic features extracted from different modality

| First Order Statistics (FOS) |
|---|
| 1. Energy |
| 2. Total Energy |
| 3. Entropy |
| 4. Minimum |
| 5. 10th percentile |
| 6. 90th percentile |
| 7. Maximum |
| 8. Mean |
| 9. Median |
| 10. Interquartile Range |
| 11. Range |
| 12. Mean Absolute Deviation (MAD) |
| 13. Robust Mean Absolute Deviation (rMAD) |
| 14. Root Mean Squared (RMS) |
| 15. Standard Deviation |
| 16. Skewness |
| 17. Kurtosis |
| 18. Variance |
| 19. Uniformity |

| Gray Level Co-occurrence Matrix (GLCM) |
|---|
| 1. Autocorrelation |
| 2. Joint Average |
| 3. Cluster Prominence |
| 4. Cluster Shade |
| 5. Cluster Tendency |
| 6. Contrast |
| 7. Correlation |
| 8. Difference Average |
| 9. Difference Entropy |
| 10. Difference Variance |
| 11. Joint Energy |
| 12. Joint Entropy |
| 13. Informal Measure of Correlation (IMC) 1 |
| 14. Informal Measure of Correlation (IMC) 2 |
| 15. Inverse Difference Moment (IDM) |
| 16. Inverse Difference Moment Normalized (IDMN) |
| 17. Inverse Difference (ID) |
| 18. Inverse Difference Normalized (IDN) |
| 19. Inverse Variance |
| 20. Maximum Probability |
| 21. Sum Average |
| 22. Sum Entropy |
| 23. Sum of Squares |

| Gray Level Run Length Matrix (GLRLM) |
|---|
| 1. Short Run Emphasis (SRE) |
| 2. Long Run Emphasis (LRE) |
| 3. Gray Level Non-Uniformity (GLN) |
| 4. Gray Level Non-Uniformity Normalized (GLNN) |
| 5. Run Length Non-Uniformity (RLN) |
| 6. Run Length Non-Uniformity Normalized (RLNN) |
| 7. Run Percentage (RP) |
| 8. Gray Level Variance (GLV) |
| 9. Run Variance (RV) |
| 10. Run Entropy (RE) |
| 11. Low Gray Level Run Emphasis (LGLRE) |
| 12. High Gray Level Run Emphasis (HGLRE) |
| 13. Short Run Low Gray Level Emphasis (SRLGLE) |
| 14. Short Run High Gray Level Emphasis (SRHGLE) |
| 15. Long Run Low Gray Level Emphasis (LRLGLE) |
| 16. Long Run High Gray Level Emphasis (LRHGLE) |

| Gray Level Dependence Matrix (GLDM) |
|---|
| 1. Small Dependence Emphasis (SDE) |
| 2. Large Dependence Emphasis (LDE) |
| 3. Gray Level Non-Uniformity (GLN) |
| 4. Dependence Non-Uniformity (DN) |
| 5. Dependence Non-Uniformity Normalized (DNN) |
| 6. Gray Level Variance (GLV) |
| 7. Dependence Variance (DV) |
| 8. Dependence Entropy (DE) |
| 9. Low Gray Level Emphasis (LGLE) |
| 10. High Gray Level Emphasis (HGLE) |
| 11. Small Dependence Low Gray Level Emphasis (SDLGLE) |
| 12. Small Dependence High Gray Level Emphasis (SDHGLE) |
| 13. Large Dependence Low Gray Level Emphasis (LDLGLE) |
| 14. Large Dependence High Gray Level Emphasis (LDHGLE) |

| Shape and Morphological Features |
|---|
| 1. Volume |
| 2. Surface Area |
| 3. Surface Area to Volume ratio |
| 4. Sphericity |
| 5. Compactness 1 |
| 6. Compactness 2 |
| 7. Spherical Disproportion |
| 8. Maximum 3D diameter |
| 9. Maximum 2D diameter (Slice) |
| 10. Maximum 2D diameter (Column) |
| 11. Maximum 2D diameter (Row) |
| 12. Major Axis |
| 13. Minor Axis |
| 14. Least Axis |
| 15. Elongation |
| 16. Flatness |

| Gray Level Size Zone Matrix (GLSZM) |
|---|
| 1. Small Area Emphasis (SAE) |
| 2. Large Area Emphasis (LAE) |
| 3. Gray Level Non-Uniformity (GLN) |
| 4. Gray Level Non-Uniformity Normalized (GLNN) |
| 5. Size-Zone Non-Uniformity (SZN) |
| 6. Size-Zone Non-Uniformity Normalized (SZNN) |
| 7. Zone Percentage (ZP) |
| 8. Gray Level Variance (GLV) |
| 9. Zone Variance (ZV) |
| 10. Zone Entropy (ZE) |
| 11. Low Gray Level Zone Emphasis (LGLZE) |
| 12. High Gray Level Zone Emphasis (HGLZE) |
| 13. Small Area Low Gray Level Emphasis (SALGLE) |
| 14. Small Area High Gray Level Emphasis (SAHGLE) |
| 15. Large Area Low Gray Level Emphasis (LALGLE) |
| 16. Large Area High Gray Level Emphasis (LAHGLE) |

| Neighboring Gray Tone Difference Matrix (NGTDM) |
|---|
| 1-Coarseness |
| 2-Contrast |
| 3-Busyness |
| 4-Complexity |
| 5- Strength |

**5- Univariate analysis**

For univariate analysis, each feature value was normalized to obtain Z-scores, followed by student t-test students for comparison. A p-value of <0.05 was used as a criterion for statistically significant results. Spearman corroleation between features were performed to eliminate high corrolated features and the Fales dicovery rate (FDR) correction were assesseed and q-value were reported. Statistical analysis was performed in R 3.5.1 (using 'pROC' and 'stats' packages). We also studied the prediction performance of conventional clinical PET parameters ($SUV_{max}$, $SUV_{peak}$, $SUL_{max,}$ $SUL_{peak}$ and MTV).

**6- Feature selection**

We implemented 6 different feature selections methods in our framework and compared their performances (see Table 3). We implement 6 feature selection including Filter Based Method (Select K Best (SKB), Variance Threshold (VT) and Select Percentile (SP)), Embedded Based Method (Select from Model (SM)) and Combine Methods (Variance Threshold & Select from Model (VT-SM) and Variance Threshold & Select K Best (VT-KB)).

**Table 3.** Feature selector algorithm characteristics

| Feature Selection Method | Feature Selection Model | Abbreviation | Algorithm of action |
|---|---|---|---|
| **Filter Based Method** | **Select K Best** | **SKB** | This method scores the features using a chi square function and then "removes all but the k highest scoring features. |
| | **Variance Threshold** | **VT** | It removes all features whose variance doesn't meet some threshold. By default, it removes all zero-variance features, i.e. features that have the same value in all sample. In this method, we used 0.8 threshold. |
| | **Select Percentile** | **SP** | It removes all but a user-specified highest scoring percentage of features using common univariate statistical tests for each feature: false positive rate or family wise error |
| **Embedded Based Method** | **Select From Model** | **SM** | It is a meta-transformer that can be used along with any estimator that has a coefficient or feature importance's attribute after fitting. The features are considered unimportant and removed, if the corresponding coefficient or feature importance's values are below the provided threshold parameter. Apart from specifying the threshold numerically, there are built-in heuristics for finding a threshold using a string argument. Available heuristics are "mean", "median" and float multiples of these like "0.1*mean". |
| **Combine Methods** | **Variance Threshold & Select From Model** | **VT-SM** | Combine of Variance Threshold & Select From Model |
| | **Variance Threshold & Select K Best** | **VT-KB** | Combine of Variance Threshold & Select K Best |

**7- Classifier**

We implemented and compared 12 classifiers (Table 4, and detail of each classifier are provided in supplemental Table 1). Different method of classifications including generalized linear models (logistic regression and stochastic gradient descent), naive Bayes models (naive Bayes and Gaussian naive Bayes), nearest neighbor's model (k-nearest neighbors), decision trees model (C5.0), quadratic discriminant analysis model (QDA), support vector machines (SVC), supervised neural network models (multi-layer perceptron) and ensemble learning methods (adaptive boost, bagging and random forest) were used in this study.

**Table 4.** Classifier methods and abbreviation

| Machine learning Methods | Machine learning Algorithm | Abbreviation | Algorithm of action |
|---|---|---|---|
| **Support Vector Classifier** | **Support Vector Machine** | **SVM** | Data items are plotted as a point in n-dimensional space (n is number of features) with the value of each feature being the value of a particular coordinate and classification is performed by finding the hyper-plane that differentiate the two classes. |
| **K-Nearest Neighbors** | **K-Nearest Neighbors** | **KNN** | As a non-parametric, lazy learning algorithm, KNN algorithm is based on feature similarity and an object is classified by a majority vote of its neighbors, with the object being assigned to the class most common among its k nearest neighbors. |
| **Decision Tree** | **Decision Tree (C 5.0)** | **DT** | As a decision support tool, uses a tree-like graph of decisions and their possible consequences, including chance event outcomes, resource costs, and utility, builds a model of decisions. |
| **Quadratic Discriminant Analysis** | **Quadratic Discriminant Analysis** | **QDA** | QDA separate measurements of two classes of objects with a quadratic decision boundary, generated by fitting class conditional densities to the data. It is a more general version of the linear classifier. |
| **Supervised Neural Network Models** | **Multilayer Perceptron** | **MLP** | MLP is a class of feed forward artificial neural network which optimizes the log-loss function using stochastic gradient descent (SGD). |
| **Generalized Linear Methods** | **Stochastic Gradient Descent** | **SGD** | Works based on the differentiable objective function optimization which is a stochastic approximation of gradient descent optimization. This classifier implements a plain SGD learning routine which supports different loss functions and penalties for classification. |
| | **Logistic Regression** | **LREG** | The probabilities describing the possible outcomes of a single trial are modeled using a logistic function. |
| **Naive Bayes Methods** | **Naive Bayes** | **NB** | As a probabilistic classifier, works based on applying Bayes' theorem with strong (naive) independence assumptions between the features. |
| | **Gaussian Naive Bayes** | **GNB** | It is a generalization of Naive Bayes Networks, which are a special case of probabilistic networks that allows treating continuous variables. This method computes conditional class probabilities and then predicts the most probable class of a vector of training data X, according to sample data D. |
| **Ensemble Learning Methods** | **Random Forest** | **RF** | It is an ensemble learning approach which is essentially a collection of decision trees, where each tree is slightly different from the others. This algorithm reduces the amount of overfitting by averaging the result of all decision trees. |
| | **Adaptive Boosting** | **AB** | This ensemble learning method, consists of very simple base classifiers (weak learners). This approach focuses on the training samples that are hard to classify, that is, to let the weak learners subsequently learn from misclassified training samples to improve the performance of the ensemble. |
| | **Bagging** | **BAG** | BAG fits base classifiers each on random subsets of the original dataset and then aggregate their individual predictions (by voting) to form a final prediction. The base estimator to fit on random subsets of the dataset is a decision tree model. |

### *8- Model evaluation*

All our analyses, including feature selection and classification, were performed using an in-house developed python framework in open-source python library Scikit-Learn (37). Cross-validation (CV) was applied to models and tuning of models were performed on testing set of 10-fold CV and this process were performed 20 times to get highest stable results. Furthermore, models evaluation was performed on independent validation sets (65 patients). The predictive power of all features was investigated using the area under the receiver operator characteristic (ROC) curve (AUC). A heatmap was prepared to compare different developed models. Mean ± standard deviation (SD) of AUC of all classifier and feature selection algorithms were calculated and depicted. The cross combination of feature selection and classification methods were depicted as a heatmap (mean AUC value in cross validation).

# Results

### Univariate analysis

- **EGFR**

Results of EGFR mutation status perdition after elimination of high correlated feature in each image sets and FDR correction were presented in supplemental Table 2 (only q-value <0.05). About ninety features from all image data set had significant correlation with EGFR mutation status (q-value < 0.05) with mean AUC 0.67±0.05 (0.52-0.75). Highest prediction performance was achieved by Short Run Emphasis feature of GLRLM (AUC: 75, q-value: 0.003) and Run Variance feature of GLRLM (AUC: 75, q-value: 0.016) from LOG preprocessed image of PET with sigma value 1.5 and 1, respectively. Also, Small Area Emphasis feature and Size Zone Non Uniformity Normalized from GLSZM of CTD data set which preprocessed by LLL of wavelet and LOG with sigma 2 respectively (AUC: 75, q-value: 0.001). More detail of univariate analysis in EGFR mutation status predation including AUC, p-value and q-value were presented in supplementary heatmap 1-3, respectively.

- **KRAS**

Results of KRAS mutation status perdition after elimination of high correlated feature in each image sets and FDR correction were presented in supplemental Table 3 (only q-value <0.05). Fourteen features from all image data set were significantly correlated with KRAS mutation status (q-value < 0.05) with mean AUC 0.67±0.02 (0.53-0.71). The Large Dependence Low Gray Level Emphasis from GLDM in LOG (sigma: 5) preprocessed image of CTD with AUC 0.71 (q-value: 0.00005) had highest performances. Also, the Minimum feature from first order of LOG preprocessed in sigma 4 (AUC: 71, q-value: 0.013), 4.5 (AUC: 71, q-value: 0.01) and 5 (AUC: 70, q-value: 0.014) showed high performances in KRAS mutation status perdition. More detail of univariate analysis in KRAS mutation status predation including AUC, p-value and q-value were presented in supplementary heatmap 4-6, respectively.

**Supplemental Table 2.** EGFR mutation status perdition after elimination of high correlated feature in each image sets and FDR correction were presented as AUC, p-value and q-value

| Image_Sets | Preprocessed | Filter | Type | Features | AUC | p.value | q-value |
|---|---|---|---|---|---|---|---|
| PET | BIN 64 | BIN64 | FO | 10Percentile | 0.64 | 0.003 | 0.024 |
| | | | GLDM | DV | 0.71 | 0.007 | 0.027 |
| | LOG | Sigma: 0.5 | firstorder | Mean | 0.68 | 0.000 | 0.001 |
| | | | | 90Percentile | 0.63 | 0.001 | 0.003 |
| | | | glcm | Idn | 0.67 | 0.000 | 0.002 |
| | | | | MaximumProbability | 0.73 | 0.001 | 0.002 |
| | | | | Imc1 | 0.63 | 0.003 | 0.008 |
| | | | gldm | DependenceVariance | 0.73 | 0.000 | 0.002 |
| | | | | SmallDependenceLowGrayLevelEmphasis | 0.54 | 0.017 | 0.037 |
| | | | glszm | SmallAreaEmphasis | 0.70 | 0.000 | 0.002 |
| | | | | GrayLevelNonUniformity | 0.63 | 0.001 | 0.004 |
| | | | | SizeZoneNonUniformity | 0.71 | 0.002 | 0.006 |
| | | Sigma: 1.0 | firstorder | 90Percentile | 0.63 | 0.001 | 0.009 |
| | | | glcm | Imc2 | 0.70 | 0.004 | 0.016 |
| | | | glrlm | RunVariance | 0.75 | 0.005 | 0.016 |
| | | Sigma: 1.5 | firstorder | Mean | 0.68 | 0.000 | 0.000 |
| | | | | 90Percentile | 0.62 | 0.001 | 0.003 |
| | | | | Median | 0.58 | 0.018 | 0.044 |
| | | | glrlm | ShortRunEmphasis | 0.75 | 0.001 | 0.003 |
| | | Sigma: 2.0 | firstorder | Mean | 0.68 | 0.000 | 0.001 |
| | | | | 90Percentile | 0.60 | 0.002 | 0.006 |
| | | | | Median | 0.61 | 0.003 | 0.008 |
| | | | | Mean | 0.69 | 0.000 | 0.000 |
| | | | | Median | 0.63 | 0.001 | 0.003 |
| | | | glszm | ZonePercentage | 0.73 | 0.000 | 0.001 |
| | | | gldm | DependenceNonUniformityNormalized | 0.73 | 0.000 | 0.000 |
| | | Sigma: 3.0 | firstorder | Mean | 0.69 | 0.000 | 0.001 |
| | | | | Median | 0.64 | 0.000 | 0.001 |
| | | | glrlm | ShortRunEmphasis | 0.73 | 0.000 | 0.001 |
| | | Sigma: 3.5 | firstorder | Median | 0.67 | 0.000 | 0.000 |
| | | | | Maximum | 0.54 | 0.005 | 0.015 |
| | | | gldm | LargeDependenceEmphasis | 0.73 | 0.001 | 0.006 |
| | | Sigma: 4.5 | firstorder | Median | 0.69 | 0.000 | 0.000 |
| | | | gldm | LargeDependenceEmphasis | 0.71 | 0.002 | 0.008 |
| | | Sigma: 4 | firstorder | Median | 0.68 | 0.000 | 0.000 |
| | | | | Maximum | 0.52 | 0.016 | 0.049 |
| | | | glrlm | RunLengthNonUniformityNormalized | 0.70 | 0.002 | 0.009 |
| | | Sigma: 5.0 | firstorder | Median | 0.70 | 0.000 | 0.000 |
| | | | gldm | LargeDependenceEmphasis | 0.69 | 0.003 | 0.017 |
| | WAV | HHH | gldm | DependenceNonUniformityNormalized | 0.71 | 0.000 | 0.000 |
| | | | glszm | SmallAreaEmphasis | 0.65 | 0.002 | 0.017 |
| | | HHL | firstorder | Mean | 0.61 | 0.010 | 0.039 |
| | | HHL | gldm | DependenceNonUniformityNormalized | 0.73 | 0.000 | 0.001 |
| | | HHL | ngtdm | Strength | 0.73 | 0.005 | 0.028 |
| | | HLH | firstorder | Median | 0.63 | 0.009 | 0.034 |
| | | | glrlm | RunLengthNonUniformityNormalized | 0.73 | 0.000 | 0.003 |
| | | | ngtdm | Strength | 0.71 | 0.008 | 0.034 |
| | | HLL | glszm | SizeZoneNonUniformityNormalized | 0.72 | 0.000 | 0.002 |
| | | LHH | glrlm | RunPercentage | 0.73 | 0.001 | 0.005 |
| | | LHL | firstorder | Median | 0.59 | 0.011 | 0.026 |
| | | | gldm | SmallDependenceEmphasis | 0.73 | 0.000 | 0.002 |
| | | | glszm | LargeAreaHighGrayLevelEmphasis | 0.66 | 0.002 | 0.010 |
| | | | ngtdm | Busyness | 0.65 | 0.009 | 0.026 |
| | | LLH | firstorder | Mean | 0.66 | 0.001 | 0.005 |
| | | | glcm | InverseVariance | 0.72 | 0.001 | 0.005 |
| | | | | ClusterShade | 0.62 | 0.017 | 0.032 |
| | | | | LargeDependenceHighGrayLevelEmphasis | 0.68 | 0.005 | 0.021 |
| | | | ngtdm | Coarseness | 0.65 | 0.016 | 0.032 |
| | | | | Busyness | 0.64 | 0.012 | 0.032 |
| | | | | Contrast | 0.63 | 0.015 | 0.032 |
| | | LLL | firstorder | 10Percentile | 0.65 | 0.002 | 0.017 |
| | | | glcm | InverseVariance | 0.66 | 0.004 | 0.018 |
| | | | glszm | ZoneVariance | 0.64 | 0.008 | 0.024 |
| CTD | BIN 64 | BIN 64 | FO | 90Percentile | 0.67 | 0.006 | 0.034 |
| | | | GLCM | IMC2 | 0.66 | 0.004 | 0.034 |
| | | | GLDM | DE | 0.66 | 0.005 | 0.034 |
| | | | GLSZM | ZE | 0.69 | 0.001 | 0.021 |
| | | | GLSZM | SAE | 0.67 | 0.003 | 0.034 |
| | LOG | Sigma: 1.0 | glszm | SizeZoneNonUniformityNormalized | 0.71 | 0.000 | 0.008 |
| | | Sigma: 1.5 | glszm | SizeZoneNonUniformityNormalized | 0.71 | 0.000 | 0.008 |
| | | Sigma: 2.0 | glszm | SizeZoneNonUniformityNormalized | 0.75 | 0.000 | 0.001 |
| | | Sigma: 3.0 | gldm | SmallDependenceHighGrayLevelEmphasis | 0.65 | 0.000 | 0.011 |
| | | | | DependenceNonUniformityNormalized | 0.60 | 0.002 | 0.026 |
| | | | glszm | SmallAreaHighGrayLevelEmphasis | 0.63 | 0.002 | 0.026 |
| | | Sigma: 3.5 | gldm | SmallDependenceHighGrayLevelEmphasis | 0.66 | 0.002 | 0.028 |
| | | | | DependenceNonUniformityNormalized | 0.60 | 0.003 | 0.028 |
| | | | glszm | SmallAreaHighGrayLevelEmphasis | 0.65 | 0.002 | 0.028 |
| | WAV | HLH | gldm | LargeDependenceHighGrayLevelEmphasis | 0.67 | 0.002 | 0.038 |
| | | HLL | firstorder | Skewness | 0.67 | 0.002 | 0.045 |
| | | LHH | firstorder | Kurtosis | 0.63 | 0.002 | 0.013 |
| | | | glcm | Idn | 0.68 | 0.001 | 0.013 |
| | | | glszm | GrayLevelNonUniformity | 0.58 | 0.002 | 0.014 |
| | | LLL | firstorder | 90Percentile | 0.68 | 0.004 | 0.034 |
| | | | glszm | SmallAreaEmphasis | 0.75 | 0.000 | 0.001 |
| | | | | ZoneEntropy | 0.67 | 0.002 | 0.028 |
| | | | | SmallAreaHighGrayLevelEmphasis | 0.68 | 0.005 | 0.034 |
| CT | BIN64 | BIN64 | GLSZM | SAE | 0.68 | 0.001 | 0.044 |
| | LOG | Sigma: 1.5 | glcm | InverseVariance | 0.69 | 0.001 | 0.022 |
| | | Sigma: 2.0 | glcm | InverseVariance | 0.71 | 0.000 | 0.003 |
| | WAV | LHL | firstorder | Skewness | 0.68 | 0.001 | 0.033 |

**Supplemental Table 3.** KRAS mutation status perdition after elimination of high correlated feature in each image sets and FDR correction were presented as AUC, p-value and q-value

| **Image_Set** | **Preporcessed** | **Filter** | **Type** | **Features** | **AUC** | **p.value** | **q-value** |
|---|---|---|---|---|---|---|---|
| CTD | LOG | Sigma: 3.0 | glszm | LowGrayLevelZoneEmphasis | 0.63 | 0.001 | 0.029 |
| | | Sigma: 4.5 | firstorder | Minimum | 0.71 | 0.001 | 0.010 |
| | | | | Skewness | 0.69 | 0.001 | 0.010 |
| | | | gldm | LargeDependenceLowGrayLevelEmphasis | 0.68 | 0.001 | 0.013 |
| | | | | LowGrayLevelEmphasis | 0.63 | 0.004 | 0.031 |
| | | Sigma: 4 | firstorder | Minimum | 0.71 | 0.001 | 0.013 |
| | | | | Skewness | 0.68 | 0.001 | 0.013 |
| | | | gldm | LargeDependenceLowGrayLevelEmphasis | 0.69 | 0.002 | 0.013 |
| | | | | LowGrayLevelEmphasis | 0.65 | 0.001 | 0.013 |
| | | | glszm | LowGrayLevelZoneEmphasis | 0.63 | 0.002 | 0.014 |
| | | Sigma: 5.0 | firstorder | Minimum | 0.70 | 0.001 | 0.014 |
| | | | | Skewness | 0.68 | 0.003 | 0.025 |
| | | | gldm | LargeDependenceLowGrayLevelEmphasis | 0.71 | 0.000 | 0.000 |
| | | | | LowGrayLevelEmphasis | 0.67 | 0.000 | 0.002 |

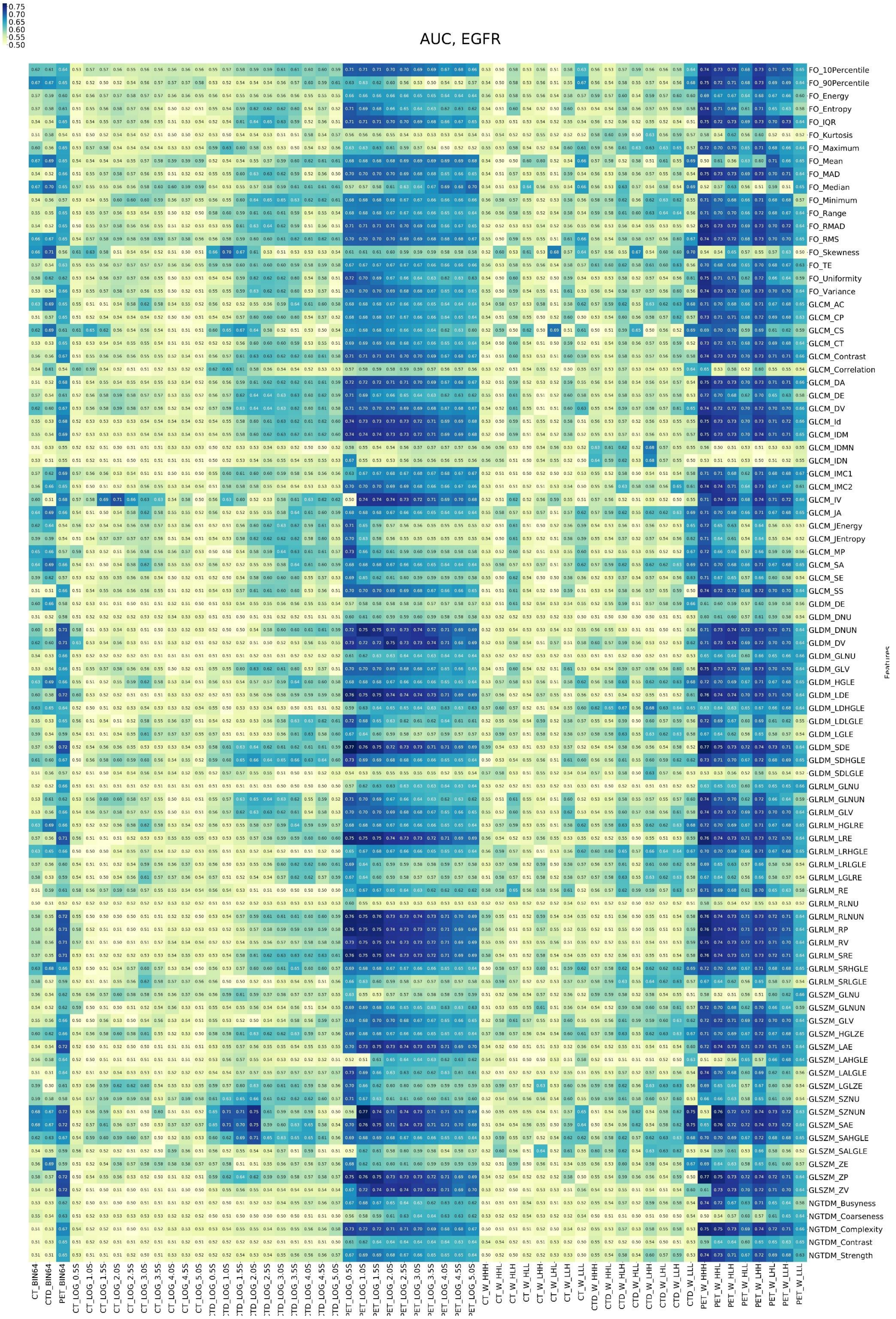

**Supplemental Fig 1.** AUC of EGFR prediction using univariate analysis of radiomics features in different date sets

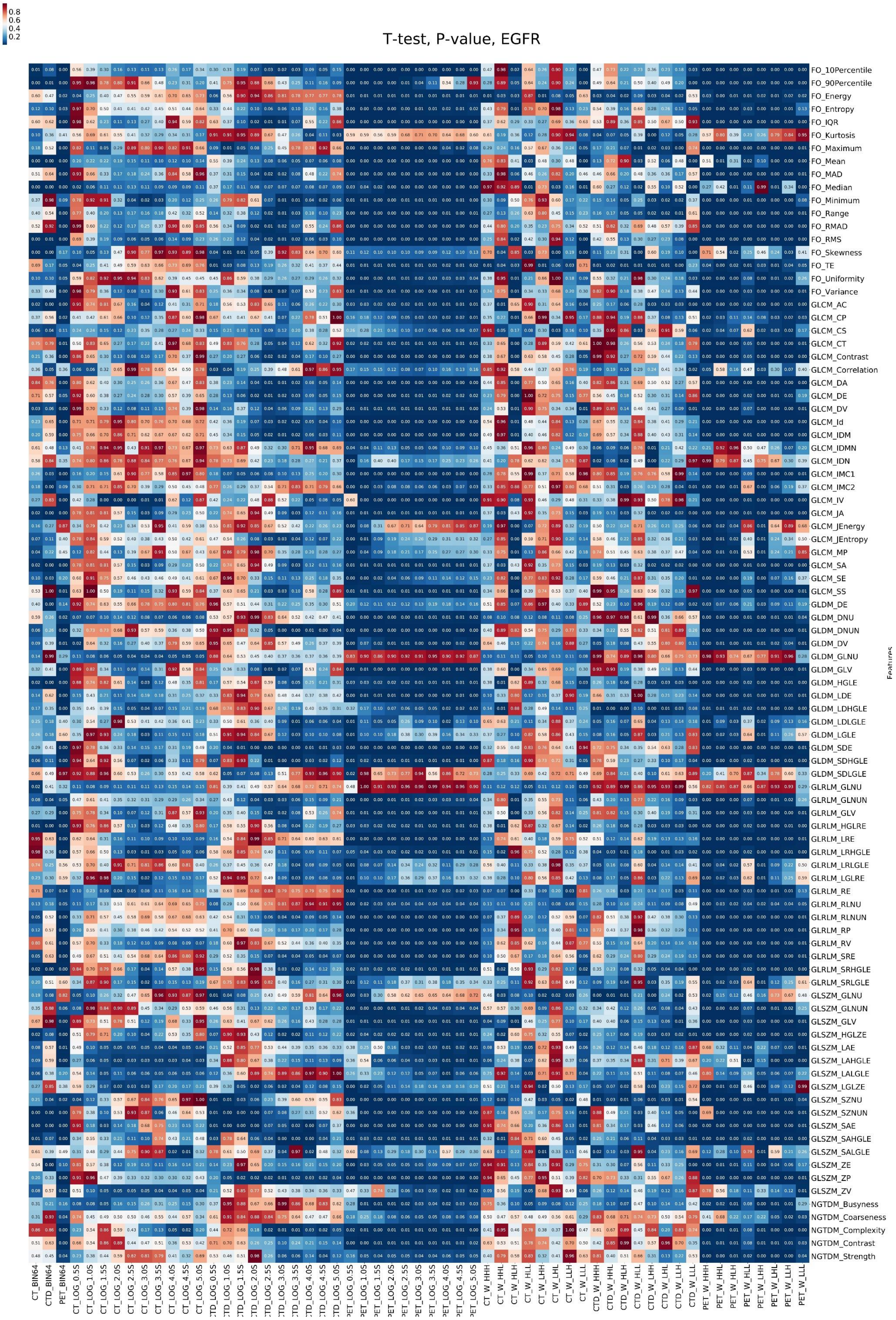

**Supplemental Fig 2.** P-value of EGFR prediction using univariate analysis of radiomics features in different date sets

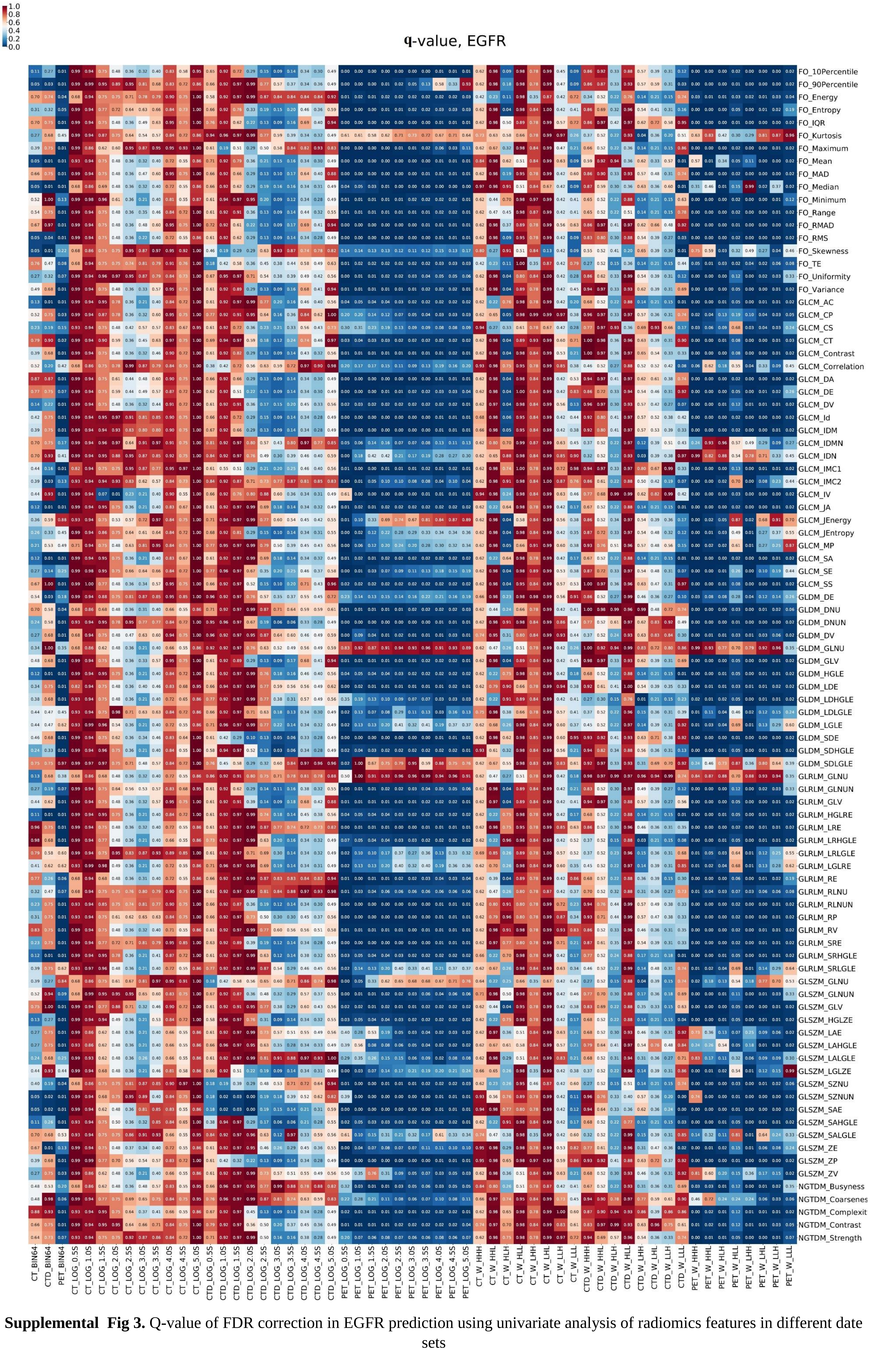

**Supplemental Fig 3.** Q-value of FDR correction in EGFR prediction using univariate analysis of radiomics features in different date sets

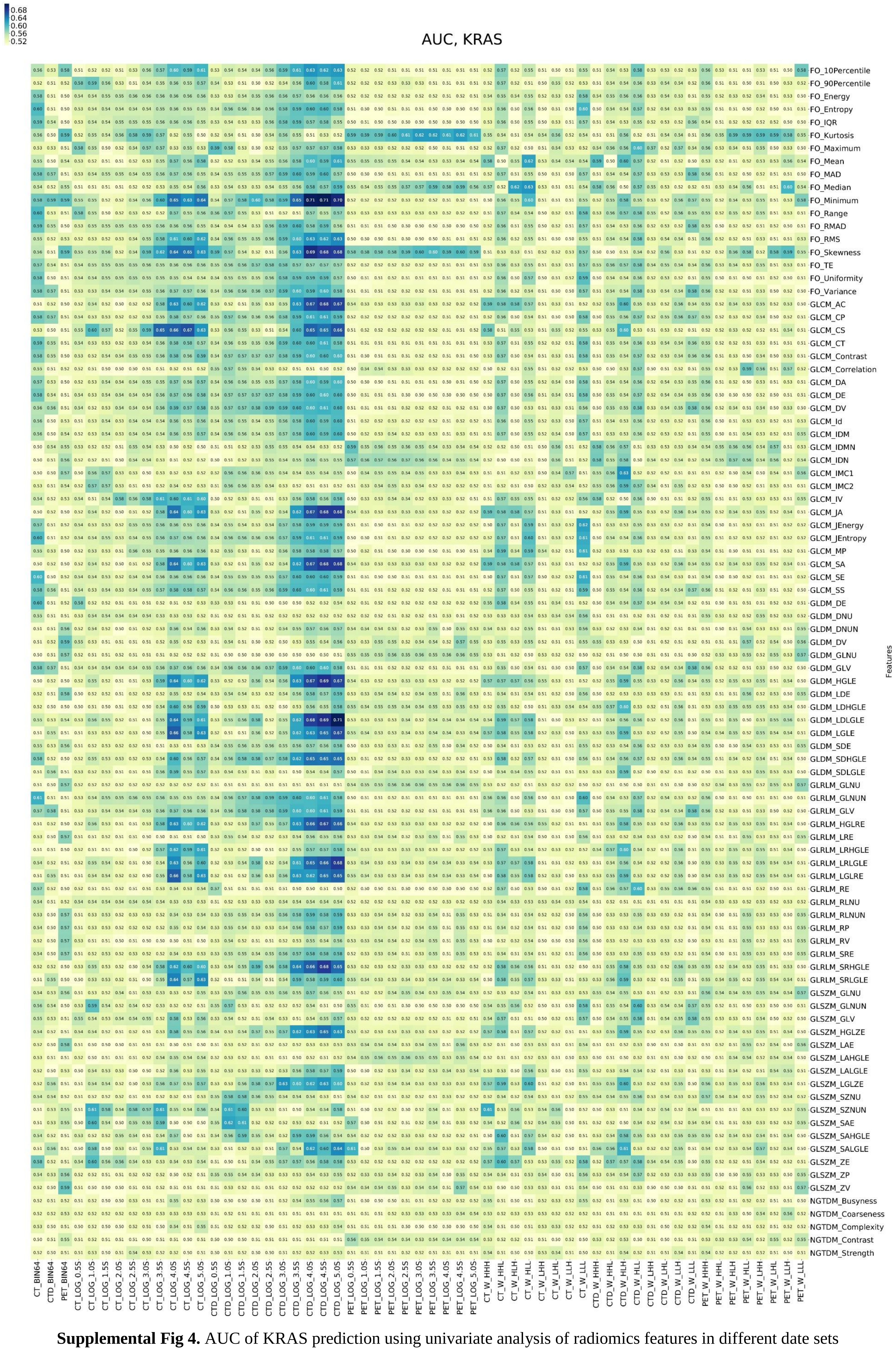

**Supplemental Fig 4.** AUC of KRAS prediction using univariate analysis of radiomics features in different date sets

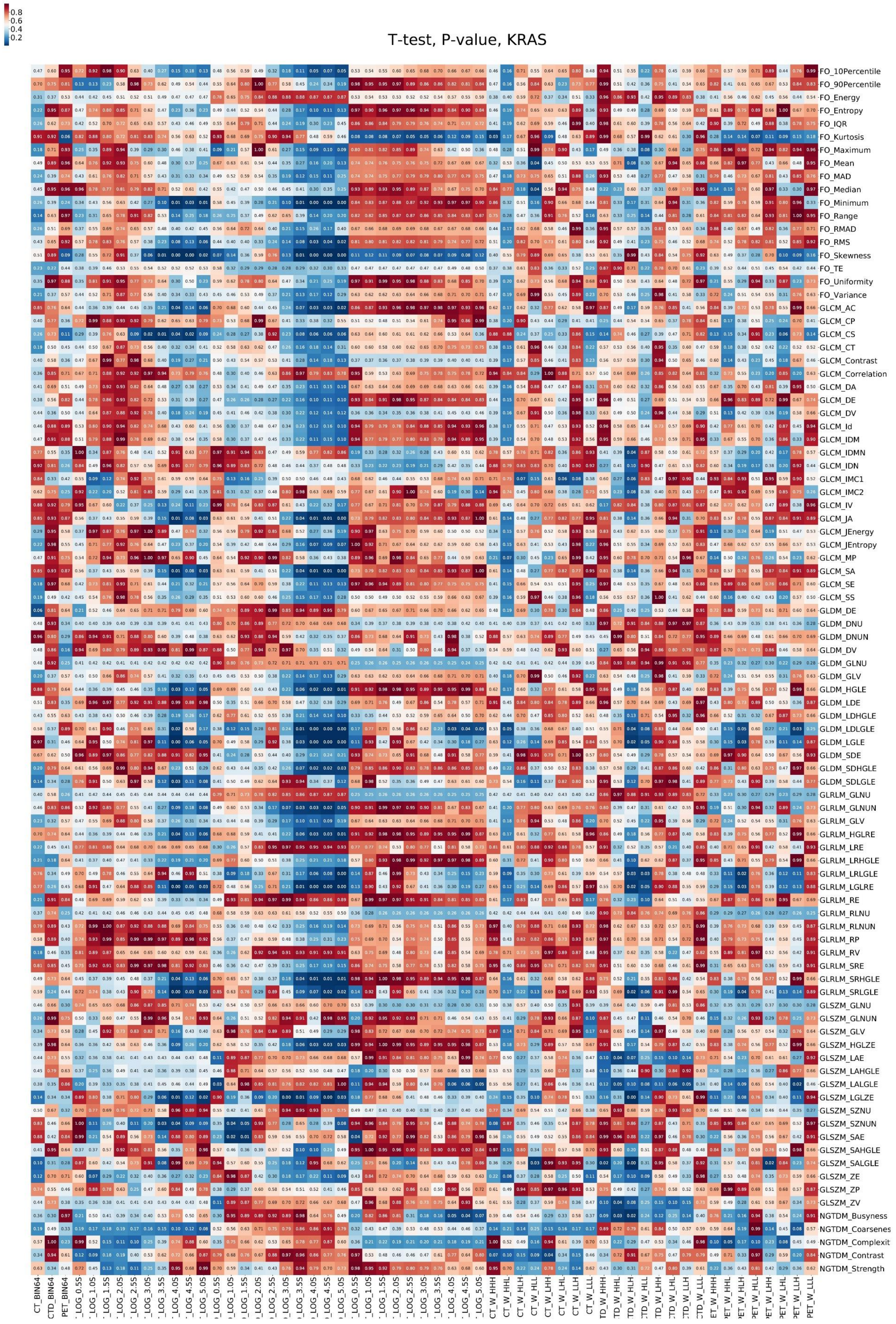

**Supplemental Fig 5.** P-value of KRAS prediction using univariate analysis of radiomics features in different date sets

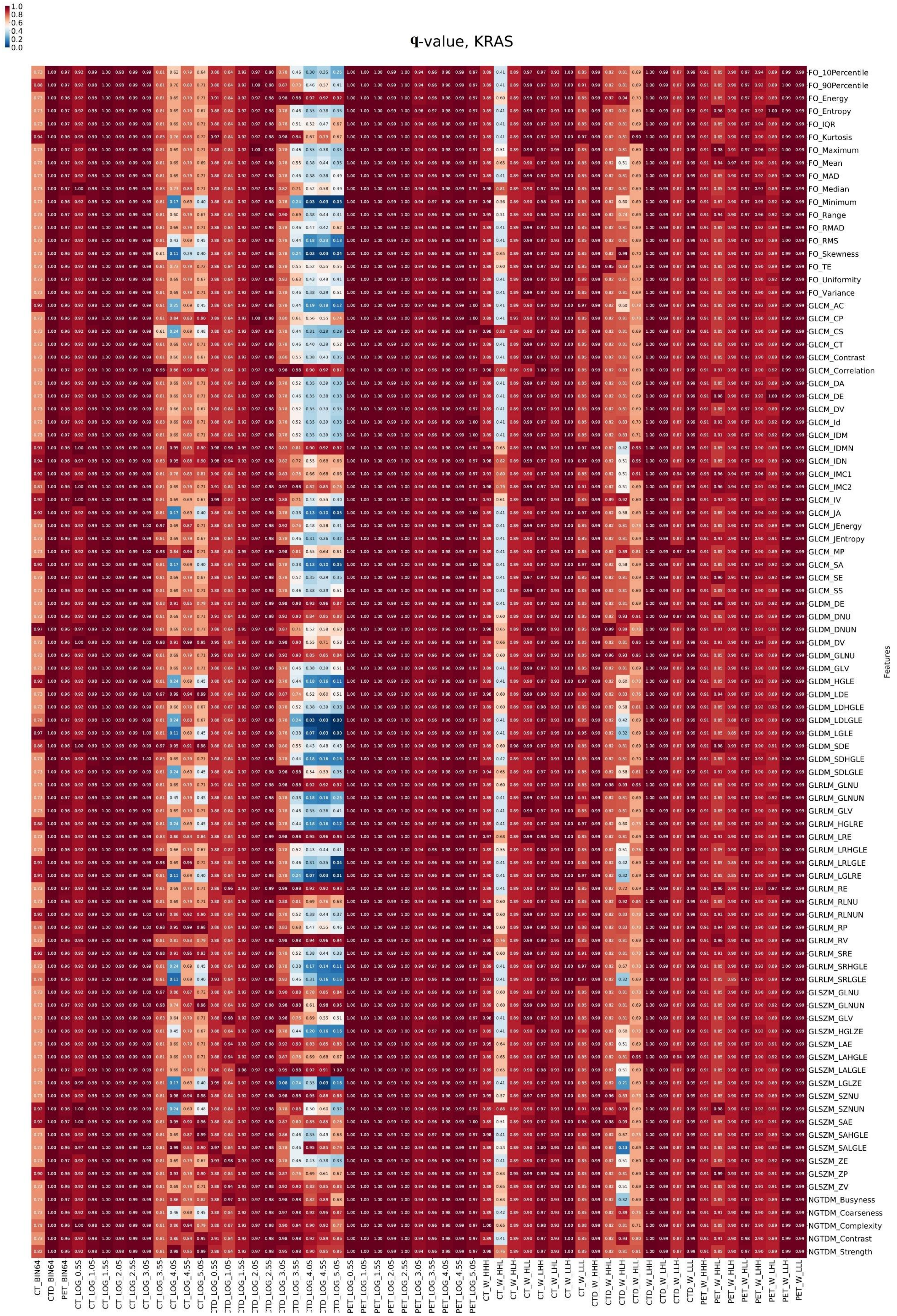

**Supplemental Fig 6.** Q-value of FDR correction in EGFR prediction using univariate analysis of radiomics features in different date sets

- **Shape features of EGFR and KRAS**

Details of EGFR and KRAS mutation status predation using shape features including AUC, p-value and q-value were presented in supplementary heatmap 7-9, respectively. The Surface Volume Ratio feature from PET images (AUC: 0.60, q-value: 0.28) and Flatness from CT images (AUC: 0.67, q-value:0.11) had highest perdition power in EGFR and KRAS respectively, however statistically not significant.

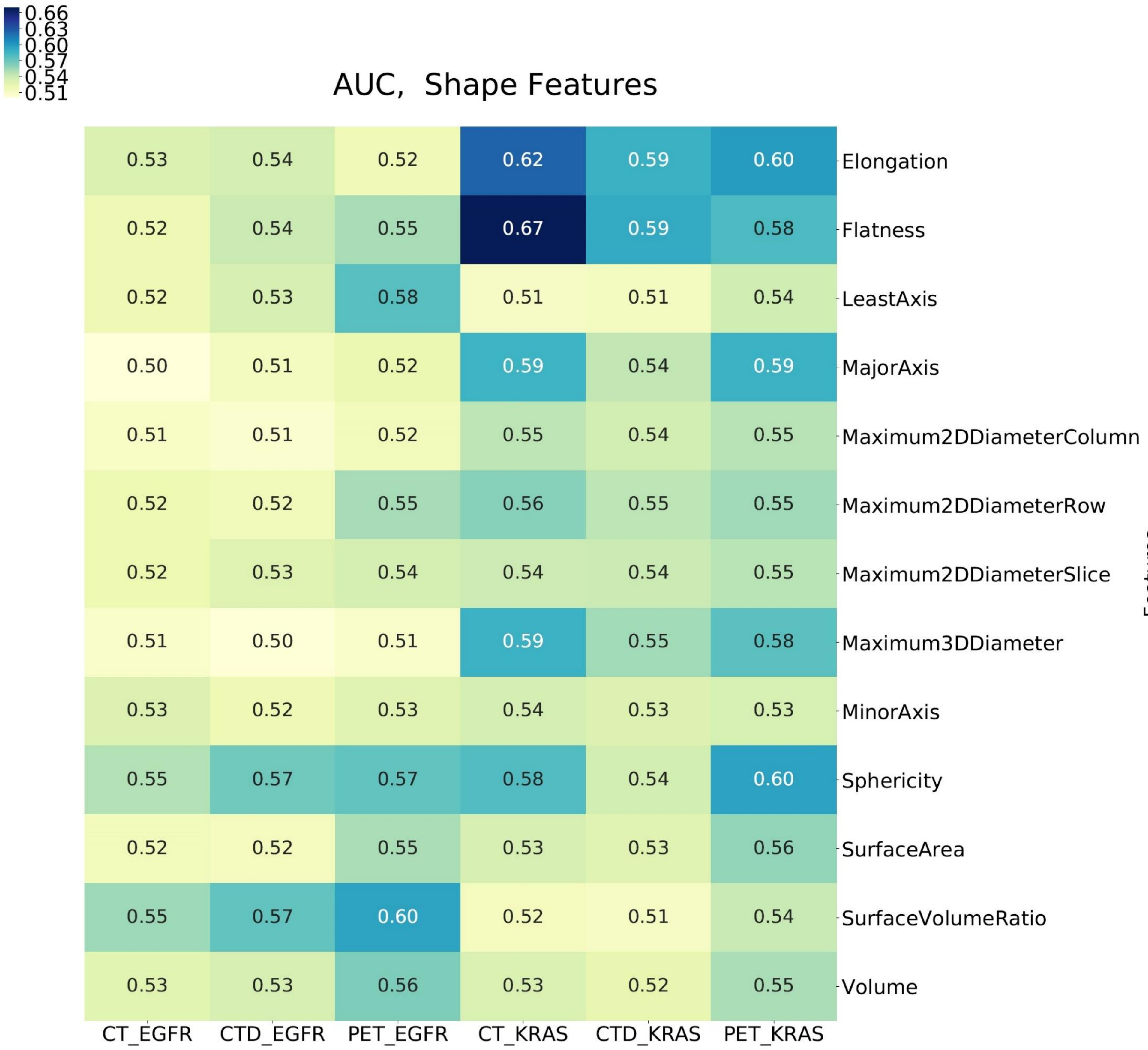

**Supplemental Fig 7.** AUC of shape features in prediction of EGFR and KRAS using univariate analysis of radiomics features in different date sets

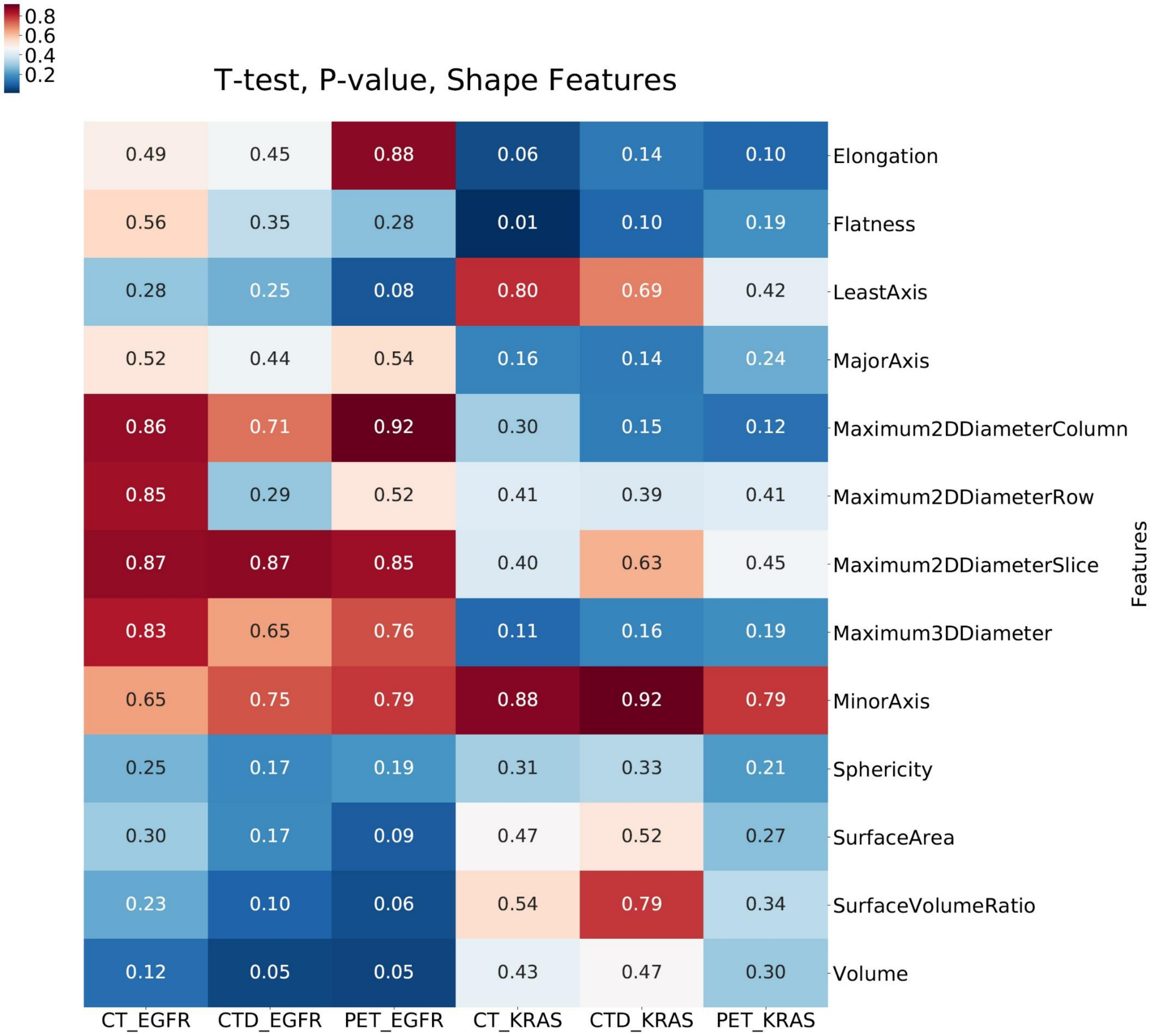

**Supplemental Fig 8.** P-value of shape features in prediction of EGFR and KRAS using univariate analysis of radiomics features in different date sets

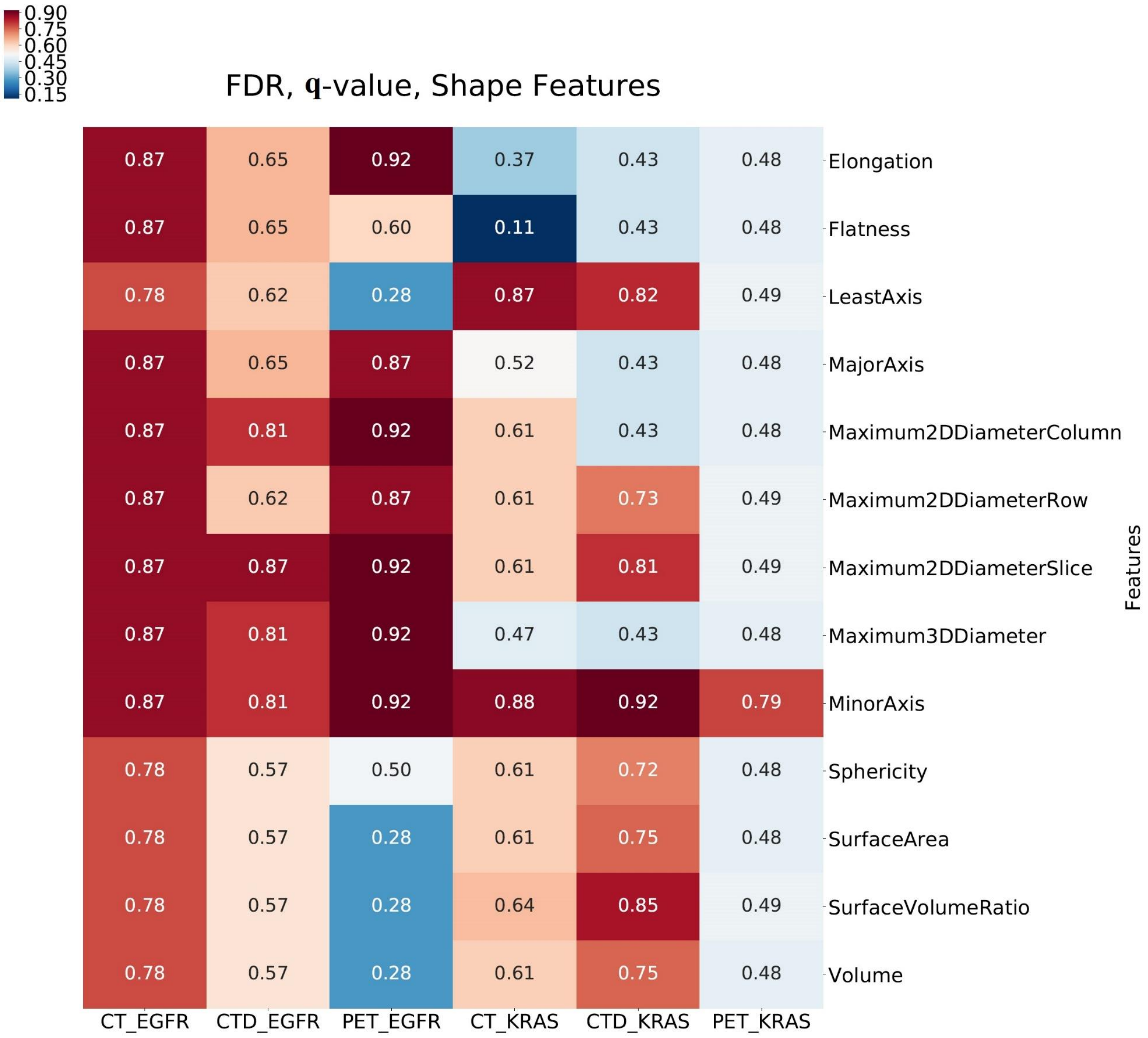

**Supplemental Fig 9.** Q-value of FDR correction of shape features in prediction of EGFR and KRAS using univariate analysis of radiomics features in different date sets

- **Conventional PET models**

Our univariate analysis on conventional PET models for EGFR mutation status prediction showed that $SUV_{peak}$ with AUC=0.69 (P-value = 0.0002) had the highest performance, and results for $SUL_{peak}$, $SUV_{max}$, $SUL_{max}$ and MTV were 0.60 (P-value = 0.0003), 0.59 (P-value < 0.0001), 0.55 (P-value < 0.0001) and 0.56 (P-value = 0.0004) respectively. Our results involving conventional features for KRAS mutation status prediction were all poor: MTV (AUC: 0.55, P-value = 0.0011), $SUV_{max}$ (AUC: 0.52, P-value < 0.0001), $SUL_{max}$ (AUC: 0.53, P-value < 0.0001), $SUV_{peak}$ (AUC: 0.52, P-value = 0.0013) and $SUL_{peak}$ (AUC: 0.51, P-value = 0.0008).

## Multivariate machine learning radiomic models

- **EGFR**

Fig 2 shows a heatmap of EGFR mutation status prediction results in different combination of feature selection, classifier and image sets. Here, we observe a wide range of performance from 0.5 to 0.82. We further observed that the combination method of LOG preprocessed of PET image set with sigma 3.5 (PET_LOG_3.5S) with VT feature selector and SGD classifier (PET_LOG_3.5S+VT+SGD, AUC: 0.82) had the highest predictive performance, followed by PET image set which preprocessed by LHH of wavelet (PET_W_LHH) with VT_SM feature selector and MLP classifier (PET_W_LHH+ VT_SM+MLP, AUC: 0.81), LOG preprocessed of PET image set with sigma 4.0 (PET_LOG_4.0S) with SM feature selector and SGD classifier (PET_LOG_4.0S+SM+SGD, AUC: 0.81), PET_W_LHH image set with SM feature selector and SGD classifier (PET_W_LHH+SM+SGD, AUC: 0.80).

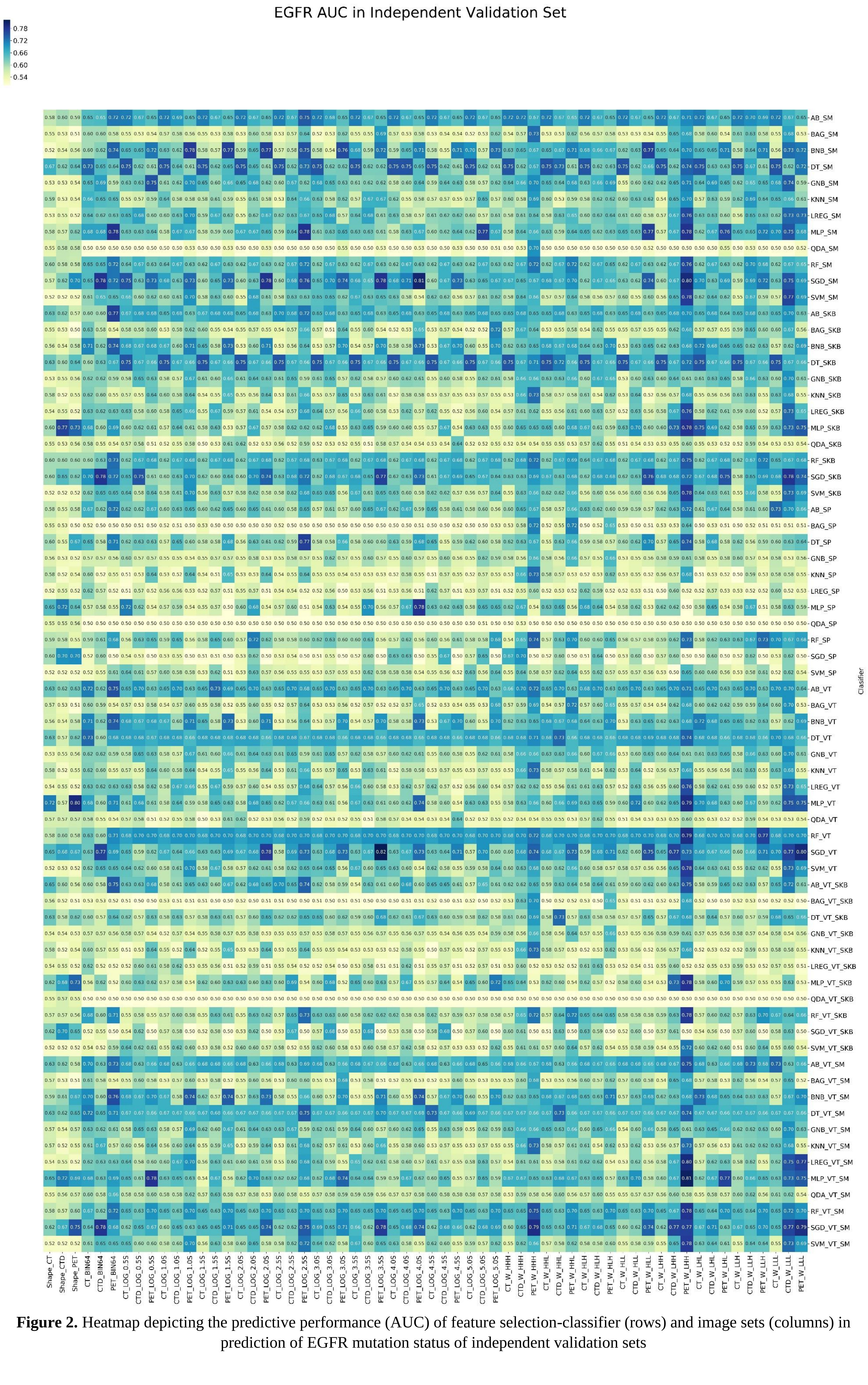

**Figure 2.** Heatmap depicting the predictive performance (AUC) of feature selection-classifier (rows) and image sets (columns) in prediction of EGFR mutation status of independent validation sets

Our results for EGFR mutation status prediction based on feature selection methods and image sets are depicted in supplemental Tables 4 and 5 (mean ± SD & Min-Max). It can be deduced from that tables selection performance had a range from 0.51 to 0.82, and the combination of VT_SM feature selection with PET_W_LHH image set had the highest performance (AUC: 0.73 ± 0.074 & 0.58 - 0.81), followed by SM feature selection with PET_W_LHH image set (AUC: 0.72 ± 0.079 & 0.5 - 0.8), VT feature selection with PET_W_LHH image set (AUC: 0.71 ± 0.065 & 0.6 - 0.79) and SKB feature selection with PET_W_LHH image set (AUC: 0.7 ± 0.06 & 0.6 - 0.78). Supplemental Fig 10 delicate the box plot of EGFR mutation status prediction based on feature selection methods.

Supplemental Table 6 and 7 (mean ± SD & Min-Max) shows our results regarding EGFR mutation status prediction based on classifier and image sets. According to these results, classifier performance has a range from 0.50 to 0.82 and combination of RF classifier with PET_W_LHH image set had the highest performance (AUC: 0.76 ± 0.024 & 0.73 - 0.79), followed by BNB classifier with PET image set which discretized into 64 Bin (PET_BIN64) (AUC: 0.75 ± 0.0096 & 0.74 - 0.76) and SVM classifier with PET_W_LHH image set (AUC: 0.75 ± 0.051 & 0.65 - 0.78) and AB classifier with PET_BIN64 image set (AUC: 0.74 ± 0.018 & 0.72 - 0.77). Supplemental figure 11 shows the box plot of EGFR mutation status prediction based on classifier method.

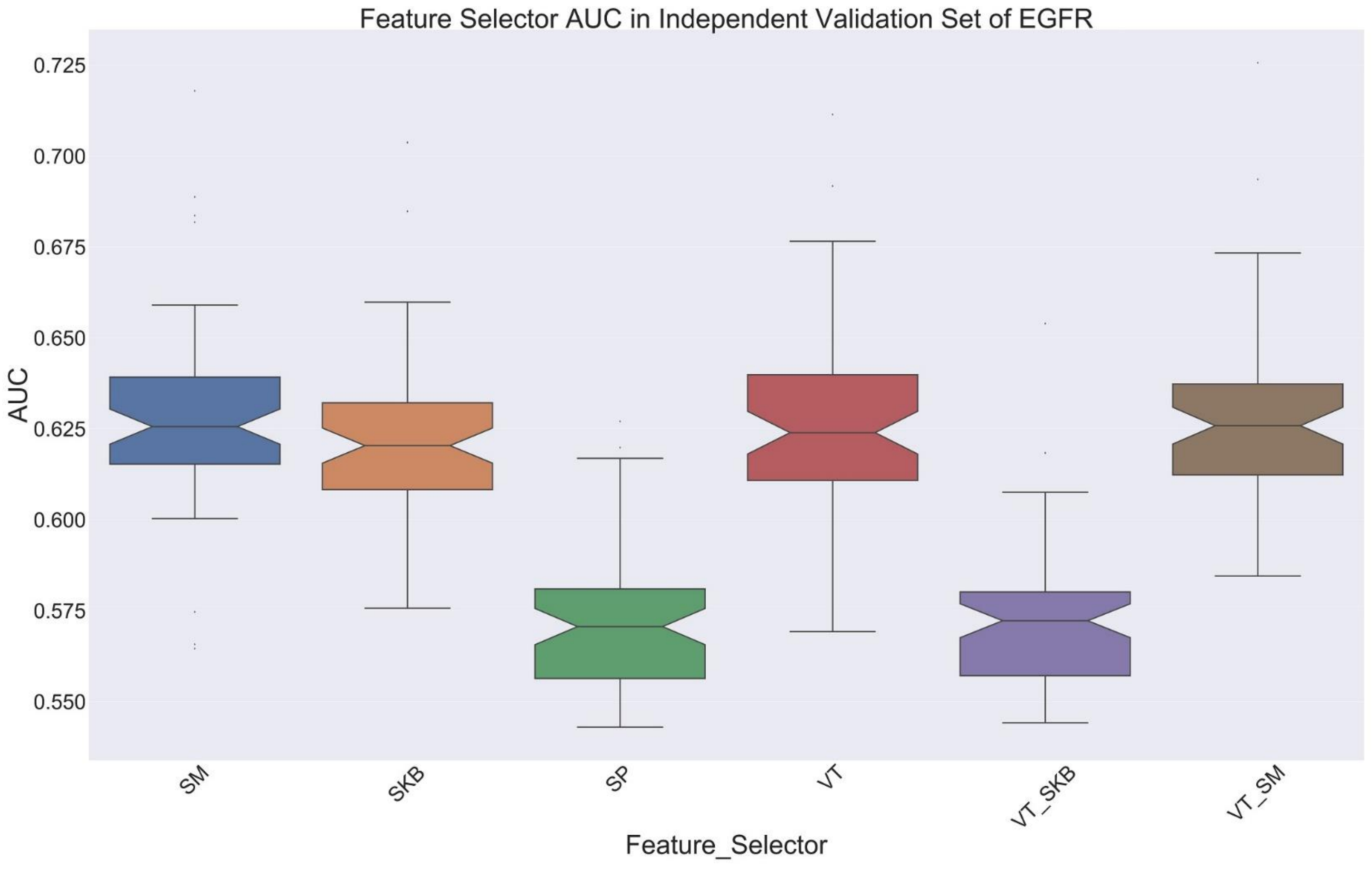

**Supplemental Fig 10.** AUC box plot of different feature selector in independent validation set of EGFR

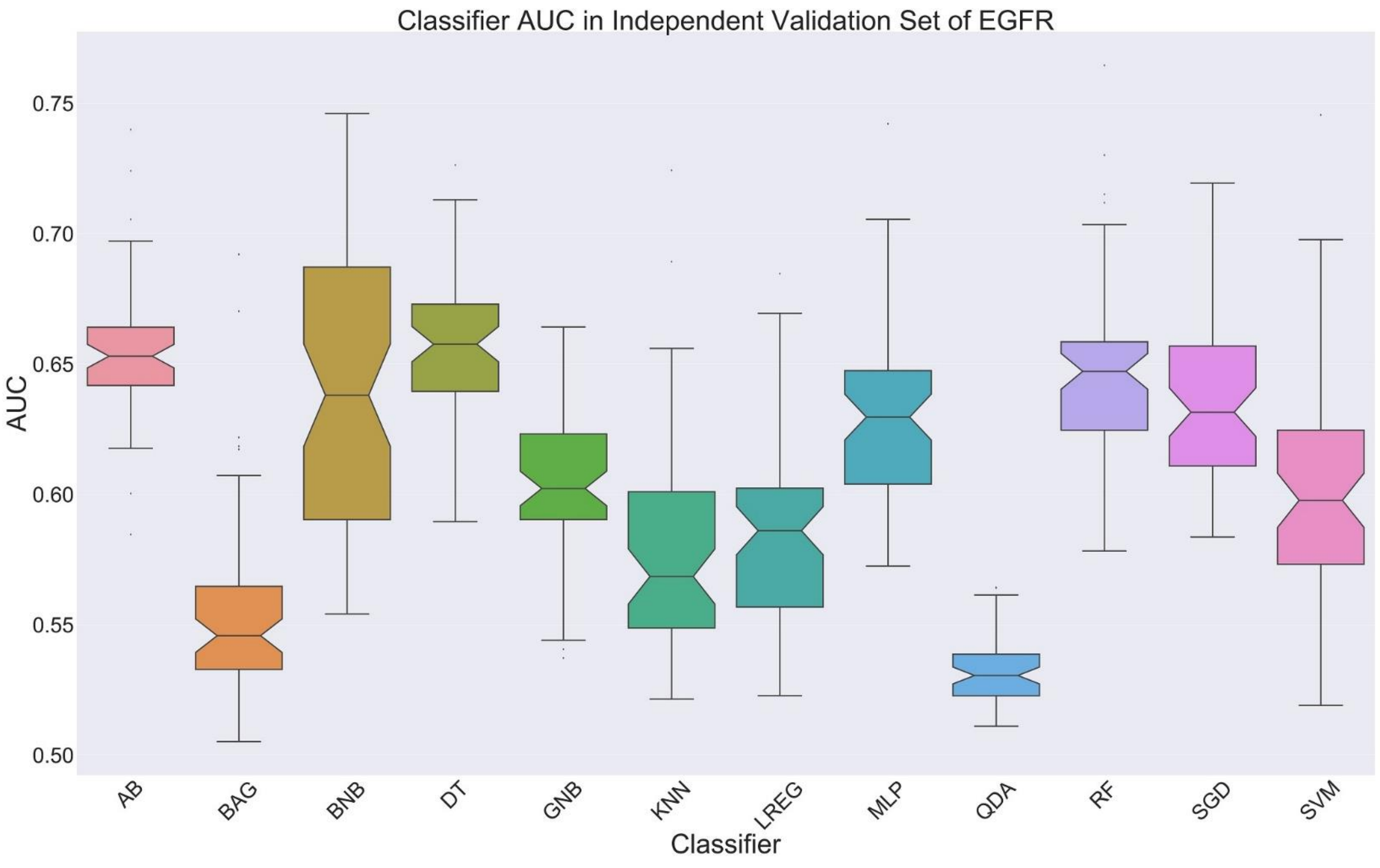

**Supplemental Fig 11.** AUC box plot of different classifier in independent validation set of EGFR

**Supplemental Table 4.** EGFR Feature Selector Min-Max in different modality, preprocessing and setting

| Modality | Preprocessed | Setting | SM | SKB | SP | VT | VT_SKB | VT_SM |
|---|---|---|---|---|---|---|---|---|
| **CT** | **Bin 64** | | 0.5 - 0.73 | 0.58 - 0.71 | 0.5 - 0.67 | 0.58 - 0.73 | 0.5 - 0.68 | 0.6 - 0.72 |
| | **LOG** | **Sigma: 0.5** | 0.5 - 0.75 | 0.57 - 0.75 | 0.5 - 0.72 | 0.57 - 0.68 | 0.5 - 0.64 | 0.55 - 0.68 |
| | | **Sigma: 1** | 0.5 - 0.75 | 0.52 - 0.75 | 0.5 - 0.6 | 0.52 - 0.68 | 0.5 - 0.58 | 0.53 - 0.68 |
| | | **Sigma: 1.5** | 0.5 - 0.75 | 0.5 - 0.75 | 0.5 - 0.6 | 0.5 - 0.68 | 0.5 - 0.63 | 0.53 - 0.68 |
| | | **sigma: 2** | 0.5 - 0.75 | 0.53 - 0.75 | 0.5 - 0.6 | 0.53 - 0.68 | 0.5 - 0.63 | 0.53 - 0.68 |
| | | **sigma: 2.5** | 0.5 - 0.75 | 0.53 - 0.75 | 0.5 - 0.62 | 0.52 - 0.68 | 0.5 - 0.7 | 0.52 - 0.68 |
| | | **sigma: 3** | 0.5 - 0.75 | 0.52 - 0.75 | 0.5 - 0.62 | 0.52 - 0.68 | 0.5 - 0.65 | 0.55 - 0.69 |
| | | **sigma: 3.5** | 0.5 - 0.75 | 0.53 - 0.75 | 0.5 - 0.6 | 0.52 - 0.68 | 0.5 - 0.65 | 0.52 - 0.68 |
| | | **sigma: 4** | 0.5 - 0.75 | 0.52 - 0.75 | 0.5 - 0.63 | 0.52 - 0.68 | 0.5 - 0.62 | 0.52 - 0.68 |
| | | **sigma: 4.5** | 0.5 - 0.75 | 0.53 - 0.75 | 0.5 - 0.65 | 0.52 - 0.68 | 0.5 - 0.65 | 0.52 - 0.73 |
| | | **Sigma: 5** | 0.5 - 0.75 | 0.52 - 0.75 | 0.5 - 0.63 | 0.52 - 0.68 | 0.5 - 0.65 | 0.53 - 0.69 |
| | **WAV** | **HHH** | 0.5 - 0.75 | 0.52 - 0.75 | 0.5 - 0.67 | 0.52 - 0.68 | 0.5 - 0.65 | 0.53 - 0.68 |
| | | **HHL** | 0.5 - 0.75 | 0.53 - 0.75 | 0.5 - 0.63 | 0.54 - 0.68 | 0.5 - 0.62 | 0.53 - 0.68 |
| | | **HLH** | 0.5 - 0.75 | 0.54 - 0.75 | 0.5 - 0.68 | 0.57 - 0.68 | 0.5 - 0.65 | 0.57 - 0.68 |
| | | **HLL** | 0.5 - 0.75 | 0.51 - 0.75 | 0.5 - 0.6 | 0.51 - 0.68 | 0.5 - 0.59 | 0.53 - 0.68 |
| | | **LHH** | 0.5 - 0.75 | 0.53 - 0.75 | 0.5 - 0.62 | 0.53 - 0.68 | 0.5 - 0.6 | 0.54 - 0.68 |
| | | **LHL** | 0.5 - 0.75 | 0.55 - 0.75 | 0.5 - 0.61 | 0.55 - 0.72 | 0.5 - 0.6 | 0.55 - 0.73 |
| | | **LLH** | 0.5 - 0.75 | 0.52 - 0.75 | 0.5 - 0.63 | 0.52 - 0.68 | 0.5 - 0.62 | 0.55 - 0.68 |
| | | **LLL** | 0.5 - 0.75 | 0.53 - 0.75 | 0.5 - 0.73 | 0.53 - 0.7 | 0.5 - 0.68 | 0.55 - 0.73 |
| | **Shape** | | 0.52 - 0.67 | 0.52 - 0.63 | 0.52 - 0.65 | 0.52 - 0.72 | 0.52 - 0.65 | 0.52 - 0.65 |
| **CTD** | **Bin 64** | | 0.5 - 0.78 | 0.55 - 0.78 | 0.5 - 0.62 | 0.55 - 0.77 | 0.5 - 0.62 | 0.58 - 0.78 |
| | **LOG** | **Sigma: 0.5** | 0.5 - 0.68 | 0.55 - 0.75 | 0.5 - 0.63 | 0.53 - 0.7 | 0.5 - 0.63 | 0.56 - 0.67 |
| | | **Sigma: 1** | 0.5 - 0.69 | 0.55 - 0.68 | 0.5 - 0.65 | 0.55 - 0.7 | 0.5 - 0.63 | 0.57 - 0.67 |
| | | **Sigma: 1.5** | 0.5 - 0.67 | 0.53 - 0.68 | 0.5 - 0.65 | 0.53 - 0.73 | 0.5 - 0.63 | 0.57 - 0.68 |
| | | **sigma: 2** | 0.5 - 0.68 | 0.52 - 0.7 | 0.5 - 0.72 | 0.52 - 0.7 | 0.5 - 0.68 | 0.58 - 0.7 |
| | | **sigma: 2.5** | 0.5 - 0.68 | 0.52 - 0.68 | 0.5 - 0.6 | 0.52 - 0.7 | 0.5 - 0.69 | 0.55 - 0.67 |
| | | **sigma: 3** | 0.5 - 0.7 | 0.51 - 0.68 | 0.5 - 0.63 | 0.53 - 0.7 | 0.5 - 0.68 | 0.53 - 0.68 |
| | | **sigma: 3.5** | 0.5 - 0.68 | 0.51 - 0.68 | 0.5 - 0.7 | 0.51 - 0.7 | 0.5 - 0.68 | 0.55 - 0.67 |
| | | **sigma: 4** | 0.5 - 0.75 | 0.53 - 0.68 | 0.5 - 0.67 | 0.54 - 0.7 | 0.5 - 0.68 | 0.55 - 0.68 |
| | | **sigma: 4.5** | 0.5 - 0.67 | 0.54 - 0.68 | 0.5 - 0.67 | 0.53 - 0.7 | 0.5 - 0.68 | 0.53 - 0.68 |
| | | **Sigma: 5** | 0.5 - 0.77 | 0.52 - 0.68 | 0.5 - 0.65 | 0.52 - 0.7 | 0.5 - 0.65 | 0.52 - 0.68 |
| | **WAV** | **HHH** | 0.52 - 0.72 | 0.54 - 0.68 | 0.52 - 0.7 | 0.54 - 0.7 | 0.5 - 0.66 | 0.59 - 0.67 |
| | | **HHL** | 0.5 - 0.73 | 0.55 - 0.72 | 0.5 - 0.65 | 0.55 - 0.73 | 0.5 - 0.73 | 0.55 - 0.73 |
| | | **HLH** | 0.5 - 0.67 | 0.54 - 0.68 | 0.5 - 0.64 | 0.54 - 0.7 | 0.5 - 0.64 | 0.54 - 0.68 |
| | | **HLL** | 0.5 - 0.67 | 0.54 - 0.7 | 0.5 - 0.62 | 0.54 - 0.72 | 0.5 - 0.6 | 0.57 - 0.7 |
| | | **LHH** | 0.5 - 0.67 | 0.55 - 0.73 | 0.5 - 0.65 | 0.55 - 0.77 | 0.5 - 0.73 | 0.57 - 0.77 |
| | | **LHL** | 0.5 - 0.69 | 0.53 - 0.69 | 0.5 - 0.68 | 0.53 - 0.7 | 0.5 - 0.64 | 0.56 - 0.71 |
| | | **LLH** | 0.52 - 0.7 | 0.59 - 0.68 | 0.5 - 0.67 | 0.59 - 0.7 | 0.5 - 0.63 | 0.56 - 0.73 |
| | | **LLL** | 0.5 - 0.77 | 0.53 - 0.78 | 0.5 - 0.7 | 0.53 - 0.77 | 0.5 - 0.72 | 0.62 - 0.77 |
| | **Shape** | | 0.52 - 0.62 | 0.52 - 0.77 | 0.52 - 0.72 | 0.52 - 0.68 | 0.52 - 0.7 | 0.52 - 0.72 |
| **PET** | **Bin 64** | | 0.5 - 0.78 | 0.54 - 0.77 | 0.5 - 0.72 | 0.54 - 0.75 | 0.5 - 0.75 | 0.54 - 0.76 |
| | **LOG** | **Sigma: 0.5** | 0.5 - 0.75 | 0.51 - 0.68 | 0.5 - 0.67 | 0.51 - 0.7 | 0.5 - 0.68 | 0.58 - 0.78 |
| | | **Sigma: 1** | 0.52 - 0.78 | 0.58 - 0.71 | 0.5 - 0.65 | 0.58 - 0.71 | 0.5 - 0.65 | 0.6 - 0.74 |
| | | **Sigma: 1.5** | 0.53 - 0.77 | 0.53 - 0.73 | 0.5 - 0.68 | 0.52 - 0.73 | 0.5 - 0.67 | 0.52 - 0.74 |
| | | **sigma: 2** | 0.53 - 0.78 | 0.53 - 0.74 | 0.5 - 0.64 | 0.53 - 0.78 | 0.5 - 0.65 | 0.53 - 0.74 |
| | | **sigma: 2.5** | 0.5 - 0.78 | 0.59 - 0.72 | 0.5 - 0.77 | 0.59 - 0.73 | 0.5 - 0.74 | 0.55 - 0.75 |
| | | **sigma: 3** | 0.55 - 0.76 | 0.52 - 0.7 | 0.5 - 0.66 | 0.52 - 0.73 | 0.5 - 0.62 | 0.55 - 0.74 |
| | | **sigma: 3.5** | 0.52 - 0.78 | 0.54 - 0.77 | 0.5 - 0.67 | 0.52 - 0.82 | 0.5 - 0.68 | 0.51 - 0.78 |
| | | **sigma: 4** | 0.52 - 0.81 | 0.54 - 0.73 | 0.5 - 0.78 | 0.54 - 0.74 | 0.5 - 0.67 | 0.53 - 0.74 |
| | | **sigma: 4.5** | 0.53 - 0.73 | 0.52 - 0.7 | 0.5 - 0.63 | 0.52 - 0.71 | 0.5 - 0.61 | 0.55 - 0.7 |
| | | **Sigma: 5** | 0.51 - 0.73 | 0.54 - 0.72 | 0.5 - 0.68 | 0.54 - 0.7 | 0.5 - 0.72 | 0.55 - 0.7 |
| | **WAV** | **HHH** | 0.64 - 0.73 | 0.54 - 0.73 | 0.5 - 0.74 | 0.54 - 0.74 | 0.5 - 0.73 | 0.58 - 0.79 |
| | | **HHL** | 0.5 - 0.72 | 0.53 - 0.69 | 0.5 - 0.72 | 0.53 - 0.73 | 0.5 - 0.72 | 0.56 - 0.71 |
| | | **HLH** | 0.5 - 0.69 | 0.55 - 0.7 | 0.5 - 0.66 | 0.55 - 0.71 | 0.5 - 0.66 | 0.54 - 0.71 |
| | | **HLL** | 0.52 - 0.77 | 0.52 - 0.76 | 0.5 - 0.7 | 0.52 - 0.75 | 0.5 - 0.65 | 0.53 - 0.74 |
| | | **LHH** | 0.5 - 0.8 | 0.6 - 0.78 | 0.5 - 0.74 | 0.6 - 0.79 | 0.5 - 0.78 | 0.58 - 0.81 |
| | | **LHL** | 0.54 - 0.76 | 0.52 - 0.75 | 0.5 - 0.64 | 0.52 - 0.7 | 0.5 - 0.7 | 0.53 - 0.77 |
| | | **LLH** | 0.5 - 0.72 | 0.52 - 0.72 | 0.5 - 0.73 | 0.52 - 0.77 | 0.5 - 0.7 | 0.54 - 0.7 |
| | | **LLL** | 0.52 - 0.73 | 0.54 - 0.75 | 0.5 - 0.68 | 0.52 - 0.8 | 0.5 - 0.66 | 0.52 - 0.79 |
| | **Shape** | | 0.51 - 0.7 | 0.5 - 0.73 | 0.5 - 0.7 | 0.51 - 0.8 | 0.51 - 0.73 | 0.51 - 0.75 |

**Supplemental Table 5.** EGFR Feature Selector Mean+Sd different modality, preprocessing and setting

| Modality | Preprocessed | Setting | SM | SKB | SP | VT | VT_SKB | VT_SM |
|---|---|---|---|---|---|---|---|---|
| **CT** | **Bin 64** | | 0.64 ± 0.056 | 0.63 ± 0.042 | 0.58 ± 0.056 | 0.65 ± 0.05 | 0.57 ± 0.053 | 0.65 ± 0.042 |
| | **LOG** | **Sigma: 0.5** | 0.64 ± 0.077 | 0.63 ± 0.054 | 0.57 ± 0.07 | 0.63 ± 0.046 | 0.56 ± 0.052 | 0.63 ± 0.042 |
| | | **Sigma: 1** | 0.63 ± 0.067 | 0.6 ± 0.062 | 0.56 ± 0.033 | 0.61 ± 0.057 | 0.56 ± 0.028 | 0.61 ± 0.047 |
| | | **Sigma: 1.5** | 0.61 ± 0.069 | 0.6 ± 0.066 | 0.55 ± 0.032 | 0.6 ± 0.06 | 0.55 ± 0.044 | 0.6 ± 0.054 |
| | | **sigma: 2** | 0.61 ± 0.077 | 0.6 ± 0.058 | 0.55 ± 0.033 | 0.61 ± 0.055 | 0.56 ± 0.043 | 0.61 ± 0.048 |
| | | **sigma: 2.5** | 0.6 ± 0.072 | 0.6 ± 0.069 | 0.55 ± 0.035 | 0.59 ± 0.058 | 0.57 ± 0.057 | 0.6 ± 0.051 |
| | | **sigma: 3** | 0.63 ± 0.072 | 0.61 ± 0.063 | 0.54 ± 0.036 | 0.61 ± 0.059 | 0.57 ± 0.053 | 0.63 ± 0.041 |
| | | **sigma: 3.5** | 0.63 ± 0.072 | 0.62 ± 0.067 | 0.55 ± 0.036 | 0.62 ± 0.064 | 0.55 ± 0.043 | 0.62 ± 0.058 |
| | | **sigma: 4** | 0.62 ± 0.069 | 0.6 ± 0.065 | 0.56 ± 0.049 | 0.6 ± 0.059 | 0.55 ± 0.045 | 0.61 ± 0.05 |
| | | **sigma: 4.5** | 0.61 ± 0.07 | 0.59 ± 0.065 | 0.57 ± 0.059 | 0.59 ± 0.061 | 0.56 ± 0.053 | 0.61 ± 0.062 |
| | | **Sigma: 5** | 0.61 ± 0.077 | 0.6 ± 0.068 | 0.56 ± 0.044 | 0.59 ± 0.056 | 0.55 ± 0.043 | 0.6 ± 0.051 |
| | **Shape** | | 0.57 ± 0.043 | 0.58 ± 0.038 | 0.57 ± 0.039 | 0.59 ± 0.057 | 0.58 ± 0.043 | 0.58 ± 0.042 |
| | **WAV** | **HHH** | 0.62 ± 0.07 | 0.6 ± 0.061 | 0.57 ± 0.051 | 0.6 ± 0.052 | 0.57 ± 0.045 | 0.6 ± 0.056 |
| | | **HHL** | 0.62 ± 0.073 | 0.62 ± 0.062 | 0.56 ± 0.04 | 0.62 ± 0.056 | 0.56 ± 0.04 | 0.62 ± 0.055 |
| | | **HLH** | 0.62 ± 0.069 | 0.62 ± 0.057 | 0.57 ± 0.061 | 0.62 ± 0.043 | 0.58 ± 0.051 | 0.63 ± 0.036 |
| | | **HLL** | 0.61 ± 0.071 | 0.58 ± 0.071 | 0.55 ± 0.033 | 0.58 ± 0.063 | 0.55 ± 0.03 | 0.59 ± 0.055 |
| | | **LHH** | 0.6 ± 0.073 | 0.61 ± 0.062 | 0.56 ± 0.041 | 0.6 ± 0.055 | 0.56 ± 0.031 | 0.6 ± 0.046 |
| | | **LHL** | 0.63 ± 0.068 | 0.64 ± 0.072 | 0.55 ± 0.046 | 0.64 ± 0.058 | 0.55 ± 0.035 | 0.63 ± 0.057 |
| | | **LLH** | 0.61 ± 0.069 | 0.61 ± 0.058 | 0.55 ± 0.05 | 0.61 ± 0.05 | 0.56 ± 0.04 | 0.62 ± 0.039 |
| | | **LLL** | 0.63 ± 0.076 | 0.62 ± 0.062 | 0.59 ± 0.075 | 0.62 ± 0.062 | 0.58 ± 0.059 | 0.62 ± 0.051 |
| **CTD** | **Bin 64** | | 0.65 ± 0.065 | 0.62 ± 0.057 | 0.56 ± 0.042 | 0.62 ± 0.052 | 0.56 ± 0.035 | 0.64 ± 0.053 |
| | **LOG** | **Sigma: 0.5** | 0.62 ± 0.056 | 0.63 ± 0.058 | 0.57 ± 0.05 | 0.63 ± 0.06 | 0.58 ± 0.049 | 0.62 ± 0.035 |
| | | **Sigma: 1** | 0.62 ± 0.051 | 0.62 ± 0.043 | 0.57 ± 0.055 | 0.63 ± 0.054 | 0.58 ± 0.047 | 0.62 ± 0.037 |
| | | **Sigma: 1.5** | 0.61 ± 0.054 | 0.61 ± 0.053 | 0.56 ± 0.052 | 0.63 ± 0.063 | 0.58 ± 0.045 | 0.63 ± 0.038 |
| | | **sigma: 2** | 0.64 ± 0.051 | 0.62 ± 0.056 | 0.6 ± 0.072 | 0.63 ± 0.06 | 0.59 ± 0.056 | 0.63 ± 0.033 |
| | | **sigma: 2.5** | 0.63 ± 0.052 | 0.62 ± 0.053 | 0.56 ± 0.035 | 0.63 ± 0.061 | 0.58 ± 0.07 | 0.62 ± 0.032 |
| | | **sigma: 3** | 0.62 ± 0.066 | 0.62 ± 0.064 | 0.57 ± 0.047 | 0.62 ± 0.064 | 0.59 ± 0.063 | 0.61 ± 0.045 |
| | | **sigma: 3.5** | 0.62 ± 0.054 | 0.62 ± 0.051 | 0.59 ± 0.062 | 0.62 ± 0.06 | 0.58 ± 0.056 | 0.61 ± 0.035 |
| | | **sigma: 4** | 0.62 ± 0.074 | 0.61 ± 0.048 | 0.59 ± 0.056 | 0.62 ± 0.054 | 0.59 ± 0.057 | 0.61 ± 0.047 |
| | | **sigma: 4.5** | 0.61 ± 0.055 | 0.62 ± 0.055 | 0.57 ± 0.048 | 0.61 ± 0.061 | 0.58 ± 0.058 | 0.62 ± 0.047 |
| | | **Sigma: 5** | 0.61 ± 0.073 | 0.61 ± 0.059 | 0.58 ± 0.05 | 0.62 ± 0.067 | 0.57 ± 0.049 | 0.61 ± 0.051 |
| | **Shape** | | 0.56 ± 0.035 | 0.58 ± 0.072 | 0.57 ± 0.07 | 0.57 ± 0.045 | 0.58 ± 0.063 | 0.58 ± 0.062 |
| | **WAV** | **HHH** | 0.63 ± 0.05 | 0.64 ± 0.038 | 0.62 ± 0.056 | 0.64 ± 0.047 | 0.61 ± 0.045 | 0.64 ± 0.027 |
| | | **HHL** | 0.62 ± 0.07 | 0.62 ± 0.056 | 0.58 ± 0.047 | 0.64 ± 0.062 | 0.59 ± 0.068 | 0.62 ± 0.053 |
| | | **HLH** | 0.62 ± 0.053 | 0.63 ± 0.043 | 0.58 ± 0.049 | 0.64 ± 0.051 | 0.58 ± 0.052 | 0.63 ± 0.043 |
| | | **HLL** | 0.6 ± 0.053 | 0.63 ± 0.048 | 0.56 ± 0.042 | 0.63 ± 0.059 | 0.56 ± 0.036 | 0.63 ± 0.04 |
| | | **LHH** | 0.64 ± 0.047 | 0.65 ± 0.051 | 0.58 ± 0.052 | 0.65 ± 0.061 | 0.6 ± 0.067 | 0.65 ± 0.047 |
| | | **LHL** | 0.63 ± 0.049 | 0.64 ± 0.055 | 0.59 ± 0.06 | 0.64 ± 0.054 | 0.57 ± 0.047 | 0.64 ± 0.043 |
| | | **LLH** | 0.66 ± 0.048 | 0.64 ± 0.028 | 0.58 ± 0.058 | 0.64 ± 0.041 | 0.57 ± 0.041 | 0.64 ± 0.039 |
| | | **LLL** | 0.69 ± 0.075 | 0.68 ± 0.065 | 0.59 ± 0.067 | 0.69 ± 0.063 | 0.6 ± 0.066 | 0.69 ± 0.048 |
| **PET** | **Bin 64** | | 0.66 ± 0.081 | 0.65 ± 0.083 | 0.58 ± 0.087 | 0.65 ± 0.079 | 0.58 ± 0.086 | 0.66 ± 0.067 |
| | **LOG** | **Sigma: 0.5** | 0.63 ± 0.075 | 0.63 ± 0.048 | 0.57 ± 0.064 | 0.63 ± 0.049 | 0.58 ± 0.064 | 0.65 ± 0.057 |
| | | **Sigma: 1** | 0.65 ± 0.074 | 0.65 ± 0.041 | 0.56 ± 0.054 | 0.65 ± 0.044 | 0.56 ± 0.052 | 0.66 ± 0.043 |
| | | **Sigma: 1.5** | 0.63 ± 0.065 | 0.63 ± 0.062 | 0.56 ± 0.071 | 0.64 ± 0.06 | 0.57 ± 0.067 | 0.64 ± 0.067 |
| | | **sigma: 2** | 0.64 ± 0.074 | 0.62 ± 0.069 | 0.56 ± 0.055 | 0.64 ± 0.074 | 0.57 ± 0.066 | 0.64 ± 0.064 |
| | | **sigma: 2.5** | 0.68 ± 0.08 | 0.66 ± 0.044 | 0.57 ± 0.086 | 0.66 ± 0.041 | 0.59 ± 0.091 | 0.67 ± 0.062 |
| | | **sigma: 3** | 0.64 ± 0.059 | 0.61 ± 0.059 | 0.55 ± 0.05 | 0.62 ± 0.069 | 0.55 ± 0.045 | 0.65 ± 0.061 |
| | | **sigma: 3.5** | 0.65 ± 0.065 | 0.63 ± 0.065 | 0.55 ± 0.053 | 0.63 ± 0.079 | 0.57 ± 0.065 | 0.64 ± 0.075 |
| | | **sigma: 4** | 0.63 ± 0.078 | 0.62 ± 0.066 | 0.57 ± 0.088 | 0.64 ± 0.073 | 0.56 ± 0.051 | 0.64 ± 0.071 |
| | | **sigma: 4.5** | 0.62 ± 0.065 | 0.61 ± 0.065 | 0.55 ± 0.049 | 0.61 ± 0.072 | 0.54 ± 0.04 | 0.61 ± 0.059 |
| | | **Sigma: 5** | 0.62 ± 0.054 | 0.62 ± 0.062 | 0.57 ± 0.063 | 0.62 ± 0.06 | 0.57 ± 0.067 | 0.62 ± 0.057 |
| | **Shape** | | 0.57 ± 0.057 | 0.58 ± 0.064 | 0.57 ± 0.069 | 0.59 ± 0.081 | 0.57 ± 0.069 | 0.6 ± 0.076 |
| | **WAV** | **HHH** | 0.68 ± 0.028 | 0.66 ± 0.051 | 0.63 ± 0.088 | 0.68 ± 0.056 | 0.62 ± 0.092 | 0.67 ± 0.059 |
| | | **HHL** | 0.64 ± 0.061 | 0.64 ± 0.052 | 0.6 ± 0.083 | 0.65 ± 0.059 | 0.57 ± 0.074 | 0.64 ± 0.049 |
| | | **HLH** | 0.62 ± 0.051 | 0.62 ± 0.054 | 0.58 ± 0.061 | 0.63 ± 0.054 | 0.58 ± 0.062 | 0.63 ± 0.055 |
| | | **HLL** | 0.64 ± 0.083 | 0.61 ± 0.071 | 0.55 ± 0.06 | 0.61 ± 0.076 | 0.55 ± 0.052 | 0.63 ± 0.065 |
| | | **LHH** | 0.72 ± 0.079 | 0.7 ± 0.06 | 0.62 ± 0.098 | 0.71 ± 0.065 | 0.65 ± 0.11 | 0.73 ± 0.074 |
| | | **LHL** | 0.64 ± 0.064 | 0.62 ± 0.063 | 0.55 ± 0.048 | 0.62 ± 0.049 | 0.57 ± 0.068 | 0.63 ± 0.067 |
| | | **LLH** | 0.64 ± 0.066 | 0.61 ± 0.06 | 0.56 ± 0.07 | 0.62 ± 0.071 | 0.56 ± 0.064 | 0.63 ± 0.055 |
| | | **LLL** | 0.65 ± 0.072 | 0.64 ± 0.07 | 0.57 ± 0.066 | 0.65 ± 0.084 | 0.56 ± 0.061 | 0.66 ± 0.091 |

**Supplemental Table 6.** EGFR Classifier Min-Max different modality, preprocessing and setting

| Modality | Preprocessed | Setting | AB | BAG | BNB | DT | GNB | KNN | LREG | MLP | QDA | RF | SGD | SVM |
|---|---|---|---|---|---|---|---|---|---|---|---|---|---|---|
| **CT** | **Bin 64** | | 0.6 - 0.72 | 0.52 - 0.63 | 0.6 - 0.71 | 0.6 - 0.73 | 0.57 - 0.65 | 0.6 - 0.66 | 0.62 - 0.64 | 0.56 - 0.68 | 0.5 - 0.6 | 0.59 - 0.68 | 0.52 - 0.7 | 0.52 - 0.62 |
| | **LOG** | **Sigma: 0.5** | 0.62 - 0.72 | 0.5 - 0.58 | 0.65 - 0.68 | 0.62 - 0.75 | 0.58 - 0.63 | 0.51 - 0.6 | 0.51 - 0.65 | 0.6 - 0.72 | 0.5 - 0.58 | 0.55 - 0.68 | 0.54 - 0.75 | 0.64 - 0.66 |
| | | **Sigma: 1** | 0.58 - 0.72 | 0.52 - 0.57 | 0.63 - 0.67 | 0.57 - 0.75 | 0.52 - 0.61 | 0.53 - 0.64 | 0.56 - 0.6 | 0.57 - 0.63 | 0.5 - 0.58 | 0.57 - 0.68 | 0.53 - 0.68 | 0.55 - 0.6 |
| | | **Sigma: 1.5** | 0.6 - 0.72 | 0.51 - 0.6 | 0.58 - 0.65 | 0.58 - 0.75 | 0.55 - 0.65 | 0.52 - 0.58 | 0.52 - 0.59 | 0.54 - 0.62 | 0.5 - 0.57 | 0.55 - 0.68 | 0.51 - 0.65 | 0.53 - 0.58 |
| | | **sigma: 2** | 0.6 - 0.72 | 0.5 - 0.55 | 0.53 - 0.59 | 0.56 - 0.75 | 0.55 - 0.65 | 0.53 - 0.55 | 0.52 - 0.6 | 0.57 - 0.67 | 0.5 - 0.62 | 0.55 - 0.68 | 0.53 - 0.67 | 0.53 - 0.6 |
| | | **sigma: 2.5** | 0.6 - 0.72 | 0.51 - 0.54 | 0.53 - 0.58 | 0.62 - 0.75 | 0.55 - 0.63 | 0.53 - 0.54 | 0.54 - 0.62 | 0.57 - 0.62 | 0.5 - 0.6 | 0.57 - 0.68 | 0.53 - 0.63 | 0.56 - 0.58 |
| | | **sigma: 3** | 0.57 - 0.72 | 0.5 - 0.57 | 0.53 - 0.6 | 0.58 - 0.75 | 0.55 - 0.68 | 0.55 - 0.63 | 0.52 - 0.65 | 0.54 - 0.63 | 0.5 - 0.57 | 0.62 - 0.68 | 0.51 - 0.69 | 0.53 - 0.65 |
| | | **sigma: 3.5** | 0.54 - 0.72 | 0.5 - 0.55 | 0.53 - 0.68 | 0.58 - 0.75 | 0.55 - 0.64 | 0.53 - 0.57 | 0.53 - 0.66 | 0.55 - 0.67 | 0.5 - 0.59 | 0.58 - 0.68 | 0.52 - 0.68 | 0.53 - 0.67 |
| | | **sigma: 4** | 0.6 - 0.72 | 0.5 - 0.57 | 0.58 - 0.6 | 0.62 - 0.75 | 0.55 - 0.6 | 0.52 - 0.62 | 0.51 - 0.63 | 0.57 - 0.6 | 0.5 - 0.57 | 0.57 - 0.68 | 0.53 - 0.68 | 0.57 - 0.63 |
| | | **sigma: 4.5** | 0.65 - 0.72 | 0.5 - 0.53 | 0.53 - 0.58 | 0.63 - 0.75 | 0.55 - 0.59 | 0.5 - 0.57 | 0.55 - 0.62 | 0.57 - 0.63 | 0.5 - 0.6 | 0.6 - 0.68 | 0.55 - 0.63 | 0.52 - 0.62 |
| | | **Sigma: 5** | 0.57 - 0.72 | 0.5 - 0.55 | 0.6 - 0.7 | 0.58 - 0.75 | 0.55 - 0.58 | 0.52 - 0.57 | 0.52 - 0.6 | 0.57 - 0.65 | 0.5 - 0.58 | 0.57 - 0.68 | 0.57 - 0.65 | 0.53 - 0.63 |
| | **Shape** | | 0.58 - 0.65 | 0.55 - 0.57 | 0.52 - 0.59 | 0.6 - 0.67 | 0.53 - 0.57 | 0.57 - 0.59 | 0.52 - 0.54 | 0.58 - 0.72 | 0.55 - 0.57 | 0.57 - 0.6 | 0.57 - 0.65 | 0.52 - 0.52 |
| | **WAV** | **HHH** | 0.6 - 0.72 | 0.53 - 0.57 | 0.62 - 0.63 | 0.61 - 0.75 | 0.58 - 0.64 | 0.53 - 0.6 | 0.52 - 0.58 | 0.58 - 0.67 | 0.5 - 0.53 | 0.54 - 0.68 | 0.6 - 0.67 | 0.55 - 0.6 |
| | | **HHL** | 0.58 - 0.72 | 0.5 - 0.54 | 0.65 - 0.68 | 0.55 - 0.75 | 0.58 - 0.65 | 0.58 - 0.6 | 0.52 - 0.58 | 0.6 - 0.65 | 0.5 - 0.56 | 0.57 - 0.68 | 0.51 - 0.68 | 0.57 - 0.62 |
| | | **HLH** | 0.58 - 0.72 | 0.5 - 0.6 | 0.64 - 0.68 | 0.59 - 0.75 | 0.57 - 0.63 | 0.52 - 0.61 | 0.6 - 0.62 | 0.62 - 0.68 | 0.5 - 0.57 | 0.6 - 0.68 | 0.51 - 0.67 | 0.55 - 0.6 |
| | | **HLL** | 0.59 - 0.72 | 0.52 - 0.57 | 0.53 - 0.62 | 0.57 - 0.75 | 0.53 - 0.55 | 0.53 - 0.6 | 0.52 - 0.61 | 0.58 - 0.65 | 0.5 - 0.55 | 0.58 - 0.68 | 0.52 - 0.63 | 0.55 - 0.6 |
| | | **LHH** | 0.6 - 0.72 | 0.52 - 0.55 | 0.62 - 0.65 | 0.57 - 0.75 | 0.58 - 0.62 | 0.54 - 0.56 | 0.5 - 0.58 | 0.53 - 0.62 | 0.5 - 0.56 | 0.59 - 0.68 | 0.57 - 0.68 | 0.53 - 0.56 |
| | | **LHL** | 0.58 - 0.72 | 0.5 - 0.6 | 0.65 - 0.73 | 0.58 - 0.75 | 0.57 - 0.64 | 0.51 - 0.57 | 0.52 - 0.63 | 0.58 - 0.75 | 0.5 - 0.55 | 0.57 - 0.68 | 0.5 - 0.7 | 0.6 - 0.64 |
| | | **LLH** | 0.58 - 0.72 | 0.5 - 0.62 | 0.58 - 0.65 | 0.6 - 0.75 | 0.58 - 0.62 | 0.5 - 0.62 | 0.53 - 0.59 | 0.58 - 0.65 | 0.5 - 0.6 | 0.57 - 0.68 | 0.52 - 0.67 | 0.51 - 0.55 |
| | | **LLL** | 0.65 - 0.73 | 0.51 - 0.6 | 0.56 - 0.57 | 0.6 - 0.75 | 0.58 - 0.68 | 0.58 - 0.65 | 0.57 - 0.62 | 0.55 - 0.7 | 0.5 - 0.61 | 0.62 - 0.7 | 0.53 - 0.7 | 0.52 - 0.57 |
| **CTD** | **Bin 65** | | 0.58 - 0.65 | 0.5 - 0.6 | 0.6 - 0.62 | 0.57 - 0.65 | 0.57 - 0.69 | 0.52 - 0.67 | 0.52 - 0.63 | 0.58 - 0.68 | 0.5 - 0.58 | 0.6 - 0.65 | 0.55 - 0.78 | 0.52 - 0.65 |
| | **LOG** | **Sigma: 0.5** | 0.62 - 0.7 | 0.5 - 0.6 | 0.65 - 0.67 | 0.57 - 0.68 | 0.57 - 0.65 | 0.53 - 0.57 | 0.57 - 0.68 | 0.61 - 0.68 | 0.5 - 0.58 | 0.58 - 0.7 | 0.53 - 0.75 | 0.58 - 0.62 |
| | | **Sigma: 1** | 0.61 - 0.7 | 0.51 - 0.58 | 0.58 - 0.62 | 0.63 - 0.68 | 0.55 - 0.62 | 0.52 - 0.6 | 0.56 - 0.67 | 0.58 - 0.67 | 0.5 - 0.58 | 0.6 - 0.7 | 0.55 - 0.65 | 0.58 - 0.62 |
| | | **Sigma: 1.5** | 0.6 - 0.73 | 0.5 - 0.58 | 0.57 - 0.58 | 0.58 - 0.68 | 0.57 - 0.6 | 0.51 - 0.61 | 0.56 - 0.67 | 0.57 - 0.67 | 0.5 - 0.63 | 0.63 - 0.7 | 0.51 - 0.65 | 0.58 - 0.67 |
| | | **sigma: 2** | 0.63 - 0.7 | 0.5 - 0.6 | 0.6 - 0.65 | 0.6 - 0.68 | 0.58 - 0.68 | 0.53 - 0.61 | 0.57 - 0.62 | 0.6 - 0.7 | 0.5 - 0.58 | 0.63 - 0.72 | 0.62 - 0.7 | 0.6 - 0.68 |
| | | **sigma: 2.5** | 0.58 - 0.7 | 0.5 - 0.6 | 0.55 - 0.58 | 0.59 - 0.68 | 0.55 - 0.67 | 0.55 - 0.64 | 0.54 - 0.63 | 0.6 - 0.69 | 0.5 - 0.58 | 0.58 - 0.7 | 0.54 - 0.69 | 0.52 - 0.63 |
| | | **sigma: 3** | 0.58 - 0.7 | 0.5 - 0.53 | 0.54 - 0.57 | 0.58 - 0.68 | 0.58 - 0.65 | 0.55 - 0.58 | 0.54 - 0.68 | 0.61 - 0.68 | 0.5 - 0.58 | 0.63 - 0.7 | 0.55 - 0.7 | 0.57 - 0.65 |
| | | **sigma: 3.5** | 0.63 - 0.7 | 0.5 - 0.6 | 0.55 - 0.59 | 0.6 - 0.68 | 0.56 - 0.62 | 0.53 - 0.67 | 0.56 - 0.68 | 0.6 - 0.7 | 0.5 - 0.59 | 0.62 - 0.7 | 0.6 - 0.68 | 0.6 - 0.63 |
| | | **sigma: 4** | 0.63 - 0.7 | 0.5 - 0.57 | 0.55 - 0.65 | 0.59 - 0.75 | 0.56 - 0.62 | 0.55 - 0.58 | 0.56 - 0.62 | 0.6 - 0.67 | 0.5 - 0.57 | 0.58 - 0.7 | 0.58 - 0.71 | 0.58 - 0.62 |
| | | **sigma: 4.5** | 0.58 - 0.7 | 0.5 - 0.57 | 0.55 - 0.67 | 0.55 - 0.68 | 0.54 - 0.64 | 0.57 - 0.57 | 0.55 - 0.62 | 0.6 - 0.67 | 0.5 - 0.58 | 0.56 - 0.7 | 0.64 - 0.68 | 0.56 - 0.62 |
| | | **Sigma: 5** | 0.6 - 0.7 | 0.5 - 0.53 | 0.55 - 0.57 | 0.6 - 0.68 | 0.54 - 0.62 | 0.57 - 0.65 | 0.57 - 0.63 | 0.6 - 0.77 | 0.5 - 0.57 | 0.58 - 0.7 | 0.6 - 0.7 | 0.52 - 0.61 |
| | **Shape** | | 0.55 - 0.62 | 0.52 - 0.53 | 0.54 - 0.61 | 0.55 - 0.62 | 0.53 - 0.55 | 0.52 - 0.53 | 0.55 - 0.55 | 0.57 - 0.77 | 0.53 - 0.58 | 0.57 - 0.6 | 0.62 - 0.7 | 0.52 - 0.53 |
| | **WAV** | **HHH** | 0.62 - 0.72 | 0.57 - 0.67 | 0.63 - 0.65 | 0.6 - 0.68 | 0.56 - 0.66 | 0.58 - 0.66 | 0.55 - 0.62 | 0.63 - 0.67 | 0.5 - 0.59 | 0.65 - 0.7 | 0.61 - 0.7 | 0.61 - 0.64 |
| | | **HHL** | 0.57 - 0.7 | 0.52 - 0.57 | 0.67 - 0.67 | 0.63 - 0.73 | 0.56 - 0.64 | 0.53 - 0.57 | 0.52 - 0.63 | 0.59 - 0.68 | 0.5 - 0.6 | 0.63 - 0.7 | 0.6 - 0.67 | 0.57 - 0.62 |
| | | **HLH** | 0.62 - 0.7 | 0.5 - 0.62 | 0.63 - 0.66 | 0.58 - 0.68 | 0.55 - 0.67 | 0.53 - 0.62 | 0.59 - 0.63 | 0.57 - 0.65 | 0.5 - 0.62 | 0.6 - 0.7 | 0.59 - 0.68 | 0.56 - 0.62 |
| | | **HLL** | 0.59 - 0.7 | 0.5 - 0.6 | 0.63 - 0.63 | 0.57 - 0.68 | 0.55 - 0.6 | 0.55 - 0.64 | 0.53 - 0.63 | 0.6 - 0.72 | 0.5 - 0.57 | 0.57 - 0.7 | 0.6 - 0.63 | 0.55 - 0.6 |
| | | **LHH** | 0.63 - 0.7 | 0.52 - 0.65 | 0.63 - 0.64 | 0.62 - 0.68 | 0.59 - 0.65 | 0.57 - 0.65 | 0.6 - 0.67 | 0.62 - 0.73 | 0.5 - 0.6 | 0.62 - 0.7 | 0.6 - 0.77 | 0.5 - 0.65 |
| | | **LHL** | 0.59 - 0.7 | 0.5 - 0.62 | 0.65 - 0.68 | 0.63 - 0.68 | 0.55 - 0.69 | 0.53 - 0.63 | 0.55 - 0.63 | 0.6 - 0.69 | 0.5 - 0.58 | 0.6 - 0.7 | 0.56 - 0.71 | 0.6 - 0.64 |
| | | **LLH** | 0.61 - 0.73 | 0.52 - 0.65 | 0.62 - 0.64 | 0.56 - 0.68 | 0.57 - 0.66 | 0.59 - 0.69 | 0.52 - 0.65 | 0.57 - 0.67 | 0.5 - 0.62 | 0.63 - 0.7 | 0.6 - 0.69 | 0.58 - 0.67 |
| | | **LLL** | 0.63 - 0.72 | 0.51 - 0.7 | 0.62 - 0.73 | 0.62 - 0.68 | 0.53 - 0.74 | 0.58 - 0.68 | 0.52 - 0.75 | 0.63 - 0.75 | 0.5 - 0.62 | 0.64 - 0.72 | 0.62 - 0.78 | 0.6 - 0.77 |
| **PET** | **Bin 66** | | 0.72 - 0.77 | 0.5 - 0.58 | 0.74 - 0.76 | 0.64 - 0.71 | 0.56 - 0.61 | 0.55 - 0.65 | 0.52 - 0.63 | 0.52 - 0.78 | 0.5 - 0.66 | 0.68 - 0.73 | 0.5 - 0.72 | 0.59 - 0.65 |
| | **LOG** | **Sigma: 0.5** | 0.63 - 0.68 | 0.5 - 0.6 | 0.68 - 0.72 | 0.61 - 0.67 | 0.54 - 0.75 | 0.59 - 0.64 | 0.52 - 0.62 | 0.54 - 0.78 | 0.5 - 0.6 | 0.55 - 0.7 | 0.5 - 0.73 | 0.57 - 0.64 |
| | | **Sigma: 1** | 0.63 - 0.66 | 0.5 - 0.62 | 0.71 - 0.78 | 0.57 - 0.66 | 0.54 - 0.7 | 0.58 - 0.64 | 0.53 - 0.7 | 0.54 - 0.67 | 0.5 - 0.62 | 0.56 - 0.7 | 0.5 - 0.73 | 0.58 - 0.7 |
| | | **Sigma: 1.5** | 0.65 - 0.69 | 0.5 - 0.58 | 0.73 - 0.77 | 0.61 - 0.68 | 0.57 - 0.67 | 0.59 - 0.65 | 0.51 - 0.62 | 0.5 - 0.63 | 0.5 - 0.61 | 0.6 - 0.7 | 0.5 - 0.73 | 0.51 - 0.6 |
| | | **sigma: 2** | 0.61 - 0.66 | 0.5 - 0.58 | 0.71 - 0.77 | 0.61 - 0.66 | 0.53 - 0.63 | 0.58 - 0.64 | 0.51 - 0.67 | 0.54 - 0.65 | 0.5 - 0.53 | 0.62 - 0.7 | 0.5 - 0.78 | 0.57 - 0.61 |
| | | **sigma: 2.5** | 0.65 - 0.75 | 0.5 - 0.66 | 0.64 - 0.75 | 0.65 - 0.77 | 0.57 - 0.62 | 0.64 - 0.68 | 0.52 - 0.68 | 0.51 - 0.78 | 0.5 - 0.59 | 0.6 - 0.73 | 0.5 - 0.76 | 0.55 - 0.72 |
| | | **sigma: 3** | 0.57 - 0.66 | 0.5 - 0.68 | 0.7 - 0.76 | 0.62 - 0.66 | 0.57 - 0.63 | 0.54 - 0.65 | 0.5 - 0.57 | 0.52 - 0.74 | 0.5 - 0.59 | 0.6 - 0.7 | 0.5 - 0.74 | 0.55 - 0.62 |
| | | **sigma: 3.5** | 0.61 - 0.67 | 0.5 - 0.69 | 0.7 - 0.72 | 0.6 - 0.7 | 0.57 - 0.62 | 0.53 - 0.67 | 0.51 - 0.61 | 0.56 - 0.63 | 0.5 - 0.58 | 0.56 - 0.7 | 0.5 - 0.82 | 0.58 - 0.65 |
| | | **sigma: 4** | 0.59 - 0.66 | 0.51 - 0.65 | 0.71 - 0.74 | 0.65 - 0.68 | 0.57 - 0.65 | 0.55 - 0.6 | 0.51 - 0.57 | 0.55 - 0.78 | 0.5 - 0.58 | 0.56 - 0.7 | 0.5 - 0.81 | 0.54 - 0.58 |
| | | **sigma: 4.5** | 0.61 - 0.66 | 0.51 - 0.6 | 0.7 - 0.71 | 0.59 - 0.66 | 0.56 - 0.63 | 0.55 - 0.55 | 0.51 - 0.62 | 0.54 - 0.64 | 0.5 - 0.64 | 0.59 - 0.7 | 0.5 - 0.73 | 0.52 - 0.57 |
| | | **Sigma: 5** | 0.56 - 0.66 | 0.52 - 0.72 | 0.7 - 0.73 | 0.58 - 0.66 | 0.59 - 0.62 | 0.55 - 0.57 | 0.51 - 0.61 | 0.55 - 0.72 | 0.5 - 0.58 | 0.58 - 0.7 | 0.5 - 0.69 | 0.62 - 0.64 |
| | **Shape** | | 0.56 - 0.63 | 0.5 - 0.51 | 0.56 - 0.67 | 0.62 - 0.67 | 0.52 - 0.57 | 0.54 - 0.55 | 0.52 - 0.52 | 0.62 - 0.8 | 0.55 - 0.58 | 0.55 - 0.6 | 0.62 - 0.75 | 0.52 - 0.52 |
| | **WAV** | **HHH** | 0.65 - 0.72 | 0.64 - 0.73 | 0.65 - 0.67 | 0.66 - 0.71 | 0.66 - 0.7 | 0.69 - 0.73 | 0.52 - 0.64 | 0.52 - 0.66 | 0.5 - 0.7 | 0.72 - 0.75 | 0.5 - 0.79 | 0.58 - 0.68 |
| | | **HHL** | 0.63 - 0.66 | 0.52 - 0.72 | 0.68 - 0.71 | 0.57 - 0.66 | 0.64 - 0.68 | 0.53 - 0.61 | 0.52 - 0.65 | 0.54 - 0.69 | 0.5 - 0.56 | 0.69 - 0.72 | 0.5 - 0.73 | 0.62 - 0.66 |
| | | **HLH** | 0.6 - 0.66 | 0.55 - 0.65 | 0.67 - 0.71 | 0.57 - 0.66 | 0.66 - 0.69 | 0.62 - 0.62 | 0.52 - 0.64 | 0.52 - 0.63 | 0.5 - 0.55 | 0.65 - 0.7 | 0.5 - 0.71 | 0.54 - 0.6 |
| | | **HLL** | 0.57 - 0.66 | 0.51 - 0.58 | 0.65 - 0.77 | 0.65 - 0.7 | 0.56 - 0.66 | 0.52 - 0.62 | 0.51 - 0.58 | 0.53 - 0.77 | 0.5 - 0.57 | 0.58 - 0.7 | 0.5 - 0.76 | 0.56 - 0.6 |
| | | **LHH** | 0.7 - 0.75 | 0.64 - 0.68 | 0.68 - 0.7 | 0.68 - 0.74 | 0.61 - 0.71 | 0.68 - 0.73 | 0.52 - 0.8 | 0.5 - 0.81 | 0.5 - 0.6 | 0.73 - 0.79 | 0.5 - 0.8 | 0.65 - 0.78 |
| | | **LHL** | 0.63 - 0.66 | 0.5 - 0.62 | 0.65 - 0.71 | 0.57 - 0.66 | 0.56 - 0.66 | 0.52 - 0.59 | 0.53 - 0.63 | 0.54 - 0.77 | 0.5 - 0.57 | 0.62 - 0.7 | 0.5 - 0.75 | 0.56 - 0.62 |
| | | **LLH** | 0.57 - 0.69 | 0.5 - 0.64 | 0.63 - 0.71 | 0.59 - 0.66 | 0.54 - 0.65 | 0.53 - 0.64 | 0.52 - 0.63 | 0.51 - 0.72 | 0.5 - 0.56 | 0.68 - 0.77 | 0.5 - 0.72 | 0.58 - 0.64 |
| | | **LLL** | 0.61 - 0.66 | 0.5 - 0.56 | 0.69 - 0.72 | 0.64 - 0.72 | 0.56 - 0.63 | 0.55 - 0.61 | 0.51 - 0.77 | 0.53 - 0.75 | 0.5 - 0.54 | 0.65 - 0.7 | 0.5 - 0.8 | 0.54 - 0.69 |

**Supplemental Table 7.** EGFR Classifier Mean+Sd different modality, preprocessing and settin

| Modality | Preprocessed | Setting | AB | BAG | BNB | DT | GNB | KNN | LREG | MLP | QDA | RF | SGD | SVM |
|---|---|---|---|---|---|---|---|---|---|---|---|---|---|---|
| **CT** | **Bin 64** | | 0.66 ± 0.049 | 0.58 ± 0.048 | 0.68 ± 0.054 | 0.67 ± 0.064 | 0.61 ± 0.035 | 0.62 ± 0.024 | 0.63 ± 0.01 | 0.64 ± 0.063 | 0.54 ± 0.049 | 0.64 ± 0.036 | 0.61 ± 0.073 | 0.59 ± 0.042 |
| | **LOG** | **Sigma: 0.5** | 0.66 ± 0.034 | 0.54 ± 0.032 | 0.67 ± 0.017 | 0.68 ± 0.06 | 0.59 ± 0.018 | 0.55 ± 0.034 | 0.6 ± 0.066 | 0.64 ± 0.044 | 0.54 ± 0.04 | 0.62 ± 0.052 | 0.62 ± 0.078 | 0.64 ± 0.0082 |
| | | **Sigma: 1** | 0.65 ± 0.05 | 0.54 ± 0.016 | 0.66 ± 0.017 | 0.67 ± 0.079 | 0.57 ± 0.031 | 0.58 ± 0.038 | 0.58 ± 0.016 | 0.58 ± 0.026 | 0.52 ± 0.032 | 0.62 ± 0.042 | 0.61 ± 0.057 | 0.58 ± 0.018 |
| | | **Sigma: 1.5** | 0.66 ± 0.04 | 0.55 ± 0.03 | 0.62 ± 0.032 | 0.67 ± 0.075 | 0.6 ± 0.04 | 0.54 ± 0.022 | 0.56 ± 0.024 | 0.57 ± 0.027 | 0.51 ± 0.027 | 0.62 ± 0.047 | 0.59 ± 0.061 | 0.56 ± 0.024 |
| | | **sigma: 2** | 0.65 ± 0.043 | 0.53 ± 0.021 | 0.56 ± 0.027 | 0.66 ± 0.084 | 0.6 ± 0.039 | 0.54 ± 0.0091 | 0.56 ± 0.025 | 0.61 ± 0.034 | 0.55 ± 0.058 | 0.61 ± 0.05 | 0.6 ± 0.056 | 0.57 ± 0.028 |
| | | **sigma: 2.5** | 0.68 ± 0.043 | 0.52 ± 0.013 | 0.55 ± 0.025 | 0.68 ± 0.06 | 0.59 ± 0.035 | 0.53 ± 0.0017 | 0.56 ± 0.03 | 0.6 ± 0.02 | 0.54 ± 0.042 | 0.62 ± 0.043 | 0.58 ± 0.042 | 0.58 ± 0.011 |
| | | **sigma: 3** | 0.65 ± 0.052 | 0.53 ± 0.026 | 0.56 ± 0.034 | 0.68 ± 0.064 | 0.6 ± 0.049 | 0.58 ± 0.039 | 0.6 ± 0.063 | 0.6 ± 0.031 | 0.52 ± 0.029 | 0.64 ± 0.025 | 0.61 ± 0.062 | 0.62 ± 0.044 |
| | | **sigma: 3.5** | 0.64 ± 0.063 | 0.52 ± 0.021 | 0.57 ± 0.073 | 0.67 ± 0.073 | 0.6 ± 0.038 | 0.54 ± 0.014 | 0.61 ± 0.062 | 0.63 ± 0.04 | 0.53 ± 0.036 | 0.62 ± 0.036 | 0.62 ± 0.073 | 0.62 ± 0.069 |
| | | **sigma: 4** | 0.65 ± 0.041 | 0.52 ± 0.025 | 0.59 ± 0.0088 | 0.68 ± 0.056 | 0.58 ± 0.025 | 0.54 ± 0.04 | 0.55 ± 0.045 | 0.58 ± 0.015 | 0.53 ± 0.037 | 0.63 ± 0.039 | 0.63 ± 0.05 | 0.61 ± 0.031 |
| | | **sigma: 4.5** | 0.67 ± 0.028 | 0.52 ± 0.011 | 0.55 ± 0.025 | 0.7 ± 0.052 | 0.56 ± 0.014 | 0.53 ± 0.024 | 0.6 ± 0.026 | 0.59 ± 0.025 | 0.53 ± 0.039 | 0.63 ± 0.031 | 0.6 ± 0.029 | 0.59 ± 0.044 |
| | | **Sigma: 5** | 0.64 ± 0.057 | 0.53 ± 0.02 | 0.62 ± 0.05 | 0.68 ± 0.068 | 0.56 ± 0.014 | 0.53 ± 0.018 | 0.56 ± 0.03 | 0.61 ± 0.031 | 0.52 ± 0.032 | 0.62 ± 0.043 | 0.6 ± 0.037 | 0.58 ± 0.034 |
| | **Shape** | | 0.62 ± 0.029 | 0.56 ± 0.0085 | 0.56 ± 0.031 | 0.63 ± 0.021 | 0.54 ± 0.015 | 0.58 ± 0.008 | 0.54 ± 0.0094 | 0.64 ± 0.048 | 0.55 ± 0.0079 | 0.59 ± 0.012 | 0.61 ± 0.027 | 0.52 ± 0.0031 |
| | **WAV** | **HHH** | 0.65 ± 0.043 | 0.55 ± 0.015 | 0.62 ± 0.0083 | 0.68 ± 0.062 | 0.6 ± 0.028 | 0.55 ± 0.027 | 0.55 ± 0.026 | 0.62 ± 0.035 | 0.51 ± 0.016 | 0.61 ± 0.052 | 0.63 ± 0.033 | 0.56 ± 0.022 |
| | | **HHL** | 0.64 ± 0.054 | 0.53 ± 0.016 | 0.67 ± 0.017 | 0.66 ± 0.083 | 0.62 ± 0.03 | 0.59 ± 0.0064 | 0.55 ± 0.022 | 0.63 ± 0.017 | 0.53 ± 0.031 | 0.62 ± 0.046 | 0.62 ± 0.083 | 0.58 ± 0.022 |
| | | **HLH** | 0.66 ± 0.047 | 0.55 ± 0.035 | 0.65 ± 0.021 | 0.68 ± 0.066 | 0.59 ± 0.024 | 0.57 ± 0.046 | 0.61 ± 0.01 | 0.65 ± 0.025 | 0.53 ± 0.037 | 0.64 ± 0.03 | 0.61 ± 0.054 | 0.57 ± 0.018 |
| | | **HLL** | 0.65 ± 0.052 | 0.54 ± 0.015 | 0.56 ± 0.039 | 0.67 ± 0.075 | 0.54 ± 0.0062 | 0.54 ± 0.027 | 0.53 ± 0.038 | 0.61 ± 0.028 | 0.51 ± 0.019 | 0.62 ± 0.039 | 0.59 ± 0.046 | 0.56 ± 0.021 |
| | | **LHH** | 0.65 ± 0.043 | 0.54 ± 0.013 | 0.63 ± 0.017 | 0.66 ± 0.081 | 0.59 ± 0.014 | 0.56 ± 0.0085 | 0.56 ± 0.03 | 0.59 ± 0.033 | 0.52 ± 0.027 | 0.62 ± 0.037 | 0.61 ± 0.045 | 0.55 ± 0.011 |
| | | **LHL** | 0.65 ± 0.049 | 0.56 ± 0.037 | 0.7 ± 0.037 | 0.67 ± 0.075 | 0.61 ± 0.027 | 0.54 ± 0.025 | 0.57 ± 0.045 | 0.64 ± 0.068 | 0.52 ± 0.027 | 0.62 ± 0.043 | 0.63 ± 0.085 | 0.62 ± 0.016 |
| | | **LLH** | 0.65 ± 0.047 | 0.57 ± 0.049 | 0.63 ± 0.033 | 0.68 ± 0.063 | 0.6 ± 0.017 | 0.58 ± 0.053 | 0.57 ± 0.025 | 0.6 ± 0.025 | 0.52 ± 0.039 | 0.63 ± 0.037 | 0.59 ± 0.049 | 0.54 ± 0.018 |
| | | **LLL** | 0.7 ± 0.039 | 0.56 ± 0.04 | 0.56 ± 0.0039 | 0.69 ± 0.057 | 0.61 ± 0.038 | 0.62 ± 0.03 | 0.59 ± 0.025 | 0.62 ± 0.051 | 0.53 ± 0.041 | 0.66 ± 0.029 | 0.63 ± 0.063 | 0.55 ± 0.016 |
| **CTD** | **Bin 64** | | 0.62 ± 0.024 | 0.56 ± 0.039 | 0.61 ± 0.0083 | 0.61 ± 0.036 | 0.61 ± 0.044 | 0.6 ± 0.055 | 0.6 ± 0.042 | 0.62 ± 0.036 | 0.53 ± 0.036 | 0.62 ± 0.019 | 0.71 ± 0.11 | 0.61 ± 0.06 |
| | **LOG** | **Sigma: 0.5** | 0.66 ± 0.033 | 0.54 ± 0.041 | 0.66 ± 0.0083 | 0.64 ± 0.042 | 0.62 ± 0.041 | 0.55 ± 0.014 | 0.6 ± 0.042 | 0.63 ± 0.027 | 0.54 ± 0.046 | 0.65 ± 0.041 | 0.63 ± 0.072 | 0.6 ± 0.013 |
| | | **Sigma: 1** | 0.66 ± 0.039 | 0.55 ± 0.035 | 0.6 ± 0.014 | 0.66 ± 0.02 | 0.57 ± 0.023 | 0.56 ± 0.037 | 0.63 ± 0.042 | 0.63 ± 0.035 | 0.53 ± 0.036 | 0.65 ± 0.034 | 0.61 ± 0.04 | 0.61 ± 0.014 |
| | | **Sigma: 1.5** | 0.66 ± 0.049 | 0.54 ± 0.038 | 0.57 ± 0.0096 | 0.65 ± 0.04 | 0.59 ± 0.014 | 0.56 ± 0.034 | 0.63 ± 0.05 | 0.62 ± 0.037 | 0.53 ± 0.052 | 0.66 ± 0.026 | 0.6 ± 0.053 | 0.63 ± 0.027 |
| | | **sigma: 2** | 0.67 ± 0.025 | 0.56 ± 0.049 | 0.62 ± 0.021 | 0.65 ± 0.03 | 0.63 ± 0.038 | 0.56 ± 0.031 | 0.6 ± 0.019 | 0.67 ± 0.035 | 0.52 ± 0.032 | 0.67 ± 0.034 | 0.65 ± 0.035 | 0.63 ± 0.03 |
| | | **sigma: 2.5** | 0.65 ± 0.041 | 0.55 ± 0.041 | 0.56 ± 0.013 | 0.64 ± 0.036 | 0.63 ± 0.05 | 0.59 ± 0.035 | 0.58 ± 0.034 | 0.64 ± 0.033 | 0.52 ± 0.032 | 0.65 ± 0.04 | 0.65 ± 0.057 | 0.59 ± 0.046 |
| | | **sigma: 3** | 0.65 ± 0.047 | 0.52 ± 0.017 | 0.56 ± 0.014 | 0.64 ± 0.041 | 0.62 ± 0.028 | 0.56 ± 0.013 | 0.59 ± 0.05 | 0.65 ± 0.033 | 0.52 ± 0.033 | 0.66 ± 0.028 | 0.66 ± 0.055 | 0.62 ± 0.034 |
| | | **sigma: 3.5** | 0.67 ± 0.024 | 0.55 ± 0.042 | 0.57 ± 0.015 | 0.64 ± 0.037 | 0.59 ± 0.018 | 0.6 ± 0.056 | 0.61 ± 0.041 | 0.64 ± 0.033 | 0.52 ± 0.035 | 0.66 ± 0.029 | 0.65 ± 0.031 | 0.61 ± 0.013 |
| | | **sigma: 4** | 0.67 ± 0.023 | 0.53 ± 0.025 | 0.59 ± 0.04 | 0.66 ± 0.055 | 0.6 ± 0.02 | 0.58 ± 0.014 | 0.6 ± 0.026 | 0.64 ± 0.029 | 0.52 ± 0.028 | 0.65 ± 0.041 | 0.65 ± 0.045 | 0.59 ± 0.014 |
| | | **sigma: 4.5** | 0.65 ± 0.041 | 0.53 ± 0.026 | 0.64 ± 0.06 | 0.63 ± 0.051 | 0.6 ± 0.033 | 0.57 ± 0 | 0.57 ± 0.023 | 0.63 ± 0.024 | 0.53 ± 0.033 | 0.63 ± 0.057 | 0.67 ± 0.018 | 0.59 ± 0.025 |
| | | **Sigma: 5** | 0.66 ± 0.035 | 0.52 ± 0.014 | 0.55 ± 0.008 | 0.64 ± 0.035 | 0.6 ± 0.031 | 0.58 ± 0.034 | 0.59 ± 0.027 | 0.66 ± 0.057 | 0.52 ± 0.025 | 0.64 ± 0.048 | 0.66 ± 0.035 | 0.57 ± 0.029 |
| | **Shape** | | 0.6 ± 0.025 | 0.53 ± 0.0041 | 0.56 ± 0.031 | 0.59 ± 0.028 | 0.54 ± 0.0086 | 0.52 ± 0.0061 | 0.55 ± 0 | 0.67 ± 0.084 | 0.56 ± 0.017 | 0.58 ± 0.015 | 0.67 ± 0.032 | 0.52 ± 0.0068 |
| | **WAV** | **HHH** | 0.67 ± 0.036 | 0.61 ± 0.036 | 0.64 ± 0.0083 | 0.65 ± 0.032 | 0.63 ± 0.05 | 0.65 ± 0.031 | 0.6 ± 0.025 | 0.65 ± 0.016 | 0.54 ± 0.031 | 0.67 ± 0.02 | 0.65 ± 0.032 | 0.63 ± 0.011 |
| | | **HHL** | 0.65 ± 0.046 | 0.55 ± 0.017 | 0.67 ± 0 | 0.71 ± 0.042 | 0.61 ± 0.04 | 0.56 ± 0.014 | 0.57 ± 0.039 | 0.63 ± 0.039 | 0.53 ± 0.041 | 0.66 ± 0.027 | 0.64 ± 0.025 | 0.6 ± 0.019 |
| | | **HLH** | 0.66 ± 0.03 | 0.56 ± 0.046 | 0.64 ± 0.017 | 0.63 ± 0.044 | 0.62 ± 0.058 | 0.55 ± 0.033 | 0.62 ± 0.015 | 0.62 ± 0.033 | 0.56 ± 0.062 | 0.65 ± 0.034 | 0.66 ± 0.039 | 0.59 ± 0.022 |
| | | **HLL** | 0.65 ± 0.046 | 0.54 ± 0.04 | 0.63 ± 0.0004 | 0.64 ± 0.043 | 0.58 ± 0.026 | 0.61 ± 0.041 | 0.59 ± 0.046 | 0.66 ± 0.05 | 0.52 ± 0.028 | 0.64 ± 0.05 | 0.61 ± 0.014 | 0.57 ± 0.017 |
| | | **LHH** | 0.66 ± 0.027 | 0.6 ± 0.059 | 0.64 ± 0.0052 | 0.66 ± 0.021 | 0.63 ± 0.026 | 0.58 ± 0.034 | 0.63 ± 0.037 | 0.68 ± 0.047 | 0.53 ± 0.041 | 0.66 ± 0.029 | 0.68 ± 0.073 | 0.61 ± 0.066 |
| | | **LHL** | 0.66 ± 0.039 | 0.57 ± 0.045 | 0.67 ± 0.017 | 0.66 ± 0.02 | 0.62 ± 0.057 | 0.56 ± 0.035 | 0.6 ± 0.035 | 0.66 ± 0.031 | 0.52 ± 0.033 | 0.65 ± 0.037 | 0.64 ± 0.055 | 0.63 ± 0.016 |
| | | **LLH** | 0.68 ± 0.045 | 0.58 ± 0.053 | 0.63 ± 0.01 | 0.64 ± 0.055 | 0.62 ± 0.045 | 0.63 ± 0.039 | 0.59 ± 0.05 | 0.64 ± 0.038 | 0.56 ± 0.054 | 0.67 ± 0.03 | 0.64 ± 0.032 | 0.63 ± 0.033 |
| | | **LLL** | 0.69 ± 0.031 | 0.62 ± 0.084 | 0.66 ± 0.053 | 0.65 ± 0.024 | 0.65 ± 0.09 | 0.64 ± 0.053 | 0.67 ± 0.11 | 0.71 ± 0.056 | 0.53 ± 0.045 | 0.68 ± 0.027 | 0.72 ± 0.074 | 0.7 ± 0.069 |
| **PET** | **Bin 64** | | 0.74 ± 0.018 | 0.53 ± 0.026 | 0.75 ± 0.0096 | 0.68 ± 0.032 | 0.59 ± 0.02 | 0.57 ± 0.039 | 0.59 ± 0.059 | 0.66 ± 0.1 | 0.54 ± 0.062 | 0.71 ± 0.017 | 0.64 ± 0.11 | 0.63 ± 0.024 |
| | **LOG** | **Sigma: 0.5** | 0.66 ± 0.021 | 0.55 ± 0.044 | 0.7 ± 0.019 | 0.64 ± 0.023 | 0.63 ± 0.073 | 0.63 ± 0.018 | 0.59 ± 0.034 | 0.63 ± 0.08 | 0.52 ± 0.039 | 0.65 ± 0.057 | 0.6 ± 0.092 | 0.61 ± 0.023 |
| | | **Sigma: 1** | 0.64 ± 0.012 | 0.56 ± 0.05 | 0.74 ± 0.03 | 0.62 ± 0.037 | 0.63 ± 0.073 | 0.63 ± 0.022 | 0.63 ± 0.079 | 0.6 ± 0.051 | 0.55 ± 0.048 | 0.64 ± 0.061 | 0.62 ± 0.099 | 0.66 ± 0.058 |
| | | **Sigma: 1.5** | 0.67 ± 0.019 | 0.52 ± 0.029 | 0.74 ± 0.019 | 0.65 ± 0.025 | 0.63 ± 0.046 | 0.64 ± 0.028 | 0.57 ± 0.048 | 0.58 ± 0.056 | 0.55 ± 0.05 | 0.65 ± 0.045 | 0.63 ± 0.1 | 0.56 ± 0.034 |
| | | **sigma: 2** | 0.64 ± 0.02 | 0.54 ± 0.03 | 0.73 ± 0.026 | 0.64 ± 0.024 | 0.6 ± 0.05 | 0.63 ± 0.025 | 0.56 ± 0.062 | 0.61 ± 0.048 | 0.52 ± 0.016 | 0.66 ± 0.039 | 0.67 ± 0.13 | 0.59 ± 0.022 |
| | | **sigma: 2.5** | 0.71 ± 0.036 | 0.59 ± 0.072 | 0.67 ± 0.053 | 0.7 ± 0.05 | 0.59 ± 0.018 | 0.66 ± 0.015 | 0.63 ± 0.085 | 0.63 ± 0.1 | 0.54 ± 0.045 | 0.69 ± 0.046 | 0.66 ± 0.12 | 0.63 ± 0.069 |
| | | **sigma: 3** | 0.62 ± 0.033 | 0.58 ± 0.072 | 0.71 ± 0.031 | 0.65 ± 0.018 | 0.58 ± 0.022 | 0.6 ± 0.047 | 0.54 ± 0.032 | 0.59 ± 0.084 | 0.53 ± 0.035 | 0.65 ± 0.047 | 0.64 ± 0.11 | 0.57 ± 0.023 |
| | | **sigma: 3.5** | 0.64 ± 0.023 | 0.54 ± 0.074 | 0.71 ± 0.012 | 0.65 ± 0.037 | 0.58 ± 0.019 | 0.6 ± 0.06 | 0.56 ± 0.048 | 0.6 ± 0.024 | 0.54 ± 0.037 | 0.65 ± 0.056 | 0.69 ± 0.15 | 0.62 ± 0.033 |
| | | **sigma: 4** | 0.63 ± 0.026 | 0.57 ± 0.068 | 0.73 ± 0.012 | 0.66 ± 0.013 | 0.61 ± 0.034 | 0.57 ± 0.022 | 0.55 ± 0.031 | 0.65 ± 0.097 | 0.53 ± 0.031 | 0.64 ± 0.061 | 0.67 ± 0.13 | 0.56 ± 0.017 |
| | | **sigma: 4.5** | 0.63 ± 0.02 | 0.54 ± 0.033 | 0.7 ± 0.0084 | 0.63 ± 0.032 | 0.58 ± 0.03 | 0.55 ± 0.0004 | 0.54 ± 0.043 | 0.57 ± 0.049 | 0.56 ± 0.064 | 0.65 ± 0.048 | 0.63 ± 0.11 | 0.55 ± 0.018 |
| | | **Sigma: 5** | 0.63 ± 0.035 | 0.6 ± 0.083 | 0.7 ± 0.016 | 0.62 ± 0.037 | 0.6 ± 0.012 | 0.55 ± 0.0088 | 0.55 ± 0.036 | 0.62 ± 0.069 | 0.53 ± 0.033 | 0.66 ± 0.046 | 0.6 ± 0.084 | 0.63 ± 0.0086 |
| | **Shape** | | 0.58 ± 0.024 | 0.51 ± 0.0047 | 0.6 ± 0.05 | 0.64 ± 0.019 | 0.55 ± 0.019 | 0.54 ± 0.0046 | 0.52 ± 0 | 0.7 ± 0.066 | 0.56 ± 0.012 | 0.58 ± 0.021 | 0.68 ± 0.046 | 0.52 ± 0.0037 |
| | **WAV** | **HHH** | 0.67 ± 0.026 | 0.69 ± 0.032 | 0.65 ± 0.0078 | 0.68 ± 0.021 | 0.66 ± 0.02 | 0.72 ± 0.016 | 0.6 ± 0.043 | 0.61 ± 0.063 | 0.56 ± 0.075 | 0.73 ± 0.013 | 0.65 ± 0.12 | 0.64 ± 0.038 |
| | | **HHL** | 0.65 ± 0.014 | 0.62 ± 0.081 | 0.69 ± 0.016 | 0.63 ± 0.038 | 0.66 ± 0.014 | 0.57 ± 0.031 | 0.59 ± 0.056 | 0.63 ± 0.066 | 0.52 ± 0.024 | 0.7 ± 0.0094 | 0.64 ± 0.11 | 0.65 ± 0.014 |
| | | **HLH** | 0.63 ± 0.022 | 0.62 ± 0.045 | 0.69 ± 0.019 | 0.62 ± 0.04 | 0.66 ± 0.014 | 0.62 ± 0 | 0.56 ± 0.043 | 0.57 ± 0.041 | 0.53 ± 0.03 | 0.67 ± 0.027 | 0.62 ± 0.092 | 0.57 ± 0.018 |
| | | **HLL** | 0.63 ± 0.03 | 0.54 ± 0.027 | 0.69 ± 0.056 | 0.67 ± 0.023 | 0.61 ± 0.041 | 0.54 ± 0.043 | 0.55 ± 0.031 | 0.6 ± 0.086 | 0.52 ± 0.025 | 0.64 ± 0.057 | 0.67 ± 0.13 | 0.58 ± 0.018 |
| | | **LHH** | 0.72 ± 0.023 | 0.67 ± 0.016 | 0.69 ± 0.011 | 0.73 ± 0.023 | 0.63 ± 0.042 | 0.69 ± 0.022 | 0.68 ± 0.13 | 0.74 ± 0.12 | 0.55 ± 0.05 | 0.76 ± 0.024 | 0.67 ± 0.13 | 0.75 ± 0.051 |
| | | **LHL** | 0.64 ± 0.012 | 0.54 ± 0.041 | 0.67 ± 0.031 | 0.62 ± 0.04 | 0.62 ± 0.042 | 0.55 ± 0.028 | 0.58 ± 0.043 | 0.67 ± 0.089 | 0.52 ± 0.028 | 0.66 ± 0.037 | 0.62 ± 0.1 | 0.6 ± 0.022 |
| | | **LLH** | 0.63 ± 0.047 | 0.56 ± 0.055 | 0.65 ± 0.039 | 0.63 ± 0.032 | 0.6 ± 0.05 | 0.57 ± 0.048 | 0.54 ± 0.047 | 0.6 ± 0.073 | 0.52 ± 0.027 | 0.72 ± 0.032 | 0.64 ± 0.11 | 0.61 ± 0.023 |
| | | **LLL** | 0.64 ± 0.018 | 0.52 ± 0.022 | 0.7 ± 0.015 | 0.66 ± 0.028 | 0.6 ± 0.028 | 0.56 ± 0.025 | 0.64 ± 0.1 | 0.67 ± 0.095 | 0.52 ± 0.019 | 0.68 ± 0.019 | 0.67 ± 0.14 | 0.64 ± 0.077 |

- **KRAS**

Fig 3 shows a heatmap of KRAS mutation status prediction different combination of feature selection, classifier and image sets. The model performance has a wide range from 0.5 to 0.83, which is as same as EGFR models. According to this figure, the combination methods with highest predictive performances are: LOG preprocessed of CT image set with sigma 3.5 with SM feature selector and SGD classifier (CT_LOG_3.5S+SM+SGD, AUC: 0.83), CT_LOG_5 S image set with VT_SM feature selector SGD classifier (CT_LOG_5 S+ VT_SM+SGD, AUC: 0.82). These models are followed by CTD_LOG_4 S image set with VT feature selector and SGD classifier (CTD_LOG_4S+VT+SGD, AUC: 0.80), PET_LOG_0 S image set with SM feature selector and SGD classifier (PET_LOG_0.5S+SM+SGD, AUC: 0.80) and PET image set which preprocessed by HHH of wavelet with SKB feature selector and BNB classifier (PET_W_HHH+SKB+BNB, AUC: 0.80).

Supplemental Tables 8 and 9 show KRAS mutation status prediction results based on feature selection methods and image sets. In these results, feature selection performance has a range from 0.5 to 0.83, and the combination of SM feature selection with PET_LOG_4.5S image set had the highest performance (AUC: 0.66 ± 0.093 & 0.51 - 0.79), followed by VT_SM feature selection with CT_LOG_3.5S image set (AUC: 0.66 ± 0.079 & 0.55 - 0.8), SM feature selection with CT_LOG_3.5S image set (AUC: 0.65 ± 0.11 & 0.5 - 0.83) and SM feature selection with PET_BIN64 image set (AUC: 0.64 ± 0.058 & 0.51 - 0.71). Supplemental figure 12 delicate the box plot of KRAS mutation status prediction based on feature selection methods.

KRAS mutation status prediction results based on classifier and image sets is presented in supplemental Tables 10 and 11 (mean ± SD & Min-Max). Here, the classifier performance has a range from 0.5 to 0.83 and combination of SVM classifier with CT_LOG_3.5S image set had the highest performance (AUC: 0.79 ± 0.021 & 0.76-0.8), followed by BNB classifier PET_W_HHH image set (AUC: 0.75 ± 0.1 & 0.6 - 0.81) and SGD classifier with CTD_LOG_4 S image set (AUC: 0.71 ± 0.08 & 0.56 - 0.81). Supplemental figure 13 delicate the box plot of KRAS mutation status prediction based on feature selection methods.

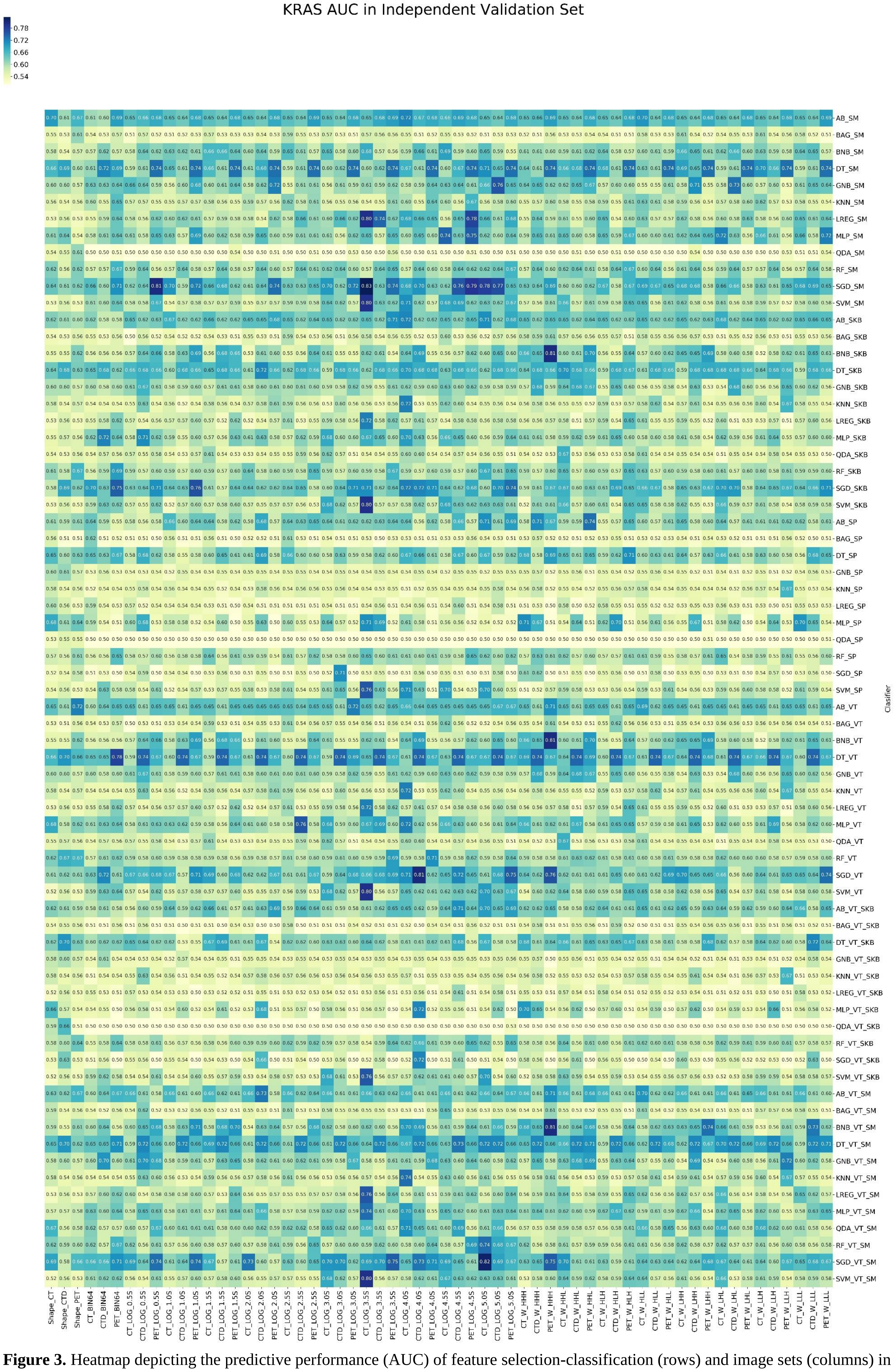

**Figure 3.** Heatmap depicting the predictive performance (AUC) of feature selection-classification (rows) and image sets (columns) in prediction of KRAS mutation status of independent validation sets

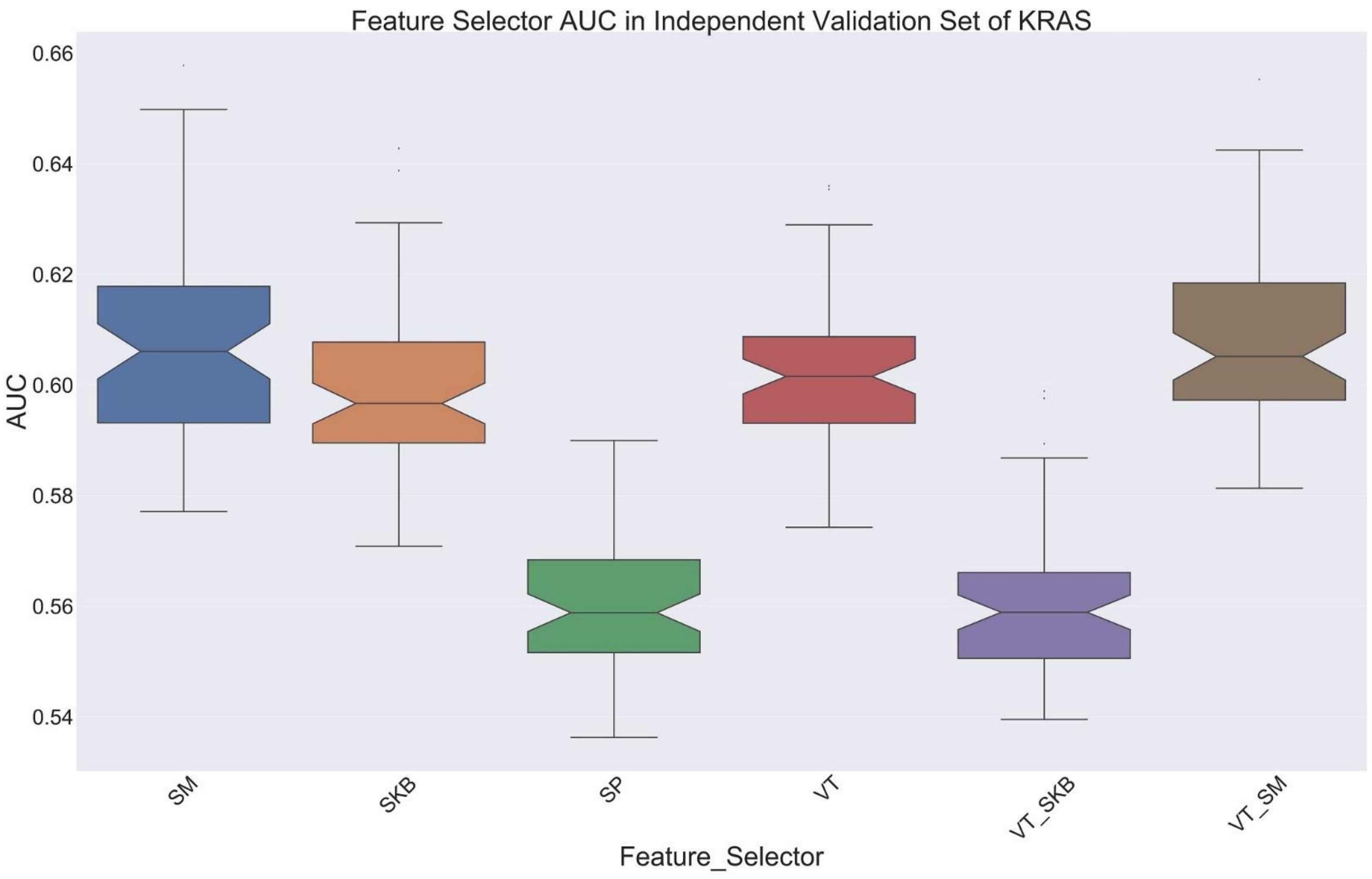

**Supplemental Fig 12.** AUC box plot of different feature selector in independent validation set of KRAS

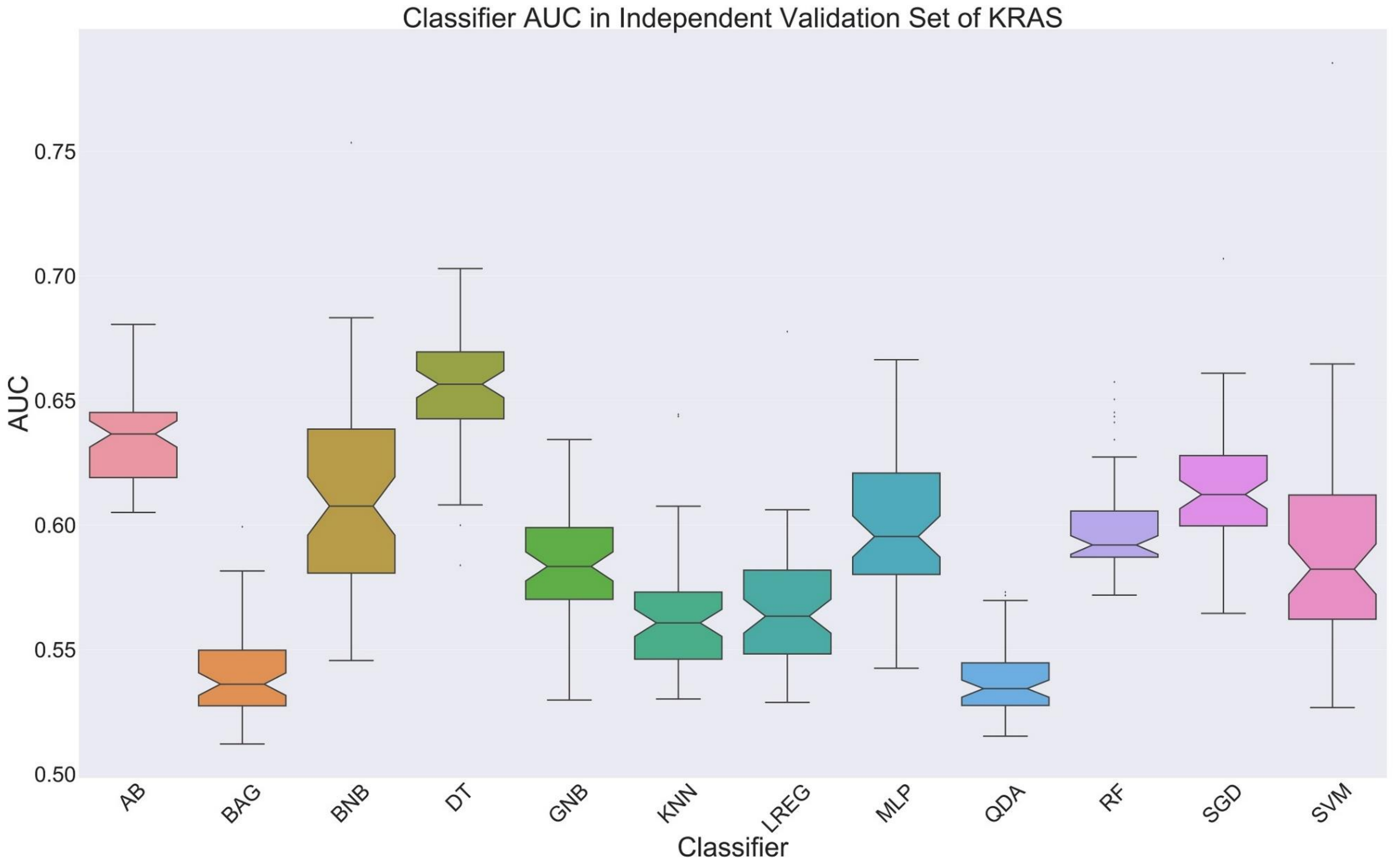

**Supplemental Fig 13.** AUC box plot of different classifier in independent validation set of KRAS

**Supplemental Table 8.** KRAS Feature Selector Min-Max different modality, preprocessing and setting

| Modality | Preprocessed | Setting | SM | SKB | SP | VT | VT_SKB | VT_SM |
|---|---|---|---|---|---|---|---|---|
| **CT** | **Bin 64** | | 0.5 - 0.66 | 0.54 - 0.7 | 0.5 - 0.65 | 0.54 - 0.65 | 0.5 - 0.6 | 0.54 - 0.66 |
| | **LOG** | **Sigma: 0.5** | 0.5 - 0.66 | 0.5 - 0.66 | 0.5 - 0.6 | 0.5 - 0.67 | 0.5 - 0.65 | 0.54 - 0.69 |
| | | **Sigma: 1** | 0.5 - 0.7 | 0.52 - 0.67 | 0.5 - 0.66 | 0.53 - 0.67 | 0.5 - 0.64 | 0.53 - 0.66 |
| | | **Sigma: 1.5** | 0.5 - 0.66 | 0.54 - 0.66 | 0.5 - 0.64 | 0.56 - 0.69 | 0.5 - 0.67 | 0.55 - 0.69 |
| | | **sigma: 2** | 0.5 - 0.65 | 0.52 - 0.65 | 0.5 - 0.61 | 0.52 - 0.65 | 0.5 - 0.61 | 0.54 - 0.73 |
| | | **sigma: 2.5** | 0.5 - 0.65 | 0.55 - 0.65 | 0.5 - 0.66 | 0.55 - 0.65 | 0.5 - 0.62 | 0.55 - 0.66 |
| | | **sigma: 3** | 0.5 - 0.7 | 0.53 - 0.68 | 0.5 - 0.64 | 0.53 - 0.68 | 0.5 - 0.68 | 0.53 - 0.7 |
| | | **sigma: 3.5** | 0.5 - 0.83 | 0.54 - 0.8 | 0.5 - 0.76 | 0.54 - 0.8 | 0.5 - 0.76 | 0.55 - 0.8 |
| | | **sigma: 4** | 0.5 - 0.72 | 0.55 - 0.72 | 0.5 - 0.71 | 0.54 - 0.72 | 0.5 - 0.65 | 0.53 - 0.74 |
| | | **sigma: 4.5** | 0.5 - 0.74 | 0.54 - 0.68 | 0.5 - 0.7 | 0.54 - 0.66 | 0.5 - 0.64 | 0.58 - 0.69 |
| | | **Sigma: 5** | 0.5 - 0.78 | 0.52 - 0.71 | 0.5 - 0.71 | 0.52 - 0.7 | 0.5 - 0.7 | 0.52 - 0.82 |
| | **Shape** | | 0.53 - 0.7 | 0.53 - 0.64 | 0.52 - 0.68 | 0.52 - 0.68 | 0.52 - 0.66 | 0.52 - 0.69 |
| | **WAV** | **HHH** | 0.5 - 0.65 | 0.52 - 0.66 | 0.5 - 0.71 | 0.54 - 0.69 | 0.5 - 0.7 | 0.53 - 0.68 |
| | | **HHL** | 0.5 - 0.66 | 0.55 - 0.7 | 0.5 - 0.65 | 0.54 - 0.67 | 0.5 - 0.66 | 0.55 - 0.7 |
| | | **HLH** | 0.5 - 0.68 | 0.53 - 0.65 | 0.5 - 0.62 | 0.53 - 0.65 | 0.5 - 0.64 | 0.55 - 0.66 |
| | | **HLL** | 0.5 - 0.7 | 0.56 - 0.66 | 0.5 - 0.61 | 0.56 - 0.69 | 0.5 - 0.65 | 0.57 - 0.7 |
| | | **LHH** | 0.5 - 0.69 | 0.54 - 0.65 | 0.5 - 0.65 | 0.54 - 0.7 | 0.5 - 0.65 | 0.54 - 0.66 |
| | | **LHL** | 0.5 - 0.72 | 0.51 - 0.7 | 0.5 - 0.66 | 0.51 - 0.66 | 0.5 - 0.64 | 0.51 - 0.7 |
| | | **LLH** | 0.5 - 0.7 | 0.52 - 0.65 | 0.5 - 0.63 | 0.52 - 0.66 | 0.5 - 0.64 | 0.53 - 0.69 |
| | | **LLL** | 0.51 - 0.68 | 0.55 - 0.65 | 0.5 - 0.7 | 0.54 - 0.65 | 0.5 - 0.66 | 0.57 - 0.66 |
| **CTD** | **Bin 64** | | 0.5 - 0.72 | 0.53 - 0.72 | 0.5 - 0.63 | 0.53 - 0.72 | 0.5 - 0.62 | 0.52 - 0.7 |
| | **LOG** | **Sigma: 0.5** | 0.5 - 0.66 | 0.54 - 0.71 | 0.5 - 0.68 | 0.52 - 0.74 | 0.5 - 0.64 | 0.54 - 0.72 |
| | | **Sigma: 1** | 0.5 - 0.64 | 0.52 - 0.68 | 0.5 - 0.6 | 0.52 - 0.74 | 0.5 - 0.6 | 0.52 - 0.72 |
| | | **Sigma: 1.5** | 0.5 - 0.68 | 0.52 - 0.68 | 0.5 - 0.65 | 0.52 - 0.74 | 0.5 - 0.69 | 0.52 - 0.72 |
| | | **sigma: 2** | 0.5 - 0.68 | 0.53 - 0.72 | 0.5 - 0.69 | 0.54 - 0.74 | 0.5 - 0.68 | 0.53 - 0.73 |
| | | **sigma: 2.5** | 0.5 - 0.66 | 0.53 - 0.68 | 0.5 - 0.63 | 0.53 - 0.76 | 0.5 - 0.66 | 0.53 - 0.72 |
| | | **sigma: 3** | 0.5 - 0.66 | 0.51 - 0.68 | 0.5 - 0.71 | 0.55 - 0.74 | 0.5 - 0.63 | 0.55 - 0.72 |
| | | **sigma: 3.5** | 0.5 - 0.74 | 0.53 - 0.68 | 0.5 - 0.69 | 0.51 - 0.74 | 0.5 - 0.62 | 0.54 - 0.72 |
| | | **sigma: 4** | 0.5 - 0.7 | 0.52 - 0.72 | 0.5 - 0.66 | 0.51 - 0.81 | 0.5 - 0.72 | 0.51 - 0.73 |
| | | **sigma: 4.5** | 0.5 - 0.76 | 0.52 - 0.68 | 0.5 - 0.67 | 0.56 - 0.74 | 0.5 - 0.71 | 0.53 - 0.73 |
| | | **Sigma: 5** | 0.5 - 0.77 | 0.55 - 0.7 | 0.5 - 0.61 | 0.54 - 0.74 | 0.5 - 0.65 | 0.55 - 0.72 |
| | **Shape** | | 0.53 - 0.69 | 0.53 - 0.69 | 0.51 - 0.61 | 0.51 - 0.7 | 0.54 - 0.7 | 0.54 - 0.7 |
| | **WAV** | **HHH** | 0.5 - 0.66 | 0.52 - 0.68 | 0.5 - 0.71 | 0.52 - 0.74 | 0.5 - 0.65 | 0.54 - 0.72 |
| | | **HHL** | 0.51 - 0.68 | 0.51 - 0.68 | 0.5 - 0.64 | 0.51 - 0.74 | 0.5 - 0.62 | 0.53 - 0.72 |
| | | **HLH** | 0.51 - 0.64 | 0.53 - 0.69 | 0.5 - 0.7 | 0.53 - 0.74 | 0.5 - 0.65 | 0.53 - 0.72 |
| | | **HLL** | 0.5 - 0.67 | 0.54 - 0.68 | 0.5 - 0.63 | 0.53 - 0.74 | 0.5 - 0.61 | 0.54 - 0.72 |
| | | **LHH** | 0.54 - 0.71 | 0.54 - 0.68 | 0.5 - 0.67 | 0.54 - 0.74 | 0.5 - 0.62 | 0.54 - 0.72 |
| | | **LHL** | 0.5 - 0.73 | 0.51 - 0.7 | 0.5 - 0.62 | 0.51 - 0.74 | 0.5 - 0.59 | 0.53 - 0.72 |
| | | **LLH** | 0.5 - 0.66 | 0.53 - 0.68 | 0.5 - 0.64 | 0.54 - 0.74 | 0.5 - 0.66 | 0.54 - 0.72 |
| | | **LLL** | 0.51 - 0.69 | 0.53 - 0.68 | 0.5 - 0.68 | 0.53 - 0.74 | 0.5 - 0.72 | 0.55 - 0.73 |
| **PET** | **Bin 64** | | 0.51 - 0.71 | 0.54 - 0.75 | 0.5 - 0.67 | 0.54 - 0.78 | 0.5 - 0.64 | 0.54 - 0.71 |
| | **LOG** | **Sigma: 0.5** | 0.5 - 0.81 | 0.52 - 0.71 | 0.5 - 0.62 | 0.51 - 0.68 | 0.5 - 0.62 | 0.52 - 0.74 |
| | | **Sigma: 1** | 0.5 - 0.74 | 0.52 - 0.76 | 0.5 - 0.64 | 0.54 - 0.71 | 0.5 - 0.62 | 0.54 - 0.74 |
| | | **Sigma: 1.5** | 0.53 - 0.74 | 0.51 - 0.66 | 0.5 - 0.62 | 0.51 - 0.68 | 0.5 - 0.61 | 0.51 - 0.7 |
| | | **sigma: 2** | 0.51 - 0.74 | 0.51 - 0.68 | 0.5 - 0.58 | 0.53 - 0.67 | 0.5 - 0.69 | 0.51 - 0.66 |
| | | **sigma: 2.5** | 0.51 - 0.74 | 0.52 - 0.66 | 0.5 - 0.65 | 0.53 - 0.67 | 0.5 - 0.64 | 0.52 - 0.68 |
| | | **sigma: 3** | 0.51 - 0.74 | 0.51 - 0.71 | 0.5 - 0.62 | 0.51 - 0.72 | 0.5 - 0.6 | 0.56 - 0.67 |
| | | **sigma: 3.5** | 0.56 - 0.74 | 0.54 - 0.71 | 0.5 - 0.64 | 0.54 - 0.69 | 0.5 - 0.65 | 0.53 - 0.75 |
| | | **sigma: 4** | 0.51 - 0.74 | 0.53 - 0.71 | 0.5 - 0.62 | 0.54 - 0.71 | 0.5 - 0.62 | 0.51 - 0.71 |
| | | **sigma: 4.5** | 0.51 - 0.79 | 0.54 - 0.68 | 0.5 - 0.65 | 0.52 - 0.67 | 0.5 - 0.65 | 0.56 - 0.69 |
| | | **Sigma: 5** | 0.51 - 0.74 | 0.51 - 0.74 | 0.5 - 0.69 | 0.51 - 0.75 | 0.5 - 0.69 | 0.55 - 0.67 |
| | **Shape** | | 0.53 - 0.67 | 0.53 - 0.67 | 0.51 - 0.64 | 0.53 - 0.72 | 0.51 - 0.64 | 0.53 - 0.67 |
| | **WAV** | **HHH** | 0.5 - 0.74 | 0.53 - 0.81 | 0.5 - 0.69 | 0.53 - 0.81 | 0.5 - 0.65 | 0.58 - 0.81 |
| | | **HHL** | 0.5 - 0.74 | 0.52 - 0.7 | 0.5 - 0.74 | 0.51 - 0.7 | 0.5 - 0.65 | 0.52 - 0.71 |
| | | **HLH** | 0.5 - 0.74 | 0.51 - 0.67 | 0.5 - 0.71 | 0.56 - 0.67 | 0.5 - 0.67 | 0.51 - 0.66 |
| | | **HLL** | 0.5 - 0.74 | 0.54 - 0.66 | 0.5 - 0.61 | 0.51 - 0.69 | 0.5 - 0.62 | 0.52 - 0.68 |
| | | **LHH** | 0.5 - 0.74 | 0.52 - 0.69 | 0.5 - 0.63 | 0.52 - 0.69 | 0.5 - 0.68 | 0.53 - 0.74 |
| | | **LHL** | 0.5 - 0.74 | 0.52 - 0.66 | 0.5 - 0.61 | 0.53 - 0.67 | 0.5 - 0.62 | 0.51 - 0.66 |
| | | **LLH** | 0.52 - 0.74 | 0.51 - 0.67 | 0.5 - 0.67 | 0.51 - 0.67 | 0.5 - 0.67 | 0.54 - 0.72 |
| | | **LLL** | 0.51 - 0.74 | 0.5 - 0.71 | 0.5 - 0.65 | 0.5 - 0.74 | 0.5 - 0.65 | 0.51 - 0.71 |

**Supplemental Table 9.** KRAS Feature Selector Mean+Sd different modality, preprocessing and setting

| Modality | Preprocessed | Setting | SM | SKB | SP | VT | VT_SKB | VT_SM |
|---|---|---|---|---|---|---|---|---|
| **CT** | **Bin 64** | | 0.59 ± 0.043 | 0.59 ± 0.049 | 0.57 ± 0.053 | 0.58 ± 0.036 | 0.55 ± 0.035 | 0.6 ± 0.039 |
| | **LOG** | **Sigma: 0.5** | 0.59 ± 0.046 | 0.59 ± 0.045 | 0.56 ± 0.033 | 0.59 ± 0.044 | 0.55 ± 0.04 | 0.6 ± 0.044 |
| | | **Sigma: 1** | 0.59 ± 0.062 | 0.58 ± 0.042 | 0.56 ± 0.044 | 0.59 ± 0.043 | 0.55 ± 0.043 | 0.58 ± 0.034 |
| | | **Sigma: 1.5** | 0.6 ± 0.05 | 0.59 ± 0.036 | 0.56 ± 0.047 | 0.6 ± 0.038 | 0.56 ± 0.055 | 0.6 ± 0.046 |
| | | **sigma: 2** | 0.58 ± 0.04 | 0.58 ± 0.045 | 0.55 ± 0.036 | 0.57 ± 0.041 | 0.55 ± 0.038 | 0.6 ± 0.054 |
| | | **sigma: 2.5** | 0.59 ± 0.043 | 0.59 ± 0.032 | 0.58 ± 0.055 | 0.59 ± 0.029 | 0.56 ± 0.041 | 0.6 ± 0.031 |
| | | **sigma: 3** | 0.61 ± 0.053 | 0.61 ± 0.046 | 0.57 ± 0.047 | 0.61 ± 0.045 | 0.57 ± 0.056 | 0.62 ± 0.051 |
| | | **sigma: 3.5** | 0.65 ± 0.11 | 0.64 ± 0.075 | 0.59 ± 0.084 | 0.64 ± 0.074 | 0.57 ± 0.075 | 0.66 ± 0.079 |
| | | **sigma: 4** | 0.62 ± 0.069 | 0.64 ± 0.064 | 0.58 ± 0.073 | 0.64 ± 0.06 | 0.56 ± 0.049 | 0.64 ± 0.067 |
| | | **sigma: 4.5** | 0.62 ± 0.068 | 0.61 ± 0.048 | 0.56 ± 0.052 | 0.61 ± 0.043 | 0.56 ± 0.046 | 0.63 ± 0.032 |
| | | **Sigma: 5** | 0.63 ± 0.079 | 0.6 ± 0.054 | 0.58 ± 0.078 | 0.6 ± 0.052 | 0.58 ± 0.076 | 0.64 ± 0.082 |
| | **Shape** | | 0.59 ± 0.056 | 0.57 ± 0.039 | 0.59 ± 0.05 | 0.59 ± 0.056 | 0.58 ± 0.046 | 0.6 ± 0.052 |
| | **WAV** | **HHH** | 0.58 ± 0.048 | 0.59 ± 0.044 | 0.57 ± 0.07 | 0.6 ± 0.056 | 0.58 ± 0.063 | 0.6 ± 0.05 |
| | | **HHL** | 0.61 ± 0.051 | 0.63 ± 0.047 | 0.56 ± 0.048 | 0.61 ± 0.039 | 0.57 ± 0.051 | 0.62 ± 0.043 |
| | | **HLH** | 0.61 ± 0.055 | 0.59 ± 0.033 | 0.56 ± 0.033 | 0.59 ± 0.033 | 0.57 ± 0.045 | 0.59 ± 0.033 |
| | | **HLL** | 0.61 ± 0.056 | 0.61 ± 0.035 | 0.56 ± 0.039 | 0.6 ± 0.04 | 0.56 ± 0.055 | 0.62 ± 0.038 |
| | | **LHH** | 0.62 ± 0.05 | 0.6 ± 0.038 | 0.56 ± 0.046 | 0.6 ± 0.048 | 0.56 ± 0.04 | 0.6 ± 0.037 |
| | | **LHL** | 0.6 ± 0.057 | 0.6 ± 0.065 | 0.57 ± 0.055 | 0.6 ± 0.047 | 0.56 ± 0.049 | 0.63 ± 0.059 |
| | | **LLH** | 0.61 ± 0.052 | 0.59 ± 0.043 | 0.56 ± 0.041 | 0.59 ± 0.048 | 0.56 ± 0.041 | 0.61 ± 0.048 |
| | | **LLL** | 0.61 ± 0.049 | 0.59 ± 0.031 | 0.57 ± 0.058 | 0.59 ± 0.033 | 0.56 ± 0.047 | 0.6 ± 0.031 |
| **CTD** | **Bin 64** | | 0.59 ± 0.055 | 0.6 ± 0.057 | 0.56 ± 0.045 | 0.6 ± 0.054 | 0.57 ± 0.045 | 0.61 ± 0.053 |
| | **LOG** | **Sigma: 0.5** | 0.6 ± 0.046 | 0.61 ± 0.057 | 0.57 ± 0.06 | 0.61 ± 0.065 | 0.57 ± 0.043 | 0.62 ± 0.058 |
| | | **Sigma: 1** | 0.59 ± 0.043 | 0.58 ± 0.05 | 0.55 ± 0.031 | 0.59 ± 0.06 | 0.55 ± 0.035 | 0.6 ± 0.052 |
| | | **Sigma: 1.5** | 0.6 ± 0.052 | 0.58 ± 0.052 | 0.56 ± 0.045 | 0.59 ± 0.063 | 0.56 ± 0.057 | 0.6 ± 0.057 |
| | | **sigma: 2** | 0.6 ± 0.055 | 0.6 ± 0.054 | 0.58 ± 0.068 | 0.6 ± 0.057 | 0.59 ± 0.065 | 0.61 ± 0.063 |
| | | **sigma: 2.5** | 0.6 ± 0.042 | 0.59 ± 0.045 | 0.56 ± 0.038 | 0.61 ± 0.073 | 0.56 ± 0.049 | 0.59 ± 0.051 |
| | | **sigma: 3** | 0.59 ± 0.049 | 0.59 ± 0.043 | 0.58 ± 0.065 | 0.6 ± 0.052 | 0.56 ± 0.04 | 0.61 ± 0.052 |
| | | **sigma: 3.5** | 0.61 ± 0.066 | 0.59 ± 0.046 | 0.58 ± 0.06 | 0.6 ± 0.071 | 0.56 ± 0.039 | 0.6 ± 0.056 |
| | | **sigma: 4** | 0.61 ± 0.063 | 0.61 ± 0.061 | 0.57 ± 0.048 | 0.63 ± 0.084 | 0.6 ± 0.08 | 0.62 ± 0.065 |
| | | **sigma: 4.5** | 0.63 ± 0.074 | 0.6 ± 0.041 | 0.57 ± 0.056 | 0.62 ± 0.058 | 0.59 ± 0.064 | 0.62 ± 0.053 |
| | | **Sigma: 5** | 0.62 ± 0.078 | 0.61 ± 0.047 | 0.57 ± 0.038 | 0.62 ± 0.06 | 0.56 ± 0.043 | 0.63 ± 0.052 |
| | **Shape** | | 0.58 ± 0.048 | 0.58 ± 0.052 | 0.57 ± 0.032 | 0.59 ± 0.055 | 0.6 ± 0.049 | 0.58 ± 0.045 |
| | **WAV** | **HHH** | 0.6 ± 0.054 | 0.61 ± 0.047 | 0.59 ± 0.066 | 0.61 ± 0.062 | 0.58 ± 0.041 | 0.61 ± 0.052 |
| | | **HHL** | 0.59 ± 0.051 | 0.59 ± 0.053 | 0.56 ± 0.046 | 0.6 ± 0.07 | 0.57 ± 0.036 | 0.6 ± 0.056 |
| | | **HLH** | 0.58 ± 0.042 | 0.59 ± 0.061 | 0.57 ± 0.052 | 0.6 ± 0.064 | 0.57 ± 0.044 | 0.59 ± 0.054 |
| | | **HLL** | 0.59 ± 0.049 | 0.59 ± 0.046 | 0.56 ± 0.041 | 0.59 ± 0.057 | 0.55 ± 0.037 | 0.6 ± 0.048 |
| | | **LHH** | 0.62 ± 0.054 | 0.6 ± 0.042 | 0.56 ± 0.045 | 0.61 ± 0.058 | 0.56 ± 0.039 | 0.62 ± 0.054 |
| | | **LHL** | 0.61 ± 0.062 | 0.6 ± 0.059 | 0.55 ± 0.04 | 0.59 ± 0.063 | 0.55 ± 0.035 | 0.6 ± 0.05 |
| | | **LLH** | 0.59 ± 0.045 | 0.59 ± 0.045 | 0.56 ± 0.047 | 0.6 ± 0.06 | 0.56 ± 0.055 | 0.59 ± 0.048 |
| | | **LLL** | 0.6 ± 0.054 | 0.6 ± 0.05 | 0.57 ± 0.056 | 0.61 ± 0.057 | 0.58 ± 0.06 | 0.62 ± 0.059 |
| **PET** | **Bin 64** | | 0.64 ± 0.058 | 0.61 ± 0.062 | 0.55 ± 0.058 | 0.62 ± 0.063 | 0.55 ± 0.05 | 0.62 ± 0.059 |
| | **LOG** | **Sigma: 0.5** | 0.63 ± 0.09 | 0.61 ± 0.054 | 0.55 ± 0.045 | 0.61 ± 0.053 | 0.55 ± 0.046 | 0.63 ± 0.059 |
| | | **Sigma: 1** | 0.62 ± 0.077 | 0.61 ± 0.067 | 0.55 ± 0.042 | 0.61 ± 0.057 | 0.54 ± 0.039 | 0.62 ± 0.056 |
| | | **Sigma: 1.5** | 0.61 ± 0.061 | 0.6 ± 0.05 | 0.54 ± 0.049 | 0.61 ± 0.058 | 0.55 ± 0.041 | 0.61 ± 0.049 |
| | | **sigma: 2** | 0.62 ± 0.08 | 0.6 ± 0.046 | 0.54 ± 0.033 | 0.6 ± 0.046 | 0.55 ± 0.059 | 0.59 ± 0.038 |
| | | **sigma: 2.5** | 0.61 ± 0.063 | 0.59 ± 0.048 | 0.55 ± 0.052 | 0.6 ± 0.049 | 0.55 ± 0.051 | 0.61 ± 0.05 |
| | | **sigma: 3** | 0.61 ± 0.071 | 0.59 ± 0.055 | 0.54 ± 0.042 | 0.6 ± 0.067 | 0.54 ± 0.035 | 0.6 ± 0.037 |
| | | **sigma: 3.5** | 0.64 ± 0.063 | 0.61 ± 0.054 | 0.55 ± 0.047 | 0.61 ± 0.057 | 0.55 ± 0.056 | 0.6 ± 0.059 |
| | | **sigma: 4** | 0.62 ± 0.064 | 0.59 ± 0.057 | 0.54 ± 0.048 | 0.6 ± 0.055 | 0.55 ± 0.047 | 0.6 ± 0.059 |
| | | **sigma: 4.5** | 0.66 ± 0.093 | 0.61 ± 0.046 | 0.55 ± 0.046 | 0.6 ± 0.049 | 0.55 ± 0.052 | 0.61 ± 0.043 |
| | | **Sigma: 5** | 0.63 ± 0.07 | 0.61 ± 0.063 | 0.56 ± 0.063 | 0.62 ± 0.066 | 0.56 ± 0.065 | 0.62 ± 0.046 |
| | **Shape** | | 0.59 ± 0.043 | 0.58 ± 0.043 | 0.57 ± 0.044 | 0.6 ± 0.06 | 0.56 ± 0.041 | 0.59 ± 0.045 |
| | **WAV** | **HHH** | 0.62 ± 0.065 | 0.62 ± 0.072 | 0.55 ± 0.071 | 0.64 ± 0.085 | 0.55 ± 0.056 | 0.64 ± 0.074 |
| | | **HHL** | 0.61 ± 0.062 | 0.61 ± 0.052 | 0.55 ± 0.076 | 0.61 ± 0.062 | 0.55 ± 0.056 | 0.61 ± 0.064 |
| | | **HLH** | 0.62 ± 0.07 | 0.62 ± 0.046 | 0.56 ± 0.07 | 0.62 ± 0.039 | 0.55 ± 0.056 | 0.61 ± 0.04 |
| | | **HLL** | 0.6 ± 0.065 | 0.59 ± 0.037 | 0.55 ± 0.042 | 0.59 ± 0.056 | 0.55 ± 0.048 | 0.6 ± 0.048 |
| | | **LHH** | 0.59 ± 0.07 | 0.58 ± 0.068 | 0.55 ± 0.044 | 0.58 ± 0.067 | 0.55 ± 0.059 | 0.59 ± 0.068 |
| | | **LHL** | 0.6 ± 0.064 | 0.58 ± 0.042 | 0.54 ± 0.044 | 0.58 ± 0.046 | 0.55 ± 0.043 | 0.58 ± 0.042 |
| | | **LLH** | 0.59 ± 0.069 | 0.6 ± 0.063 | 0.56 ± 0.057 | 0.6 ± 0.06 | 0.56 ± 0.06 | 0.61 ± 0.057 |
| | | **LLL** | 0.62 ± 0.075 | 0.59 ± 0.059 | 0.55 ± 0.048 | 0.59 ± 0.068 | 0.55 ± 0.053 | 0.6 ± 0.056 |

**Supplemental Table 10.** KRAS Classifier Min-Max different modality, preprocessing and setting

| Modality | Preprocessed | Setting | AB | BAG | BNB | DT | GNB | KNN | LREG | MLP | QDA | RF | SGD | SVM |
|---|---|---|---|---|---|---|---|---|---|---|---|---|---|---|
| CT | Bin 64 | | 0.58 - 0.64 | 0.51 - 0.62 | 0.56 - 0.6 | 0.6 - 0.65 | 0.53 - 0.63 | 0.51 - 0.6 | 0.55 - 0.59 | 0.54 - 0.63 | 0.5 - 0.62 | 0.55 - 0.62 | 0.51 - 0.7 | 0.54 - 0.61 |
| | LOG | Sigma: 0.5 | 0.56 - 0.66 | 0.5 - 0.54 | 0.57 - 0.6 | 0.58 - 0.66 | 0.54 - 0.66 | 0.55 - 0.57 | 0.57 - 0.6 | 0.56 - 0.61 | 0.5 - 0.6 | 0.58 - 0.62 | 0.54 - 0.69 | 0.54 - 0.58 |
| | | Sigma: 1 | 0.64 - 0.67 | 0.51 - 0.56 | 0.58 - 0.63 | 0.58 - 0.65 | 0.52 - 0.58 | 0.55 - 0.56 | 0.53 - 0.6 | 0.54 - 0.63 | 0.5 - 0.6 | 0.56 - 0.59 | 0.54 - 0.7 | 0.52 - 0.55 |
| | | Sigma: 1.5 | 0.64 - 0.66 | 0.51 - 0.59 | 0.56 - 0.66 | 0.59 - 0.69 | 0.54 - 0.6 | 0.55 - 0.59 | 0.53 - 0.57 | 0.54 - 0.6 | 0.5 - 0.61 | 0.57 - 0.64 | 0.53 - 0.69 | 0.53 - 0.57 |
| | | sigma: 2 | 0.58 - 0.66 | 0.52 - 0.55 | 0.53 - 0.6 | 0.61 - 0.61 | 0.55 - 0.58 | 0.53 - 0.59 | 0.51 - 0.58 | 0.53 - 0.61 | 0.5 - 0.6 | 0.57 - 0.64 | 0.51 - 0.73 | 0.53 - 0.58 |
| | | sigma: 2.5 | 0.59 - 0.66 | 0.58 - 0.61 | 0.55 - 0.65 | 0.59 - 0.66 | 0.53 - 0.6 | 0.54 - 0.62 | 0.51 - 0.61 | 0.55 - 0.6 | 0.5 - 0.62 | 0.57 - 0.62 | 0.51 - 0.64 | 0.56 - 0.61 |
| | | sigma: 3 | 0.61 - 0.66 | 0.51 - 0.58 | 0.61 - 0.65 | 0.58 - 0.63 | 0.53 - 0.59 | 0.53 - 0.56 | 0.54 - 0.6 | 0.58 - 0.68 | 0.5 - 0.65 | 0.57 - 0.62 | 0.52 - 0.7 | 0.61 - 0.68 |
| | | sigma: 3.5 | 0.61 - 0.66 | 0.52 - 0.58 | 0.62 - 0.68 | 0.58 - 0.65 | 0.53 - 0.61 | 0.54 - 0.57 | 0.53 - 0.8 | 0.56 - 0.74 | 0.5 - 0.66 | 0.54 - 0.65 | 0.52 - 0.83 | 0.76 - 0.8 |
| | | sigma: 4 | 0.64 - 0.72 | 0.52 - 0.55 | 0.59 - 0.7 | 0.61 - 0.7 | 0.53 - 0.62 | 0.51 - 0.74 | 0.53 - 0.68 | 0.54 - 0.72 | 0.5 - 0.71 | 0.57 - 0.62 | 0.51 - 0.72 | 0.57 - 0.71 |
| | | sigma: 4.5 | 0.58 - 0.66 | 0.53 - 0.65 | 0.56 - 0.69 | 0.58 - 0.63 | 0.52 - 0.63 | 0.53 - 0.63 | 0.54 - 0.59 | 0.55 - 0.74 | 0.5 - 0.6 | 0.58 - 0.59 | 0.51 - 0.69 | 0.61 - 0.7 |
| | | Sigma: 5 | 0.65 - 0.71 | 0.51 - 0.54 | 0.6 - 0.65 | 0.62 - 0.72 | 0.52 - 0.66 | 0.55 - 0.61 | 0.54 - 0.66 | 0.54 - 0.62 | 0.5 - 0.61 | 0.59 - 0.74 | 0.51 - 0.82 | 0.63 - 0.7 |
| | Shape | | 0.61 - 0.7 | 0.53 - 0.59 | 0.55 - 0.59 | 0.62 - 0.66 | 0.58 - 0.6 | 0.56 - 0.58 | 0.52 - 0.6 | 0.55 - 0.68 | 0.53 - 0.67 | 0.57 - 0.62 | 0.52 - 0.69 | 0.52 - 0.54 |
| | WAV | HHH | 0.58 - 0.66 | 0.52 - 0.58 | 0.61 - 0.68 | 0.64 - 0.69 | 0.55 - 0.64 | 0.55 - 0.56 | 0.51 - 0.56 | 0.59 - 0.71 | 0.5 - 0.57 | 0.57 - 0.62 | 0.58 - 0.64 | 0.51 - 0.57 |
| | | HHL | 0.58 - 0.66 | 0.51 - 0.57 | 0.59 - 0.6 | 0.64 - 0.7 | 0.55 - 0.64 | 0.55 - 0.58 | 0.57 - 0.64 | 0.54 - 0.65 | 0.5 - 0.67 | 0.61 - 0.65 | 0.51 - 0.7 | 0.59 - 0.66 |
| | | HLH | 0.55 - 0.66 | 0.51 - 0.55 | 0.56 - 0.64 | 0.59 - 0.68 | 0.55 - 0.57 | 0.54 - 0.59 | 0.58 - 0.65 | 0.57 - 0.65 | 0.5 - 0.56 | 0.57 - 0.61 | 0.55 - 0.67 | 0.53 - 0.61 |
| | | HLL | 0.6 - 0.7 | 0.5 - 0.61 | 0.57 - 0.6 | 0.6 - 0.63 | 0.55 - 0.57 | 0.58 - 0.63 | 0.52 - 0.63 | 0.54 - 0.6 | 0.5 - 0.66 | 0.6 - 0.63 | 0.5 - 0.69 | 0.54 - 0.65 |
| | | LHH | 0.65 - 0.66 | 0.54 - 0.61 | 0.63 - 0.66 | 0.58 - 0.69 | 0.54 - 0.58 | 0.55 - 0.59 | 0.53 - 0.62 | 0.55 - 0.62 | 0.5 - 0.56 | 0.58 - 0.62 | 0.53 - 0.7 | 0.54 - 0.62 |
| | | LHL | 0.64 - 0.68 | 0.51 - 0.56 | 0.58 - 0.66 | 0.59 - 0.7 | 0.54 - 0.58 | 0.54 - 0.57 | 0.51 - 0.66 | 0.54 - 0.72 | 0.5 - 0.68 | 0.59 - 0.65 | 0.53 - 0.7 | 0.52 - 0.66 |
| | | LLH | 0.59 - 0.66 | 0.51 - 0.61 | 0.52 - 0.63 | 0.63 - 0.7 | 0.54 - 0.57 | 0.57 - 0.62 | 0.53 - 0.58 | 0.54 - 0.66 | 0.5 - 0.68 | 0.57 - 0.61 | 0.52 - 0.64 | 0.58 - 0.61 |
| | | LLL | 0.62 - 0.66 | 0.51 - 0.58 | 0.59 - 0.64 | 0.56 - 0.6 | 0.54 - 0.61 | 0.51 - 0.58 | 0.55 - 0.65 | 0.56 - 0.7 | 0.5 - 0.61 | 0.55 - 0.59 | 0.52 - 0.68 | 0.58 - 0.61 |
| CTD | Bin 64 | | 0.58 - 0.64 | 0.51 - 0.53 | 0.56 - 0.64 | 0.62 - 0.72 | 0.56 - 0.7 | 0.54 - 0.55 | 0.53 - 0.6 | 0.54 - 0.72 | 0.5 - 0.57 | 0.56 - 0.61 | 0.53 - 0.72 | 0.6 - 0.63 |
| | LOG | Sigma: 0.5 | 0.56 - 0.66 | 0.51 - 0.62 | 0.64 - 0.65 | 0.61 - 0.74 | 0.53 - 0.7 | 0.57 - 0.63 | 0.52 - 0.56 | 0.58 - 0.71 | 0.5 - 0.57 | 0.56 - 0.64 | 0.56 - 0.66 | 0.54 - 0.58 |
| | | Sigma: 1 | 0.59 - 0.64 | 0.5 - 0.54 | 0.62 - 0.63 | 0.53 - 0.74 | 0.55 - 0.6 | 0.51 - 0.57 | 0.53 - 0.62 | 0.55 - 0.62 | 0.5 - 0.61 | 0.56 - 0.64 | 0.54 - 0.64 | 0.54 - 0.57 |
| | | Sigma: 1.5 | 0.58 - 0.64 | 0.5 - 0.53 | 0.66 - 0.68 | 0.61 - 0.74 | 0.55 - 0.63 | 0.52 - 0.56 | 0.51 - 0.59 | 0.56 - 0.62 | 0.5 - 0.58 | 0.56 - 0.64 | 0.53 - 0.68 | 0.57 - 0.58 |
| | | sigma: 2 | 0.63 - 0.73 | 0.5 - 0.55 | 0.61 - 0.64 | 0.67 - 0.74 | 0.55 - 0.62 | 0.57 - 0.58 | 0.51 - 0.55 | 0.6 - 0.68 | 0.5 - 0.62 | 0.58 - 0.64 | 0.55 - 0.66 | 0.53 - 0.57 |
| | | sigma: 2.5 | 0.62 - 0.66 | 0.5 - 0.55 | 0.56 - 0.61 | 0.6 - 0.74 | 0.55 - 0.62 | 0.56 - 0.58 | 0.51 - 0.66 | 0.55 - 0.76 | 0.5 - 0.62 | 0.56 - 0.64 | 0.54 - 0.65 | 0.53 - 0.59 |
| | | sigma: 3 | 0.59 - 0.64 | 0.51 - 0.55 | 0.55 - 0.58 | 0.62 - 0.74 | 0.55 - 0.61 | 0.55 - 0.6 | 0.51 - 0.66 | 0.56 - 0.62 | 0.5 - 0.62 | 0.57 - 0.64 | 0.56 - 0.71 | 0.61 - 0.65 |
| | | sigma: 3.5 | 0.62 - 0.68 | 0.5 - 0.56 | 0.57 - 0.61 | 0.62 - 0.74 | 0.55 - 0.59 | 0.53 - 0.55 | 0.56 - 0.74 | 0.57 - 0.69 | 0.5 - 0.61 | 0.58 - 0.64 | 0.55 - 0.7 | 0.56 - 0.63 |
| | | sigma: 4 | 0.56 - 0.67 | 0.5 - 0.52 | 0.66 - 0.69 | 0.61 - 0.74 | 0.55 - 0.59 | 0.51 - 0.55 | 0.56 - 0.66 | 0.58 - 0.72 | 0.5 - 0.65 | 0.57 - 0.66 | 0.56 - 0.81 | 0.62 - 0.63 |
| | | sigma: 4.5 | 0.61 - 0.71 | 0.51 - 0.56 | 0.57 - 0.59 | 0.67 - 0.74 | 0.55 - 0.64 | 0.55 - 0.6 | 0.58 - 0.65 | 0.56 - 0.65 | 0.5 - 0.69 | 0.57 - 0.64 | 0.51 - 0.76 | 0.54 - 0.69 |
| | | Sigma: 5 | 0.61 - 0.67 | 0.51 - 0.57 | 0.59 - 0.66 | 0.58 - 0.74 | 0.55 - 0.76 | 0.56 - 0.58 | 0.58 - 0.61 | 0.52 - 0.63 | 0.5 - 0.66 | 0.55 - 0.68 | 0.54 - 0.77 | 0.54 - 0.63 |
| | Shape | | 0.59 - 0.62 | 0.51 - 0.55 | 0.54 - 0.55 | 0.6 - 0.7 | 0.6 - 0.61 | 0.54 - 0.54 | 0.56 - 0.56 | 0.57 - 0.64 | 0.53 - 0.66 | 0.56 - 0.67 | 0.54 - 0.69 | 0.56 - 0.56 |
| | WAV | HHH | 0.61 - 0.71 | 0.51 - 0.59 | 0.61 - 0.65 | 0.58 - 0.74 | 0.57 - 0.68 | 0.57 - 0.58 | 0.53 - 0.64 | 0.61 - 0.67 | 0.5 - 0.55 | 0.56 - 0.64 | 0.56 - 0.65 | 0.52 - 0.61 |
| | | HHL | 0.58 - 0.64 | 0.51 - 0.56 | 0.61 - 0.64 | 0.61 - 0.74 | 0.56 - 0.68 | 0.53 - 0.55 | 0.5 - 0.55 | 0.6 - 0.67 | 0.5 - 0.58 | 0.56 - 0.64 | 0.55 - 0.61 | 0.57 - 0.59 |
| | | HLH | 0.57 - 0.64 | 0.51 - 0.62 | 0.55 - 0.58 | 0.61 - 0.74 | 0.54 - 0.65 | 0.53 - 0.56 | 0.54 - 0.55 | 0.59 - 0.7 | 0.5 - 0.58 | 0.56 - 0.64 | 0.55 - 0.69 | 0.54 - 0.59 |
| | | HLL | 0.57 - 0.64 | 0.51 - 0.55 | 0.6 - 0.62 | 0.58 - 0.74 | 0.51 - 0.56 | 0.54 - 0.62 | 0.52 - 0.57 | 0.59 - 0.61 | 0.5 - 0.58 | 0.56 - 0.64 | 0.54 - 0.67 | 0.53 - 0.58 |
| | | LHH | 0.58 - 0.64 | 0.51 - 0.57 | 0.65 - 0.65 | 0.57 - 0.74 | 0.52 - 0.71 | 0.54 - 0.61 | 0.54 - 0.58 | 0.58 - 0.67 | 0.5 - 0.63 | 0.57 - 0.64 | 0.53 - 0.68 | 0.55 - 0.58 |
| | | LHL | 0.57 - 0.64 | 0.52 - 0.56 | 0.6 - 0.66 | 0.57 - 0.74 | 0.51 - 0.73 | 0.51 - 0.59 | 0.51 - 0.58 | 0.59 - 0.65 | 0.5 - 0.6 | 0.54 - 0.64 | 0.51 - 0.7 | 0.53 - 0.59 |
| | | LLH | 0.6 - 0.64 | 0.51 - 0.57 | 0.56 - 0.59 | 0.6 - 0.74 | 0.51 - 0.61 | 0.53 - 0.57 | 0.5 - 0.57 | 0.6 - 0.69 | 0.5 - 0.62 | 0.55 - 0.64 | 0.53 - 0.65 | 0.57 - 0.59 |
| | | LLL | 0.58 - 0.66 | 0.5 - 0.55 | 0.65 - 0.73 | 0.61 - 0.74 | 0.56 - 0.65 | 0.53 - 0.55 | 0.53 - 0.62 | 0.58 - 0.65 | 0.5 - 0.58 | 0.56 - 0.64 | 0.56 - 0.69 | 0.58 - 0.6 |
| PET | Bin 64 | | 0.55 - 0.69 | 0.52 - 0.57 | 0.56 - 0.57 | 0.62 - 0.78 | 0.53 - 0.64 | 0.54 - 0.65 | 0.51 - 0.64 | 0.5 - 0.64 | 0.5 - 0.57 | 0.62 - 0.69 | 0.5 - 0.75 | 0.58 - 0.64 |
| | LOG | Sigma: 0.5 | 0.58 - 0.68 | 0.5 - 0.52 | 0.62 - 0.68 | 0.62 - 0.74 | 0.54 - 0.68 | 0.54 - 0.6 | 0.54 - 0.62 | 0.51 - 0.65 | 0.5 - 0.67 | 0.56 - 0.61 | 0.5 - 0.81 | 0.61 - 0.67 |
| | | Sigma: 1 | 0.6 - 0.68 | 0.51 - 0.56 | 0.61 - 0.71 | 0.55 - 0.74 | 0.54 - 0.68 | 0.54 - 0.57 | 0.5 - 0.61 | 0.52 - 0.69 | 0.5 - 0.61 | 0.56 - 0.6 | 0.5 - 0.76 | 0.57 - 0.6 |
| | | Sigma: 1.5 | 0.57 - 0.68 | 0.5 - 0.54 | 0.64 - 0.7 | 0.61 - 0.74 | 0.51 - 0.65 | 0.54 - 0.56 | 0.5 - 0.64 | 0.5 - 0.64 | 0.5 - 0.6 | 0.56 - 0.61 | 0.5 - 0.68 | 0.55 - 0.6 |
| | | sigma: 2 | 0.57 - 0.69 | 0.5 - 0.53 | 0.59 - 0.64 | 0.54 - 0.74 | 0.54 - 0.72 | 0.56 - 0.61 | 0.51 - 0.62 | 0.5 - 0.6 | 0.5 - 0.59 | 0.54 - 0.61 | 0.5 - 0.74 | 0.57 - 0.58 |
| | | sigma: 2.5 | 0.64 - 0.69 | 0.51 - 0.56 | 0.57 - 0.68 | 0.6 - 0.74 | 0.54 - 0.61 | 0.56 - 0.61 | 0.51 - 0.61 | 0.5 - 0.61 | 0.5 - 0.58 | 0.6 - 0.65 | 0.5 - 0.67 | 0.53 - 0.58 |
| | | sigma: 3 | 0.58 - 0.72 | 0.51 - 0.56 | 0.55 - 0.6 | 0.59 - 0.74 | 0.54 - 0.67 | 0.54 - 0.56 | 0.51 - 0.62 | 0.51 - 0.62 | 0.5 - 0.6 | 0.58 - 0.6 | 0.5 - 0.72 | 0.53 - 0.57 |
| | | sigma: 3.5 | 0.62 - 0.71 | 0.51 - 0.56 | 0.54 - 0.56 | 0.58 - 0.74 | 0.54 - 0.62 | 0.55 - 0.61 | 0.51 - 0.64 | 0.5 - 0.65 | 0.5 - 0.57 | 0.6 - 0.69 | 0.5 - 0.75 | 0.56 - 0.62 |
| | | sigma: 4 | 0.59 - 0.68 | 0.51 - 0.56 | 0.55 - 0.61 | 0.61 - 0.74 | 0.54 - 0.68 | 0.55 - 0.64 | 0.51 - 0.65 | 0.51 - 0.6 | 0.5 - 0.61 | 0.6 - 0.71 | 0.5 - 0.71 | 0.51 - 0.61 |
| | | sigma: 4.5 | 0.57 - 0.68 | 0.52 - 0.56 | 0.6 - 0.64 | 0.56 - 0.74 | 0.55 - 0.64 | 0.54 - 0.67 | 0.51 - 0.78 | 0.51 - 0.75 | 0.5 - 0.56 | 0.6 - 0.69 | 0.5 - 0.79 | 0.53 - 0.62 |
| | | Sigma: 5 | 0.67 - 0.69 | 0.51 - 0.56 | 0.55 - 0.6 | 0.58 - 0.74 | 0.55 - 0.65 | 0.54 - 0.6 | 0.51 - 0.68 | 0.5 - 0.64 | 0.5 - 0.56 | 0.62 - 0.67 | 0.5 - 0.75 | 0.55 - 0.67 |
| | Shape | | 0.59 - 0.72 | 0.51 - 0.61 | 0.57 - 0.62 | 0.6 - 0.66 | 0.57 - 0.57 | 0.56 - 0.57 | 0.53 - 0.53 | 0.54 - 0.64 | 0.51 - 0.61 | 0.6 - 0.67 | 0.53 - 0.66 | 0.53 - 0.53 |
| | WAV | HHH | 0.65 - 0.71 | 0.51 - 0.61 | 0.6 - 0.81 | 0.64 - 0.74 | 0.5 - 0.62 | 0.52 - 0.61 | 0.5 - 0.61 | 0.51 - 0.62 | 0.5 - 0.58 | 0.56 - 0.61 | 0.5 - 0.76 | 0.57 - 0.62 |
| | | HHL | 0.62 - 0.74 | 0.51 - 0.56 | 0.62 - 0.7 | 0.65 - 0.74 | 0.51 - 0.69 | 0.52 - 0.58 | 0.51 - 0.6 | 0.5 - 0.59 | 0.5 - 0.57 | 0.59 - 0.62 | 0.5 - 0.63 | 0.53 - 0.62 |
| | | HLH | 0.61 - 0.68 | 0.51 - 0.56 | 0.57 - 0.64 | 0.66 - 0.74 | 0.52 - 0.64 | 0.57 - 0.61 | 0.51 - 0.65 | 0.51 - 0.67 | 0.5 - 0.61 | 0.61 - 0.67 | 0.5 - 0.67 | 0.53 - 0.65 |
| | | HLL | 0.61 - 0.68 | 0.51 - 0.56 | 0.59 - 0.63 | 0.61 - 0.74 | 0.54 - 0.61 | 0.54 - 0.58 | 0.52 - 0.57 | 0.5 - 0.61 | 0.5 - 0.65 | 0.55 - 0.61 | 0.5 - 0.69 | 0.56 - 0.61 |
| | | LHH | 0.6 - 0.68 | 0.52 - 0.56 | 0.61 - 0.74 | 0.63 - 0.74 | 0.51 - 0.55 | 0.54 - 0.54 | 0.51 - 0.56 | 0.5 - 0.65 | 0.5 - 0.6 | 0.56 - 0.61 | 0.5 - 0.68 | 0.53 - 0.56 |
| | | LHL | 0.61 - 0.68 | 0.51 - 0.53 | 0.58 - 0.6 | 0.58 - 0.74 | 0.52 - 0.6 | 0.56 - 0.6 | 0.51 - 0.63 | 0.5 - 0.56 | 0.5 - 0.58 | 0.57 - 0.61 | 0.5 - 0.61 | 0.57 - 0.6 |
| | | LLH | 0.61 - 0.68 | 0.51 - 0.56 | 0.58 - 0.62 | 0.58 - 0.74 | 0.52 - 0.72 | 0.53 - 0.67 | 0.51 - 0.6 | 0.51 - 0.56 | 0.5 - 0.6 | 0.56 - 0.61 | 0.5 - 0.67 | 0.52 - 0.54 |
| | | LLL | 0.6 - 0.69 | 0.51 - 0.56 | 0.57 - 0.62 | 0.64 - 0.74 | 0.53 - 0.64 | 0.54 - 0.54 | 0.52 - 0.64 | 0.52 - 0.72 | 0.5 - 0.56 | 0.57 - 0.6 | 0.5 - 0.74 | 0.54 - 0.58 |

**Supplemental Table 11.** KRAS Classifier Mean+Sd different modality, preprocessing and setting

| Modality | Preprocessed | Setting | AB | BAG | BNB | DT | GNB | KNN | LREG | MLP | QDA | RF | SGD | SVM |
|---|---|---|---|---|---|---|---|---|---|---|---|---|---|---|
| **CT** | **Bin 64** | | 0.61 ± 0.02 | 0.55 ± 0.038 | 0.57 ± 0.019 | 0.63 ± 0.022 | 0.58 ± 0.04 | 0.54 ± 0.03 | 0.56 ± 0.017 | 0.6 ± 0.033 | 0.53 ± 0.048 | 0.57 ± 0.024 | 0.61 ± 0.08 | 0.59 ± 0.023 |
| | **LOG** | **Sigma: 0.5** | 0.62 ± 0.044 | 0.51 ± 0.015 | 0.58 ± 0.012 | 0.61 ± 0.035 | 0.6 ± 0.049 | 0.55 ± 0.0076 | 0.58 ± 0.013 | 0.59 ± 0.018 | 0.54 ± 0.045 | 0.59 ± 0.014 | 0.62 ± 0.061 | 0.57 ± 0.015 |
| | | **Sigma: 1** | 0.66 ± 0.011 | 0.53 ± 0.019 | 0.6 ± 0.024 | 0.61 ± 0.023 | 0.56 ± 0.028 | 0.56 ± 0.0034 | 0.56 ± 0.028 | 0.59 ± 0.033 | 0.53 ± 0.039 | 0.57 ± 0.015 | 0.62 ± 0.066 | 0.54 ± 0.016 |
| | | **Sigma: 1.5** | 0.65 ± 0.0071 | 0.56 ± 0.027 | 0.59 ± 0.048 | 0.64 ± 0.038 | 0.56 ± 0.024 | 0.57 ± 0.019 | 0.55 ± 0.021 | 0.57 ± 0.023 | 0.56 ± 0.06 | 0.59 ± 0.023 | 0.61 ± 0.07 | 0.56 ± 0.017 |
| | | **sigma: 2** | 0.63 ± 0.032 | 0.54 ± 0.014 | 0.55 ± 0.032 | 0.61 ± 0 | 0.57 ± 0.018 | 0.57 ± 0.022 | 0.54 ± 0.026 | 0.58 ± 0.034 | 0.53 ± 0.039 | 0.59 ± 0.024 | 0.61 ± 0.077 | 0.56 ± 0.02 |
| | | **sigma: 2.5** | 0.64 ± 0.027 | 0.6 ± 0.011 | 0.58 ± 0.051 | 0.62 ± 0.026 | 0.57 ± 0.031 | 0.59 ± 0.028 | 0.56 ± 0.04 | 0.58 ± 0.018 | 0.54 ± 0.048 | 0.6 ± 0.018 | 0.59 ± 0.062 | 0.58 ± 0.018 |
| | | **sigma: 3** | 0.64 ± 0.019 | 0.55 ± 0.025 | 0.62 ± 0.019 | 0.6 ± 0.018 | 0.57 ± 0.027 | 0.54 ± 0.013 | 0.57 ± 0.027 | 0.64 ± 0.04 | 0.57 ± 0.072 | 0.59 ± 0.019 | 0.61 ± 0.08 | 0.66 ± 0.032 |
| | | **sigma: 3.5** | 0.64 ± 0.02 | 0.55 ± 0.025 | 0.64 ± 0.027 | 0.62 ± 0.025 | 0.57 ± 0.033 | 0.56 ± 0.01 | 0.68 ± 0.12 | 0.67 ± 0.062 | 0.54 ± 0.062 | 0.59 ± 0.035 | 0.66 ± 0.12 | 0.79 ± 0.021 |
| | | **sigma: 4** | 0.68 ± 0.033 | 0.54 ± 0.015 | 0.64 ± 0.044 | 0.65 ± 0.035 | 0.58 ± 0.04 | 0.64 ± 0.096 | 0.57 ± 0.056 | 0.64 ± 0.073 | 0.57 ± 0.084 | 0.59 ± 0.019 | 0.63 ± 0.094 | 0.63 ± 0.064 |
| | | **sigma: 4.5** | 0.64 ± 0.033 | 0.58 ± 0.049 | 0.61 ± 0.061 | 0.61 ± 0.021 | 0.6 ± 0.055 | 0.59 ± 0.042 | 0.55 ± 0.023 | 0.64 ± 0.07 | 0.53 ± 0.04 | 0.59 ± 0.0053 | 0.61 ± 0.066 | 0.66 ± 0.037 |
| | | **Sigma: 5** | 0.68 ± 0.029 | 0.52 ± 0.0095 | 0.61 ± 0.026 | 0.68 ± 0.035 | 0.58 ± 0.05 | 0.57 ± 0.023 | 0.57 ± 0.045 | 0.57 ± 0.036 | 0.54 ± 0.045 | 0.65 ± 0.053 | 0.64 ± 0.13 | 0.66 ± 0.034 |
| | **Shape** | | 0.64 ± 0.033 | 0.55 ± 0.022 | 0.57 ± 0.022 | 0.65 ± 0.016 | 0.59 ± 0.007 | 0.57 ± 0.0068 | 0.54 ± 0.029 | 0.62 ± 0.059 | 0.57 ± 0.052 | 0.61 ± 0.024 | 0.6 ± 0.063 | 0.53 ± 0.0072 |
| | **WAV** | **HHH** | 0.64 ± 0.03 | 0.54 ± 0.023 | 0.65 ± 0.03 | 0.66 ± 0.023 | 0.57 ± 0.032 | 0.56 ± 0.0014 | 0.55 ± 0.019 | 0.65 ± 0.052 | 0.53 ± 0.03 | 0.59 ± 0.02 | 0.61 ± 0.027 | 0.53 ± 0.022 |
| | | **HHL** | 0.63 ± 0.034 | 0.54 ± 0.022 | 0.6 ± 0.0049 | 0.66 ± 0.023 | 0.6 ± 0.046 | 0.56 ± 0.012 | 0.59 ± 0.024 | 0.59 ± 0.039 | 0.57 ± 0.086 | 0.62 ± 0.016 | 0.61 ± 0.071 | 0.64 ± 0.025 |
| | | **HLH** | 0.63 ± 0.04 | 0.54 ± 0.018 | 0.59 ± 0.034 | 0.62 ± 0.035 | 0.55 ± 0.0092 | 0.58 ± 0.022 | 0.6 ± 0.025 | 0.61 ± 0.024 | 0.52 ± 0.024 | 0.59 ± 0.014 | 0.6 ± 0.043 | 0.58 ± 0.03 |
| | | **HLL** | 0.67 ± 0.039 | 0.56 ± 0.04 | 0.58 ± 0.013 | 0.62 ± 0.01 | 0.56 ± 0.0081 | 0.59 ± 0.019 | 0.59 ± 0.043 | 0.57 ± 0.021 | 0.56 ± 0.071 | 0.61 ± 0.013 | 0.6 ± 0.081 | 0.62 ± 0.046 |
| | | **LHH** | 0.65 ± 0.0059 | 0.56 ± 0.025 | 0.65 ± 0.011 | 0.63 ± 0.04 | 0.55 ± 0.017 | 0.57 ± 0.02 | 0.58 ± 0.044 | 0.6 ± 0.032 | 0.53 ± 0.03 | 0.6 ± 0.016 | 0.63 ± 0.059 | 0.6 ± 0.035 |
| | | **LHL** | 0.65 ± 0.014 | 0.54 ± 0.02 | 0.61 ± 0.036 | 0.65 ± 0.042 | 0.55 ± 0.016 | 0.56 ± 0.014 | 0.59 ± 0.05 | 0.62 ± 0.06 | 0.53 ± 0.07 | 0.62 ± 0.018 | 0.63 ± 0.069 | 0.61 ± 0.055 |
| | | **LLH** | 0.64 ± 0.027 | 0.56 ± 0.033 | 0.55 ± 0.053 | 0.66 ± 0.031 | 0.56 ± 0.011 | 0.59 ± 0.022 | 0.55 ± 0.025 | 0.6 ± 0.046 | 0.54 ± 0.068 | 0.6 ± 0.012 | 0.6 ± 0.055 | 0.6 ± 0.011 |
| | | **LLL** | 0.65 ± 0.015 | 0.55 ± 0.03 | 0.61 ± 0.021 | 0.58 ± 0.014 | 0.58 ± 0.033 | 0.56 ± 0.029 | 0.6 ± 0.04 | 0.61 ± 0.058 | 0.53 ± 0.042 | 0.57 ± 0.015 | 0.61 ± 0.069 | 0.59 ± 0.011 |
| **CTD** | **Bin 64** | | 0.61 ± 0.023 | 0.52 ± 0.01 | 0.59 ± 0.04 | 0.66 ± 0.036 | 0.61 ± 0.051 | 0.54 ± 0.005 | 0.57 ± 0.029 | 0.61 ± 0.066 | 0.53 ± 0.037 | 0.58 ± 0.018 | 0.62 ± 0.069 | 0.62 ± 0.015 |
| | **LOG** | **Sigma: 0.5** | 0.61 ± 0.034 | 0.56 ± 0.039 | 0.64 ± 0.0068 | 0.68 ± 0.049 | 0.62 ± 0.075 | 0.61 ± 0.028 | 0.54 ± 0.014 | 0.64 ± 0.054 | 0.53 ± 0.031 | 0.58 ± 0.027 | 0.62 ± 0.038 | 0.55 ± 0.016 |
| | | **Sigma: 1** | 0.61 ± 0.015 | 0.52 ± 0.014 | 0.63 ± 0.0025 | 0.64 ± 0.088 | 0.58 ± 0.017 | 0.53 ± 0.021 | 0.57 ± 0.033 | 0.59 ± 0.027 | 0.53 ± 0.043 | 0.58 ± 0.027 | 0.59 ± 0.041 | 0.56 ± 0.016 |
| | | **Sigma: 1.5** | 0.61 ± 0.019 | 0.52 ± 0.012 | 0.67 ± 0.0068 | 0.68 ± 0.048 | 0.59 ± 0.035 | 0.55 ± 0.022 | 0.53 ± 0.027 | 0.59 ± 0.027 | 0.53 ± 0.033 | 0.58 ± 0.029 | 0.58 ± 0.05 | 0.57 ± 0.0076 |
| | | **sigma: 2** | 0.66 ± 0.038 | 0.53 ± 0.019 | 0.62 ± 0.011 | 0.7 ± 0.027 | 0.59 ± 0.033 | 0.58 ± 0.0069 | 0.54 ± 0.014 | 0.64 ± 0.027 | 0.53 ± 0.047 | 0.59 ± 0.021 | 0.61 ± 0.038 | 0.55 ± 0.014 |
| | | **sigma: 2.5** | 0.63 ± 0.012 | 0.53 ± 0.019 | 0.57 ± 0.027 | 0.66 ± 0.059 | 0.59 ± 0.03 | 0.57 ± 0.0076 | 0.55 ± 0.054 | 0.62 ± 0.074 | 0.55 ± 0.058 | 0.59 ± 0.026 | 0.6 ± 0.044 | 0.56 ± 0.027 |
| | | **sigma: 3** | 0.61 ± 0.016 | 0.53 ± 0.019 | 0.56 ± 0.014 | 0.67 ± 0.052 | 0.58 ± 0.02 | 0.57 ± 0.02 | 0.57 ± 0.052 | 0.58 ± 0.024 | 0.54 ± 0.051 | 0.59 ± 0.026 | 0.65 ± 0.054 | 0.62 ± 0.013 |
| | | **sigma: 3.5** | 0.63 ± 0.024 | 0.53 ± 0.025 | 0.6 ± 0.02 | 0.67 ± 0.054 | 0.56 ± 0.016 | 0.54 ± 0.0059 | 0.6 ± 0.073 | 0.63 ± 0.056 | 0.53 ± 0.042 | 0.6 ± 0.02 | 0.63 ± 0.056 | 0.59 ± 0.038 |
| | | **sigma: 4** | 0.62 ± 0.038 | 0.51 ± 0.0063 | 0.68 ± 0.018 | 0.67 ± 0.056 | 0.58 ± 0.02 | 0.53 ± 0.013 | 0.6 ± 0.035 | 0.64 ± 0.044 | 0.56 ± 0.065 | 0.61 ± 0.034 | 0.71 ± 0.08 | 0.62 ± 0.0055 |
| | | **sigma: 4.5** | 0.66 ± 0.04 | 0.53 ± 0.016 | 0.58 ± 0.0078 | 0.69 ± 0.034 | 0.59 ± 0.034 | 0.57 ± 0.027 | 0.61 ± 0.023 | 0.61 ± 0.037 | 0.55 ± 0.074 | 0.59 ± 0.025 | 0.64 ± 0.086 | 0.62 ± 0.05 |
| | | **Sigma: 5** | 0.64 ± 0.02 | 0.54 ± 0.02 | 0.64 ± 0.032 | 0.66 ± 0.066 | 0.6 ± 0.079 | 0.57 ± 0.011 | 0.59 ± 0.0098 | 0.6 ± 0.039 | 0.55 ± 0.064 | 0.61 ± 0.045 | 0.66 ± 0.083 | 0.61 ± 0.039 |
| | **Shape** | | 0.61 ± 0.0091 | 0.53 ± 0.017 | 0.55 ± 0.0055 | 0.68 ± 0.04 | 0.6 ± 0.0055 | 0.54 ± 0 | 0.56 ± 0 | 0.59 ± 0.028 | 0.57 ± 0.044 | 0.59 ± 0.042 | 0.61 ± 0.049 | 0.56 ± 0 |
| | **WAV** | **HHH** | 0.64 ± 0.042 | 0.56 ± 0.03 | 0.64 ± 0.02 | 0.66 ± 0.066 | 0.63 ± 0.054 | 0.58 ± 0.0055 | 0.57 ± 0.041 | 0.64 ± 0.025 | 0.52 ± 0.021 | 0.59 ± 0.033 | 0.62 ± 0.03 | 0.58 ± 0.032 |
| | | **HHL** | 0.61 ± 0.019 | 0.53 ± 0.021 | 0.61 ± 0.014 | 0.67 ± 0.056 | 0.63 ± 0.061 | 0.54 ± 0.0082 | 0.53 ± 0.022 | 0.63 ± 0.025 | 0.53 ± 0.031 | 0.59 ± 0.033 | 0.6 ± 0.024 | 0.58 ± 0.011 |
| | | **HLH** | 0.61 ± 0.021 | 0.57 ± 0.038 | 0.56 ± 0.016 | 0.67 ± 0.052 | 0.6 ± 0.051 | 0.54 ± 0.014 | 0.54 ± 0.0067 | 0.64 ± 0.041 | 0.53 ± 0.031 | 0.58 ± 0.028 | 0.6 ± 0.054 | 0.57 ± 0.019 |
| | | **HLL** | 0.61 ± 0.021 | 0.53 ± 0.015 | 0.61 ± 0.01 | 0.66 ± 0.065 | 0.55 ± 0.018 | 0.59 ± 0.037 | 0.55 ± 0.017 | 0.6 ± 0.0084 | 0.53 ± 0.034 | 0.59 ± 0.027 | 0.62 ± 0.05 | 0.56 ± 0.019 |
| | | **LHH** | 0.61 ± 0.019 | 0.54 ± 0.017 | 0.65 ± 0 | 0.66 ± 0.07 | 0.62 ± 0.082 | 0.59 ± 0.029 | 0.56 ± 0.014 | 0.64 ± 0.032 | 0.54 ± 0.049 | 0.59 ± 0.024 | 0.61 ± 0.059 | 0.57 ± 0.01 |
| | | **LHL** | 0.61 ± 0.022 | 0.55 ± 0.016 | 0.62 ± 0.03 | 0.65 ± 0.069 | 0.62 ± 0.096 | 0.55 ± 0.032 | 0.54 ± 0.029 | 0.61 ± 0.024 | 0.53 ± 0.041 | 0.58 ± 0.033 | 0.61 ± 0.071 | 0.55 ± 0.023 |
| | | **LLH** | 0.61 ± 0.013 | 0.53 ± 0.021 | 0.58 ± 0.015 | 0.67 ± 0.055 | 0.56 ± 0.042 | 0.54 ± 0.012 | 0.54 ± 0.033 | 0.64 ± 0.033 | 0.55 ± 0.055 | 0.58 ± 0.03 | 0.59 ± 0.04 | 0.58 ± 0.01 |
| | | **LLL** | 0.61 ± 0.03 | 0.53 ± 0.018 | 0.67 ± 0.039 | 0.69 ± 0.047 | 0.61 ± 0.031 | 0.55 ± 0.014 | 0.58 ± 0.041 | 0.61 ± 0.028 | 0.53 ± 0.032 | 0.59 ± 0.027 | 0.64 ± 0.046 | 0.59 ± 0.0066 |
| **PET** | **Bin 64** | | 0.62 ± 0.057 | 0.55 ± 0.019 | 0.56 ± 0.0082 | 0.68 ± 0.058 | 0.58 ± 0.042 | 0.56 ± 0.045 | 0.59 ± 0.057 | 0.6 ± 0.068 | 0.53 ± 0.031 | 0.66 ± 0.024 | 0.63 ± 0.11 | 0.62 ± 0.026 |
| | **LOG** | **Sigma: 0.5** | 0.62 ± 0.043 | 0.51 ± 0.0079 | 0.65 ± 0.027 | 0.66 ± 0.043 | 0.6 ± 0.052 | 0.55 ± 0.022 | 0.57 ± 0.031 | 0.6 ± 0.059 | 0.56 ± 0.068 | 0.6 ± 0.019 | 0.66 ± 0.13 | 0.62 ± 0.022 |
| | | **Sigma: 1** | 0.64 ± 0.029 | 0.53 ± 0.017 | 0.68 ± 0.044 | 0.64 ± 0.068 | 0.59 ± 0.054 | 0.55 ± 0.011 | 0.57 ± 0.044 | 0.59 ± 0.065 | 0.54 ± 0.048 | 0.58 ± 0.015 | 0.66 ± 0.12 | 0.59 ± 0.0095 |
| | | **Sigma: 1.5** | 0.63 ± 0.036 | 0.52 ± 0.015 | 0.67 ± 0.026 | 0.66 ± 0.046 | 0.59 ± 0.057 | 0.55 ± 0.0063 | 0.58 ± 0.054 | 0.58 ± 0.062 | 0.53 ± 0.036 | 0.59 ± 0.018 | 0.59 ± 0.075 | 0.59 ± 0.019 |
| | | **sigma: 2** | 0.64 ± 0.053 | 0.52 ± 0.01 | 0.61 ± 0.023 | 0.64 ± 0.069 | 0.6 ± 0.068 | 0.57 ± 0.029 | 0.56 ± 0.041 | 0.56 ± 0.041 | 0.53 ± 0.036 | 0.59 ± 0.025 | 0.6 ± 0.094 | 0.57 ± 0.004 |
| | | **sigma: 2.5** | 0.65 ± 0.022 | 0.52 ± 0.018 | 0.63 ± 0.043 | 0.65 ± 0.052 | 0.58 ± 0.038 | 0.57 ± 0.023 | 0.55 ± 0.034 | 0.57 ± 0.043 | 0.53 ± 0.035 | 0.63 ± 0.022 | 0.59 ± 0.078 | 0.55 ± 0.02 |
| | | **sigma: 3** | 0.64 ± 0.051 | 0.54 ± 0.019 | 0.57 ± 0.031 | 0.66 ± 0.056 | 0.6 ± 0.053 | 0.55 ± 0.0089 | 0.55 ± 0.042 | 0.57 ± 0.045 | 0.53 ± 0.037 | 0.59 ± 0.0057 | 0.62 ± 0.1 | 0.56 ± 0.015 |
| | | **sigma: 3.5** | 0.66 ± 0.032 | 0.54 ± 0.021 | 0.55 ± 0.013 | 0.65 ± 0.055 | 0.59 ± 0.036 | 0.56 ± 0.024 | 0.59 ± 0.06 | 0.58 ± 0.057 | 0.54 ± 0.03 | 0.64 ± 0.036 | 0.64 ± 0.11 | 0.58 ± 0.024 |
| | | **sigma: 4** | 0.63 ± 0.032 | 0.54 ± 0.019 | 0.57 ± 0.028 | 0.66 ± 0.044 | 0.6 ± 0.054 | 0.56 ± 0.037 | 0.55 ± 0.056 | 0.55 ± 0.03 | 0.53 ± 0.043 | 0.63 ± 0.043 | 0.61 ± 0.094 | 0.58 ± 0.039 |
| | | **sigma: 4.5** | 0.64 ± 0.038 | 0.54 ± 0.02 | 0.62 ± 0.017 | 0.65 ± 0.063 | 0.6 ± 0.04 | 0.56 ± 0.05 | 0.6 ± 0.099 | 0.59 ± 0.086 | 0.53 ± 0.024 | 0.64 ± 0.033 | 0.62 ± 0.11 | 0.6 ± 0.042 |
| | | **Sigma: 5** | 0.68 ± 0.012 | 0.54 ± 0.021 | 0.58 ± 0.024 | 0.65 ± 0.053 | 0.59 ± 0.04 | 0.55 ± 0.022 | 0.58 ± 0.068 | 0.6 ± 0.067 | 0.52 ± 0.02 | 0.65 ± 0.016 | 0.63 ± 0.11 | 0.62 ± 0.043 |
| | **Shape** | | 0.64 ± 0.051 | 0.56 ± 0.033 | 0.61 ± 0.026 | 0.63 ± 0.019 | 0.57 ± 0 | 0.56 ± 0.0069 | 0.53 ± 0.0019 | 0.58 ± 0.042 | 0.56 ± 0.033 | 0.63 ± 0.029 | 0.61 ± 0.044 | 0.53 ± 0.0022 |
| | **WAV** | **HHH** | 0.68 ± 0.026 | 0.57 ± 0.047 | 0.75 ± 0.1 | 0.68 ± 0.034 | 0.57 ± 0.047 | 0.56 ± 0.038 | 0.56 ± 0.044 | 0.58 ± 0.048 | 0.52 ± 0.033 | 0.59 ± 0.018 | 0.65 ± 0.12 | 0.6 ± 0.019 |
| | | **HHL** | 0.67 ± 0.04 | 0.53 ± 0.02 | 0.67 ± 0.036 | 0.68 ± 0.034 | 0.62 ± 0.083 | 0.53 ± 0.024 | 0.56 ± 0.036 | 0.56 ± 0.041 | 0.53 ± 0.036 | 0.61 ± 0.013 | 0.58 ± 0.062 | 0.58 ± 0.039 |
| | | **HLH** | 0.64 ± 0.028 | 0.52 ± 0.017 | 0.63 ± 0.036 | 0.68 ± 0.031 | 0.58 ± 0.046 | 0.58 ± 0.017 | 0.6 ± 0.068 | 0.59 ± 0.068 | 0.55 ± 0.052 | 0.63 ± 0.025 | 0.59 ± 0.073 | 0.61 ± 0.06 |
| | | **HLL** | 0.64 ± 0.025 | 0.52 ± 0.017 | 0.62 ± 0.018 | 0.66 ± 0.047 | 0.58 ± 0.03 | 0.55 ± 0.017 | 0.55 ± 0.022 | 0.57 ± 0.038 | 0.54 ± 0.059 | 0.58 ± 0.027 | 0.59 ± 0.081 | 0.58 ± 0.023 |
| | | **LHH** | 0.63 ± 0.029 | 0.53 ± 0.014 | 0.68 ± 0.054 | 0.68 ± 0.036 | 0.53 ± 0.016 | 0.54 ± 0.0012 | 0.53 ± 0.02 | 0.55 ± 0.055 | 0.53 ± 0.037 | 0.58 ± 0.02 | 0.6 ± 0.082 | 0.54 ± 0.014 |
| | | **LHL** | 0.64 ± 0.024 | 0.52 ± 0.0095 | 0.59 ± 0.0087 | 0.65 ± 0.058 | 0.57 ± 0.035 | 0.56 ± 0.017 | 0.54 ± 0.045 | 0.54 ± 0.023 | 0.53 ± 0.034 | 0.59 ± 0.017 | 0.56 ± 0.045 | 0.58 ± 0.011 |
| | | **LLH** | 0.64 ± 0.025 | 0.55 ± 0.02 | 0.61 ± 0.016 | 0.65 ± 0.055 | 0.6 ± 0.088 | 0.64 ± 0.055 | 0.53 ± 0.037 | 0.54 ± 0.018 | 0.53 ± 0.036 | 0.59 ± 0.023 | 0.61 ± 0.083 | 0.53 ± 0.0099 |
| | | **LLL** | 0.64 ± 0.035 | 0.52 ± 0.018 | 0.6 ± 0.022 | 0.68 ± 0.036 | 0.57 ± 0.04 | 0.54 ± 0.0011 | 0.56 ± 0.043 | 0.61 ± 0.074 | 0.52 ± 0.024 | 0.59 ± 0.011 | 0.63 ± 0.1 | 0.57 ± 0.016 |

# Discussion

Accurate detection of EGFR/KRAS mutation status in NSCLCs allows improved selection of patients for effective therapeutic strategies (2). Although bio-techniques provide intra- and inter-tumor EGFR/KRAS mutation status in patients, they are not entirely feasible and suffer from limitations (38). In the present study, we developed a comprehensive radiomics framework that calculates radiomic features from low-dose CT, diagnostic CT, and PET images to predict EGFR and KRAS mutation status using univariate and multivariate ML algorithms in NSCLCs patients. In our radiogenomics study, several radiomics models were found as predictive for EGFR and KRAS mutation status, which may be utilized for non-invasive cost-effective assessment.

In recent years, radiogenomics—the correlation of imaging features with genomic parameters—have garnered significant interest, as non-invasive assessment frameworks to tailor therapy based on imaging biomarkers and underlying biological pathways (7, 8, 39). In our work, we observed a wide range of predictive model performances in different radiomics features, feature selection and classification algorithms, and for different imaging modalities. For EGFR and KRAS mutation status prediction univariate analysis of radiomics features and cross combination of imaging/validation /feature selection/Classification as multivariate analysis methods resulted in a wide range of performance (AUC: 0.5~0.83). The best predictive power of conventional PET parameters was achieved by $SUV_{peak}$ (AUC: 0.69, P-value = 0.0002) and MTV (AUC: 0.55, P-value = 0.0011) for EGFR and KRAS. Univariate analysis of extracted radiomics features improved prediction power up to AUC: 75 (q-value: 0.003, Short Run Emphasis feature of GLRLM from LOG preprocessed image of PET with sigma value 1.5) and AUC: 0.71 (q-value 0.00005, The Large Dependence Low Gray Level Emphasis from GLDM in LOG preprocessed image of CTD sigma value 5) for EGFR and KRAS, respectively. Furthermore, the machine learning algorithm improved the perdition power of gen status up to AUC: 0.82 for EGFR (LOG preprocessed of PET image set with sigma 3 with VT feature selector and SGD classifier) and AUC: 0.83 for KRAS (CT image set with sigma 3.5 with SM feature selector and SGD classifier). Using radiomics feature and machine learning algorithm outperformed conventional methods in perdition of EGFR and KRAS gene status in NSCLC patients.

The role of feature selection is more prominent in cross-combination of feature selection and image-sets for EGFR mutation status prediction, and the Random Forest (RF) and Support Vector Machine (SVM) method has shown to yield a higher performance in EGFR and KRAS predation. For classification, the selection of an appropriate algorithm has a more prominent role. Zhao *et al.* (40) developed a conventional radiomics model and 3D deep learning system to predict EGFR-mutant pulmonary adenocarcinoma in CT images, the best AUC were reported by 75.8 using deep learning, however, in present study the of EGFR status achieved to AUC: 0.82 which shows the role of different combination of image set, feature selector and classifier. Parmer et al. (24) compared 14 feature selection and 12 classification methods in terms of their performance and stability, and found that the Wilcoxon rank-sum test and random forest (RF) had the highest prognostic performance for feature selection and classification, respectively, while demonstrating high stability against data perturbation. Moreover, Zhang *et al.* (41) built a 54-cross combination ML algorithms framework including six feature selections and nine classification methods for survival prediction of advanced nasopharyngeal carcinoma. They found that using RF for both feature selection and classification had the highest prognostic performance, followed by RF+AB and sure independence screening (SIS) + linear support vector machines (SVMs). Abdollahi *et al. (42)* studied different combination of featrue selector and classifier for predition of intensity-modulated radiation therapy response, Gleason score and stage in prostate cancer and they showed different task of classification can be achived by different combination of feature selector and classifiers. In addition, we compared the predictive power of conventional PET parameters and radiomic features. In our radiogenomics study, we showed that radiomics features and machine learning based radiomic models are more predictive than conventional parameters.

Radiomics/genomics studies suffer from several challenges, and a robust framework for clinical decision making is highly desired (7, 43, 44). As an approved guideline, standardization efforts (IBSI in particular)(36) have sought to address the challenge of reproducing and validating reported findings by comparing and standardizing definitions and implementation of several image-feature sets between participating institutions. The IBSI also provides image biomarker nomenclature and definitions, benchmark data sets, and benchmark values to verify image processing and image biomarker calculations, as well as reporting guidelines, for high-throughput image analysis (36). In this study, we followed the IBSI protocol to resample all images to a constant voxel size to reduce feature sensitivity to variable image generation parameters (36), also ensuring the robustness and reproducibility of the study (45). Having a dataset of images generated using varying data acquisition (46, 47), reconstruction (48),processing (49) and segmentation (50) results produce inconsistencies in feature evaluation.

The limited size of a dataset is a limiting factor in radiomics studies, as the analysis with many radiomic features and only a few imaging data is prone to over-fitting. We tuned our models using 10-fold cross-validation to reduce the sensitivity of our results to input data and repeated for 20 times to make our results more reliable and all model evaluation were performed on 63 patients' independent validation set which never used in training process.  Our presented results were developed using a standardized radiomics analysis workflow that showed strong and significant predictability of gene mutations. Specifically, in the feature selection phase, we used ML algorithms to reduce the number of features in order to decrease the dimensionality and reduce over-fitting.

# Conclusion

We demonstrated that EGFR and KRAS mutation status in NSCLC patients can be predicted via a non-invasive and reliable radiomics analysis. We evaluated several different machine learning methods to find optimal methods for radiogenomics analyses. Overall, we showed that radiomic features extracted from different imaging modalities (alone or in combination) could be used for successful prediction of EGFR and KRAS mutation status, having more predictive power than conventional imaging biomarkers.